\documentclass[pra,twocolumn,showpacs,a4paper,superscriptaddress]{revtex4-1}
\usepackage{epsfig}
\usepackage{amsmath}
\usepackage{graphicx}
\usepackage{hyperref}
\usepackage{bbm}
\usepackage{xcolor}
\usepackage{latexsym}
\usepackage{amssymb}
\usepackage{braket}

\definecolor{mygreen}{rgb}{0,0.5,0}
\definecolor{myblue}{rgb}{0,0,0.75}
\definecolor{mymagenta}{cmyk}{0,1,0,0.12}

\usepackage{umoline}
\newcommand{\removed}[1]{}

\begin{document}

\title{Analog Quantum Simulation of $(1+1)$D Lattice QED with Trapped Ions}

\author{Dayou Yang}
\email{dayou.yang@uibk.ac.at}
\affiliation{Institute for Quantum Optics and Quantum Information, 
Austrian Academy of Sciences, Technikerstrasse 21a, 6020 Innsbruck, Austria}
\affiliation{Institute for Theoretical Physics, University of Innsbruck, Technikerstrasse 25, 6020 Innsbruck, Austria}

\author{Gouri Shankar Giri}
\affiliation{Department Physik, Naturwissenschaftlich-Technische Fakult{\"a}t, Universit{\"a}t Siegen, 57068 Siegen, Germany}

\author{Michael Johanning}
\affiliation{Department Physik, Naturwissenschaftlich-Technische Fakult{\"a}t, Universit{\"a}t Siegen, 57068 Siegen, Germany}

\author{Christof Wunderlich}
\affiliation{Department Physik, Naturwissenschaftlich-Technische Fakult{\"a}t, Universit{\"a}t Siegen, 57068 Siegen, Germany}

\author{Peter Zoller}
\affiliation{Institute for Quantum Optics and Quantum Information, 
Austrian Academy of Sciences, Technikerstrasse 21a, 6020 Innsbruck, Austria}
\affiliation{Institute for Theoretical Physics, University of Innsbruck, Technikerstrasse 25, 6020 Innsbruck, Austria}

\author{Philipp Hauke}
\affiliation{Institute for Quantum Optics and Quantum Information, 
Austrian Academy of Sciences, Technikerstrasse 21a, 6020 Innsbruck, Austria}
\affiliation{Institute for Theoretical Physics, University of Innsbruck, Technikerstrasse 25, 6020 Innsbruck, Austria}

\date{\today}

\begin{abstract}
The prospect of quantum simulating lattice gauge theories opens exciting possibilities for understanding fundamental forms of matter. Here, we show that trapped ions represent a promising platform in this context when simultaneously exploiting internal pseudo-spins and external phonon vibrations. We illustrate our ideas with two complementary proposals for simulating lattice-regularized quantum electrodynamics (QED) in $(1+1)$ space-time dimensions. The first scheme replaces the gauge fields by local vibrations with a high occupation number. By numerical finite-size scaling, we demonstrate that this model recovers Wilson's lattice gauge theory in a controlled way.  Its implementation can be scaled up to tens of ions in an array of micro-traps. The second scheme represents the gauge fields by spins 1/2, and thus simulates a quantum link model. As we show, this allows the fermionic matter to be replaced by bosonic degrees of freedom, permitting small-scale implementations in a linear Paul trap. Both schemes work on energy scales significantly larger than typical decoherence rates in experiments, thus enabling the investigation of phenomena such as string breaking, Coleman's quantum phase transition, and false-vacuum decay. The underlying ideas of the proposed analog simulation schemes may also be adapted to other platforms, such as superconducting qubits.   
\end{abstract}


\maketitle

\section{Introduction}
\label{sec1}

Quantum-optical setups, with their high controllability, provide an ideal means to realize quantum simulators \cite{Cirac2012}, i.e., engineered quantum-mechanical systems that mimic a desired dynamics which would be difficult to access on a classical computer. 
Particularly attractive targets for quantum simulation are lattice gauge theories (LGTs), which constitute a central framework of theoretical many-body physics. 
They describe not only the fundamental interactions between elementary particles \cite{Montvay1994,Gattringer2010} but also exotic phases of matter such as quantum spin liquids \cite{Kogut1979,Lee2006,Lacroix2010}. 
However, especially when their real-time dynamics is concerned, gauge theories are notoriously challenging to tackle on a classical computer \cite{Gattringer2010,Montvay1994}. 
For these reasons, recent years have seen a surge of interest for studying gauge theories in engineered quantum simulators \cite{Wiese2013,Wiese2014,Zohar2016,Dalmonte2016,Doucot2004,Marcos2013,Marcos2014,Mezzacapo2015,GarciaAlvarez2015,Hauke2013b}.

There exist two main difficulties to realize a gauge theory in a synthetic system, besides the design of the correct interaction terms.  
First, a gauge theory is characterized by local conservation laws, which need to be imposed on the physical quantum simulator. 
Second, to simulate the interaction between elementary particles, both bosonic and fermionic degrees of freedom (DOFs) need to be realized simultaneously (for a case where this requirement can be circumvented, see Refs.~\cite{Hamer1997,MartinezInPreparation,MuschikInPreparation}). 
Both challenges can be addressed in digital or analog quantum simulation, and in diverse quantum-optical setups, with proposals existing especially for ultracold atoms in optical lattices (see Refs.~\cite{Wiese2013,Wiese2014,Zohar2016,Dalmonte2016} for recent reviews) and superconducting qubits \cite{Doucot2004,Marcos2013,Marcos2014,Mezzacapo2015,GarciaAlvarez2015}. 
In a recent work, it has been shown how these challenges can be tackled in the well-controlled platform provided by trapped ions \cite{Hauke2013b}.  
There, it has been proposed to encode both fermionic matter and gauge degrees of freedom in pseudo-spins formed by the internal electronic states of the ions. 
The local conservation laws and correct interactions can then be transmitted between the pseudo-spins by the collective vibrational DOFs of the ions, which are eliminated in perturbation theory. 
In the present work, we pursue an alternative route where the vibrational modes are included as active DOFs, rather than eliminated in perturbation theory. 
In this way, we obtain a larger number of useful DOFs per ion, and as an additional benefit improve the relevant time scales. 
Our proposal is complementary to efforts for realizing LGTs in trapped ions via digital quantum simulation \cite{MuschikInPreparation,MartinezInPreparation}, and to recent progresses in simulating LGTs via tensor networks on classical computers~\cite{Banuls2013,Rico2014,Buyens2014,1367-2630-16-10-103015,Tagliacozzo2014,Haegeman2015,Pichler2015}.

We illustrate the versatility of our approach by two complementary schemes, which simulate a simple LGT, namely (1+1)D lattice quantum electrodynamics (QED), i.e., the massive Schwinger model \cite{Schwinger1962,Coleman1975,Coleman1976}. 
First, we introduce a representation of the gauge fields via local vibrational DOFs at high occupation number. Additionally, in one spatial dimension, the fermions are mapped to spins via the Jordan-Wigner transformation. This scheme, which we call the highly occupied boson model (HOBM), simulates a strict LGT and is scalable to large ion numbers. 
By analyzing the real-time dynamics after a quantum quench as well as by performing finite-size scalings for the ground-state phase diagram, we demonstrate that it approximates a usual Wilson LGT in a well-controlled manner. As such, it may also be interesting for realizations in other platforms, such as superconducting qubits. 
In the second scheme, which relies on the quantum link model (QLM)~\cite{Horn1981,Orland1990,Chandrasekharan1997,Brower1999}, we propose a different mapping which represents the gauge fields by internal pseudo-spins, and maps the fermionic matter to bosonic DOFs. Because this scheme has excellent time scales, but acquires systematic deviations at moderate ion numbers, it is especially suitable for small-scale proof-of-principle experiments. 
Both schemes can be implemented with one-dimensional ion chains in linear Paul traps or arrays of micro-traps, exploit only experimentally realistic abilities for controlling and coupling spins and phonons~\cite{Porras2008,Ivanov2009,Casanova2011,Porras2012,Mezzacapo2012b,Nevado2013,Ivanov2013,Kurcz2015,Nevado2015,Gerritsma2010,Gerritsma2011}, and are robust against the most common sources of imperfections. 

This paper is organized as follows. 
First, in Sec.~\ref{sec2}, we lay the background for our proposals. We describe the LGT that we aim at simulating, as well as two phenomena that we use for illustrating our ideas (Sec.~\ref{sec2A: lattice Schwinger model}). 
We also discuss on a general, abstract level two truncation schemes for the gauge fields, which are convenient for physical implementations in quantum simulators: a formalism introduced in this work, the HOBM (Sec.~\ref{sec2C: HOBM}), as well as the better known QLMs (Sec.~\ref{QLMformulation}). 
Afterwards, we discuss feasible implementation schemes as well as relevant error sources, first for the HOBM (Sec.~\ref{sec3}), and then for the QLM proposal (Sec.~\ref{sec4: QLM}). 
We conclude our work with a short summary in Sec.~\ref{sec6}.

\section{$(1+1)$D lattice QED and truncation schemes}
\label{sec2}
To illustrate our ideas, we concentrate on the simulation of QED in $1+1$ dimensional space-time. $(1+1)$D QED, also known as the Schwinger model, describes interactions between a $U(1)$ Abelian gauge field and fundamental charges (single-species fermions). Despite its simplicity, it shares much physics with $(3+1)$D $SU(3)$ quantum chromodynamics, which describes the strong interactions of the Standard Model, such as confinement, nontrivial $\theta$ vacuum, or chiral symmetry breaking and anomaly~\cite{Coleman1975,Coleman1976}. The Schwinger model has thus become a test bed for new techniques devised to study gauge theories~\cite{Banks1976,Hamer1982,Hamer1997,Byrnes2002}. We shall present first an introduction to its standard formulation on a lattice. Afterwards, we describe two truncation schemes for the gauge fields, which approach lattice QED in a given limit and are specifically suited for building analog quantum simulators.

\begin{figure*}[t]
\begin{center}
\includegraphics[width=0.95\textwidth]{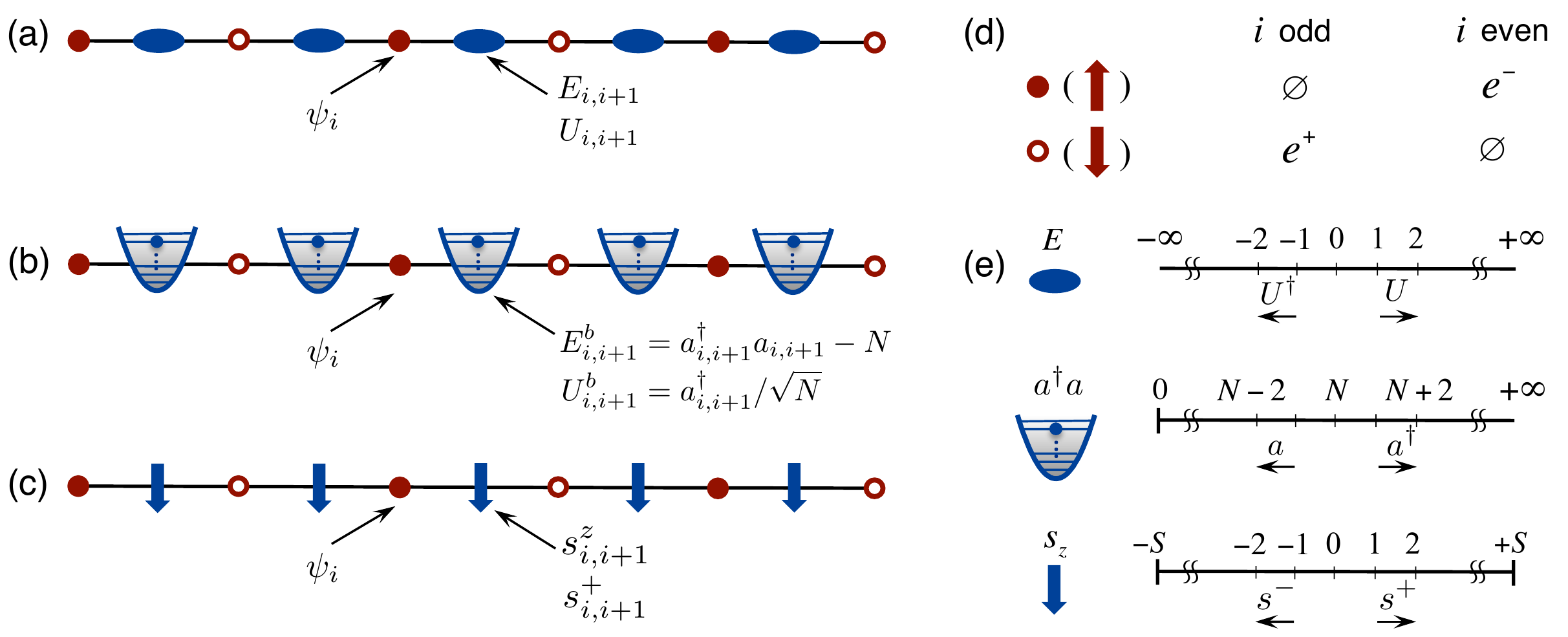}
\caption{$(1+1)$D QED on a lattice and truncation schemes. (a) The lattice representation of $(1+1)$D QED considers a system of free fermions living on sites ($\psi_i$), coupled to a $U(1)$ Abelian gauge field living on the links connecting neighboring sites ($E_{i.i+1},U_{i,i+1}$). In the standard Wilson LGT, the gauge-field degrees of freedom on each link span a 2D quantum rotor Hilbert space. 
(b) We introduce here a complementary LGT, the highly occupied boson model, by representing the gauge fields by local \emph{boson} modes ($a_{i,i+1},a_{i,i+1}^\dagger$), illustrated as harmonic oscillators. In the limit of large occupation number, $N\to \infty$, the Wilson LGT is recovered exactly. 
(c) An alternative scheme, the quantum link model \cite{Wiese2013,Wiese2014}, represents the gauge fields by spin operators ($s_{i,i+1}^z,s_{i,i+1}^+$). It approaches the Wilson formulation in the limit of large spins, $S\to\infty$. 
(d) In one spatial dimension, the lattice fermions can be mapped to spins via the Jordan--Wigner transformation. They represent the Dirac spinor of charges ($e^+$) and anti-charges ($e^-$) in a staggered fashion, i.e., on an even (odd) site, the presence (absence, denoted by $\varnothing$ in the figure) of a staggered fermion corresponds to the presence of $e^+(e^-)$.
(e) The gauge-field Hilbert space on a single link, from top to bottom for $(1+1)$D QED, the highly occupied boson model, and the quantum link model.
} 
\label{overallillustration}
\end{center}
\end{figure*}

\subsection{$(1+1)$D lattice QED}
\label{sec2A: lattice Schwinger model}
We adopt the Kogut--Susskind Hamiltonian formulation of the Wilson LGT~\cite{Kogut1975}, in which the time dimension is kept continuous while the space dimensions are discretized into a lattice geometry. 
Originally, this lattice discretization has been developed to facilitate numerical calculations, but it also provides a very convenient framework for implementations in physical quantum simulators with discrete DOFs \cite{Wiese2013,Wiese2014,Zohar2016,Dalmonte2016,Doucot2004,Marcos2013,Marcos2014,Mezzacapo2015,GarciaAlvarez2015,Hauke2013b}. 
Following Ref.~\cite{Banks1976}, $(1+1)$D QED can be represented as hopping of `staggered fermions' on a lattice with sites $i=1\dots L$, with the dynamical gauge fields sitting on the links connecting neighboring sites [see Fig.~\ref{overallillustration}(a)]. 
In this representation, the fermion operator $\psi_i$ on an even (odd) site corresponds to the upper (lower) component of the original Dirac spinor of the continuum theory, which are connected by the discrete chiral transformation~\cite{Montvay1994}. 
The corresponding Hamiltonian reads (we set $\hbar=c=1$)
\begin{equation}
\begin{split}
\label{SchwingerModelHamiltonian}
H = &-J \sum_i (\psi_i^\dagger U_{i,i+1} \psi_{i+1} + \textrm{h.c.}) \\
&+ \mu \sum_i (-1)^i \psi_i^\dagger\psi_i + V \sum_i E_{i,i+1}^2.
\end{split}
\end{equation}
Here, $J$ is the kinetic energy term, which couples matter and gauge field, $\mu$ is the fermionic rest mass, and $V$ measures the electric field energy.   
The link electric field $E_{i,i+1}$ and parallel transporter $U_{i,i+1}$ satisfy $[E_{i,i+1}, U_{i,i+1}]=U_{i,i+1}$. In the standard Kogut--Susskind formulation, the Hilbert space of the gauge field is the same as that of a 2D rotor, of which the basis states are labelled as $\ket{n}_{i,i+1}$, with $E_{i,i+1}\ket{n}_{i,i+1}=n\ket{n}_{i,i+1}$, $n\in\mathbb{Z}$. 
The parameters in Eq.~\eqref{SchwingerModelHamiltonian} are connected to the model parameters of the continuum theory, by $J=1/2a$, $\mu=m$ and $V=g^2 a/2$,
in which $g$ and $m$ are respectively the fermion--matter coupling strength and fermion rest mass in the continuum model, and $a$ the lattice constant. 
By taking the limit $a\to 0$, lattice QED recovers conventional QED in continuous space-time.

The model in Eq.~\eqref{SchwingerModelHamiltonian} becomes a gauge theory through gauge fixing the Hilbert space into the physical Coulomb sector $\mathcal{G}$ by enforcing the Gauss law $G_i \mathcal{G} = 0$, where we assumed a vanishing background charge. Here, $G_i$ is the local $U(1)$ gauge generator
 at lattice site $i$,
\begin{equation}
\label{SchwingerModelGauseLaw}
G_i = \psi_i^\dag \psi_i + \frac{1}{2}[ (-1)^i - 1 ] - (E_{i,i+1} - E_{i-1,i}).
\end{equation}
This formula is the lattice equivalent of the familiar Gauss law in the continuum, ${\rm div}\, E=\rho$, with $\rho$ the charge density.

In the later Secs.~\ref{sec3} and \ref{sec4: QLM}, we are interested in constructing analog quantum simulators for $(1+1)$D lattice QED in a trapped-ion setup, where no fermionic DOFs appear naturally. A convenient workaround, valid in one spatial dimension, is to map the single-species fermions in the original $1+1$D lattice QED to $S=1/2$ spins by the Jordan--Wigner transformation [see Fig.~\ref{overallillustration}(d)] 
\begin{subequations}
\begin{alignat}{3}
&\psi_i^\dag &&= (-1)^{\sum_{j=1}^{i-1}(\tau_j^z+1)/2}\tau_i^+,\\
&\psi_i &&= (-1)^{\sum_{j=1}^{i-1}(\tau_j^z+1)/2}\tau_i^-,\\
&\psi_i^\dag \psi_i&&=(\tau_i^z+1)/2.
\end{alignat}
\end{subequations}
Here, we use $\tau_i$ to denote the $S=1/2$ Pauli matrices on site $i$, while reserving $\sigma_i$ for the two level systems describing the ionic internal states.  
In the spin language, the Hamiltonian of $(1+1)$D lattice QED becomes
\begin{equation}
\label{SchwingerHamiltonianJordanWigner}
\begin{split}
H = &-J\sum_i (\tau_i^+ U_{i,i+1} \tau_{i+1}^- + \textrm{h.c.}) \\
&+ \frac{\mu}{2} \sum_i (-1)^i \tau_i^z + V\sum_i E_{i,i+1}^2,
\end{split}
\end{equation}
while the local gauge generator is converted to
\begin{equation} 
\label{GaussLawJordanWigner}
G_i = \frac{1}{2}[\tau_i^z+(-1)^i] - (E_{i,i+1} - E_{i-1,i}).
\end{equation}

To demonstrate the viability of the proposed quantum-simulator schemes, we will in later sections study two prominent physical physical phenomena contained in $(1+1)$D QED. The first one, called string breaking, appears in its dynamical evolution after a quantum quench. The second, Coleman's quantum phase transition, is a property of its ground-state behavior. We now explain the basic physics related to these phenomena. 

\subsubsection{String-breaking dynamics}

$(1+1)$D QED is confined at all energy scales~\cite{Coleman1975,Coleman1976}, meaning there exist no free fundamental charges in the energy spectrum. Thus, thanks to the large energy contained in the electric-field string connecting opposite charges, a sufficiently far separated charge--anti-charge pair is highly unstable. If such a pair is prepared initially, it will evolve by spontaneous creation of charge--anti-charge pairs in the space in between, and thus break the electric-field string. 
The dynamical evolution of this `string breaking' can be analyzed by monitoring the space-averaged electric field, $\langle E \rangle = \int d\mathbf{r} \langle E(\mathbf{r})\rangle/\int d\mathbf{r}$, which takes a large nonzero value in the initial state and decreases in the subsequent dynamics. 

A simple quantitative analysis for when string breaking occurs can be made in the $J\to 0$ limit. Consider an initial pair of charge--anti-charge sitting at the boundaries of a $1$D lattice of size $L$. The Gauss law, Eq.~\eqref{SchwingerModelGauseLaw}, prescribes that the gauge fields are in the state $\ket{-1}_{i,i+1}$ on the $L-1$ links of the lattice. Thus, the average electric field of this initial state  is $\langle E\rangle = \sum_{i=1}^{L-1} \langle E_{i,i+1}\rangle/(L-1)=-1$, and the energy is $(L-1)V+2\mu$. As the field energy grows linearly with $L$, for sufficiently large $L$ it becomes advantageous for the system to spontaneously create a charge--anti-charge pair, and it will thus evolve into a so-called two-meson state. 
The average electric field of this state is $\langle E\rangle=-2/(L-1)$ and its energy $2V+4\mu$. 
Comparing the energies of these two states, one finds the following lower bound of the pair separation for string breaking: $L=3+\left \lceil{2\mu/V}\right \rceil$ (where $\left \lceil{X}\right \rceil$ is the smallest integer not less than $X$). 
For $J>0$, the string-breaking dynamics will show more complex behavior, as more high-lying states are involved in the dynamics, but the general physics remains the same.

\subsubsection{Ground-state phase transition}

The second phenomenon of $(1+1)$D QED that we are interested in is a ground-state phase transition that breaks spontaneously parity symmetry, as predicted by Coleman~\cite{Coleman1976}. 
$(1+1)$D QED possesses a non-trivial vacuum angle $\theta\in [-\pi,\pi]$, among which $\theta=\pm\pi$ are gauge equivalent configurations. The vacuum state with nonzero $\theta$ contains a background electric field, $E_0 = \theta/2\pi$. 
At vacuum angle $\theta = \pi$, there are two vacuum candidates if the gauge-matter coupling is zero, which have no excitation of fermionic matter nor the gauge field, but nonzero background field $E_0 = \pm 1/2$. The two possibilities are energetically degenerate and gauge equivalent, and are connected to each other by the parity transformation. 

In the limit of large fermion rest mass, $m/g\to\infty$, the ground state of the model in the thermodynamic limit $L\to\infty$ will be one of these two states as they contain no fermionic excitations. 
Thus parity symmetry is spontaneously broken, leading to a non-zero order parameter $\langle E\rangle = \pm 1/2$. 
On the other hand, for sufficiently small $m/g>0$, proliferation of charge--anti-charge pairs  due to nonzero fermion--gauge coupling will restore the broken symmetry, and the ground state of the model is a parity-invariant disordered state, with $\langle E\rangle = 0$. This parity-symmetry breaking phase transition appears in both the continuum model and its lattice counterpart. On a lattice, it is convenient to use the parameter of the continuum model, $m/g$, as the indicator of the phase transition, which is related to the parameters of the LGT by $g=2\sqrt{JV}$ and $m=\mu$. Such a quantum phase transition has been studied extensively through various numerical methods~\cite{Hamer1982,Byrnes2002}, which show that it lies in the universality class of the $1$D transverse-field Ising model, with critical exponents $\nu=1$ for the correlation length and $\beta=1/8$ for the order parameter.

\subsection{Highly occupied boson model (HOBM)}
\label{sec2C: HOBM}

The direct quantum simulation of Eqs.~\eqref{SchwingerHamiltonianJordanWigner} and \eqref{GaussLawJordanWigner} in quantum-optical platforms is challenging, as the representation of $U_{i,i+1}$ requires the proper designation of a 2D quantum-rotor Hilbert space on each link. 
Instead, motivated by experimental possibilities in trapped ions as will be discussed below in Sec.~\ref{sec3}, we introduce the following LGT. 

Its essence, illustrated in Fig.~\ref{overallillustration}(b), is a replacement of the rotor gauge-field operators of the Kogut--Susskind formalism by bosonic DOFs, $U_{i,i+1}\to U_{i,i+1}^b =  a_{i,i+1}^\dag/\sqrt{N}$ and $E_{i,i+1}\to E_{i,i+1}^b = a_{i,i+1}^\dag a_{i,i+1} - N$.  
The resulting interaction Hamiltonian reads 
\begin{equation}
\label{HOBMHamiltonian}
\begin{split}
H_{\mathrm{HOBM}} &= -\frac{J}{\sqrt{N}} \sum_i (\tau_i^+ a_{i,i+1}^\dag \tau_{i+1}^- +\textrm{h.c.})\\
& +  \frac{\mu}{2} \sum_i (-1)^i \tau_i^z + V\sum_i (a_{i,i+1}^\dag a_{i,i+1} - N)^2.
\end{split}
\end{equation}
In this replacement, the electric field is taken to evolve around some offset value $N$, such that the bosonic Fock state $\ket{N}_{i,i+1}^b$ is mapped to the zero electric field state $\ket{0}_{i,i+1}$. 
The local gauge generator now takes the form 
\begin{equation}
\label{BosonAppGauss}
G_i = \frac{1}{2}[\tau_i^z+(-1)^i] - ( a_{i,i+1}^\dag a_{i,i+1} - a_{i-1,i}^\dag a_{i-1,i})\,.
\end{equation}
As long as the system is initialized in a bosonic Fock state with the Gauss law implemented, the interaction Hamiltonian $H_{\mathrm{HOBM}}$ will keep the system in the gauge-invariant subspace satisfying the Gauss law $G_i\mathcal{G}=0$. 

This model preserves strict gauge symmetry, as the bosonic commutation rules lead to the required commutation relation $[E_{i,i+1}^b, U_{i,i+1}^b]=U_{i,i+1}^b$. 
However, replacing the rotor by the more easily controllable bosonic DOFs comes with the price of sacrificing unitarity of the parallel transporter, $[U^{b\dag}_{i,i+1}, U_{i,i+1}^{b}]=1/N$. 
Unitarity is recovered in the limit $N\to \infty$. 

Importantly, the differences between the highly occupied boson model (HOBM), Eqs.~\eqref{HOBMHamiltonian} and \eqref{BosonAppGauss}, with the Kogut--Susskind formalism, Eqs.~\eqref{SchwingerHamiltonianJordanWigner} and \eqref{GaussLawJordanWigner}, are controlled by the parameter $N$, which can be tuned in experiment. 
For systems of finite size $L$ under open boundary conditions, the largest theoretically possible boson-number deviation from $N$ is $1+\left \lfloor{L/4}\right \rfloor$ on a single link, where $\left \lfloor{X}\right \rfloor$ is the largest integer not larger than $X$. 
Thus, for $N\gg L$, we can expect the HOBM to recover the finite-size $(1+1)$D lattice QED. 
To go to the thermodynamic limit, the correct order of limits is $\lim_{L\to\infty}\lim_{N\to\infty}$. 
Subsequently, we will quantify these statements in more detail, first in the string-breaking dynamics, and then in the scaling behavior of the ground-state phase transition.

\subsubsection{String-breaking dynamics}
\label{sec:HOBMStringBreakingDynamics}

\begin{figure}[t]
\begin{center}
\includegraphics[width=0.99\columnwidth]{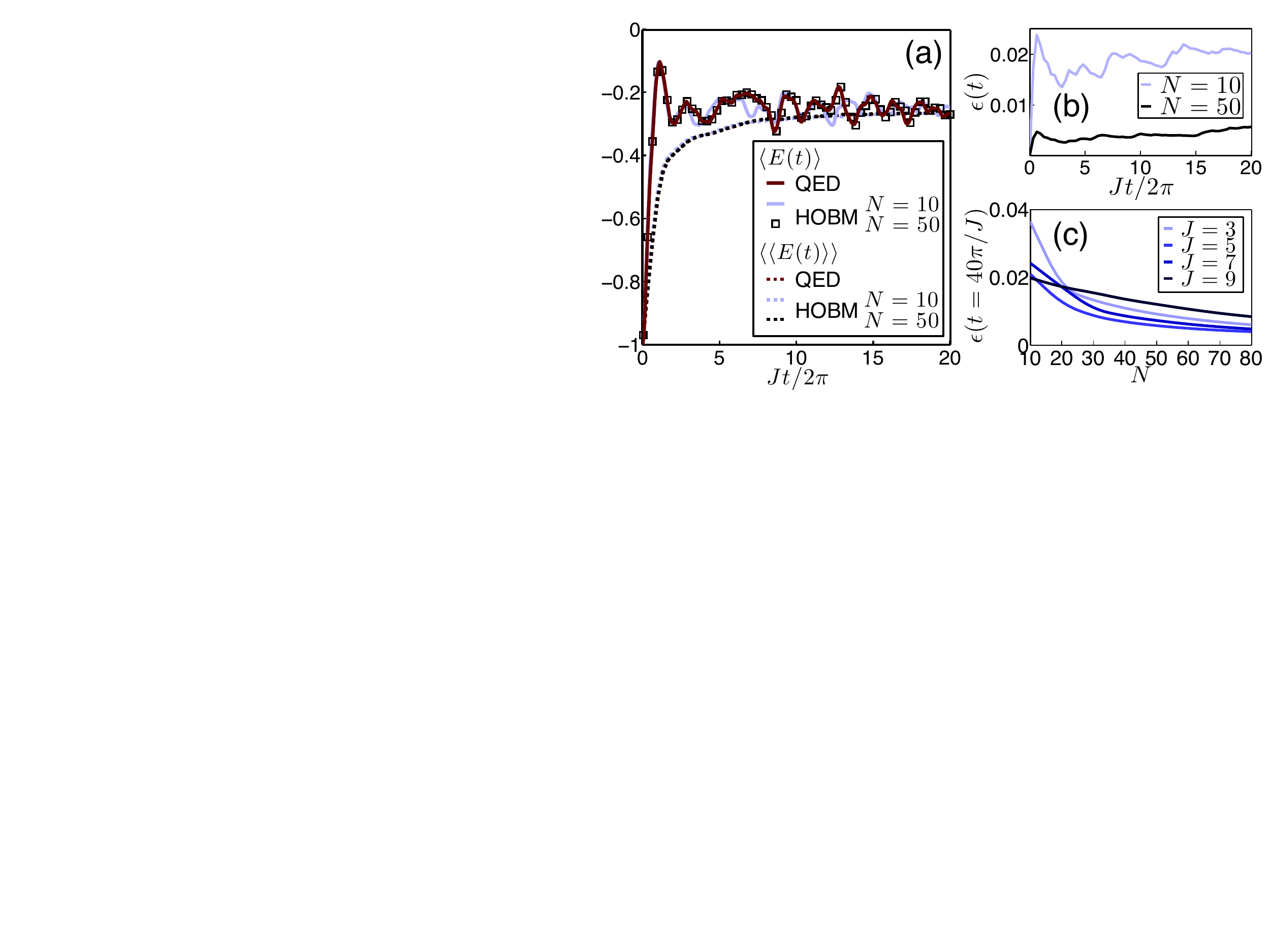}
\caption{String-breaking dynamics on a lattice with $L=12$ sites, as a quantitative measure of how reliably the HOBM approximates the lattice QED. Parameters are $\mu=V=0.2J$, and the initial state is a charge--anti-charge pair at the boundary of the lattice connected by a string of electric field. 
(a) The electric-field string breaks by spontaneous creation of charge--anti-charge pairs, as indicated by a reduction of the absolute value of the space-averaged electric-field strength, $\langle E(t)\rangle$. The HOBM (thin solid line and squares) manifests similar string-breaking dynamics as lattice QED and reaches quantitative agreement over the considered times for moderate $N$. 
After additional time averaging (dashed), the curves for both models are hard to discern. 
(b) The time averaged error, $\epsilon(t)$, quantifying the difference between the HOBM and lattice QED, remains bounded during time evolution and is suppressed by higher boson number $N$. (c) Also under varying the matter--gauge-field coupling $J$, $\epsilon(t)$ remains always bounded (here shown at $t=40\pi/J$ as an example), and is suppressed with larger boson number $N$. 
}
\label{stringbreaking}
\end{center}
\end{figure}

Since string-breaking dynamics will in principle involve all the physical states, it serves as a good measure of how well the HOBM approximates the original lattice QED across the energy spectrum. We analyze the real-time string-breaking dynamics on a lattice of size $L=12$ with open boundary conditions. 
The initial state is a highly-unstable meson state, in which an anti-charge (down-spin) sits at the left boundary of the lattice while a charge (up-spin) sits at the right boundary, with a string of electric field connecting them. 

In Fig.~\ref{stringbreaking}(a), we show the exact evolution of the space-averaged electric field, $\langle E(t) \rangle= \sum_{i=1}^{L-1} \langle E_{i,i+1}(t)\rangle/(L-1)$.
Already for $N=10$ a rough agreement between the HOBM and the lattice QED is reached, which becomes excellent for $N=50$. 
We also plot the space- and time-averaged electric field, defined as $\langle\langle E(t) \rangle\rangle= \int_0^t dt' \langle E(t') \rangle /t$, which allows us to quantify how the long-time limit is approached. 
The curves for the two models are hardly discernible, even for only $N=10$. 
Since $\langle\langle E(t) \rangle\rangle$ extracts low-frequency components in the time-evolution, this agreement indicates that already with $N=10$ the low-lying spectrum of the two models is practically the same. 
When $N\gg L$, the full spectrum of the HOBM recovers that of the lattice QED. 

To further quantify the difference between HOBM and lattice QED, we introduce the space- and time-averaged electric-field difference $\epsilon(t)=\frac{1}{t}\int_0^t dt' |\langle E(t')\rangle_{\textrm{HOBM}} - \langle E(t')\rangle_{\textrm{QED}}|$. As shown in Figs.~\ref{stringbreaking}(b) and (c), $\epsilon(t)$ remains bounded during time evolution and decreases with increasing $N$. These findings indicate that the HOBM can faithfully represent the dynamics of (1+1)D lattice QED.

\subsubsection{Ground-state phase transition}
\label{sec:HOBMGroundStatePhaseTransition}

\begin{figure}[t]
\begin{center}
\includegraphics[width=0.99\columnwidth]{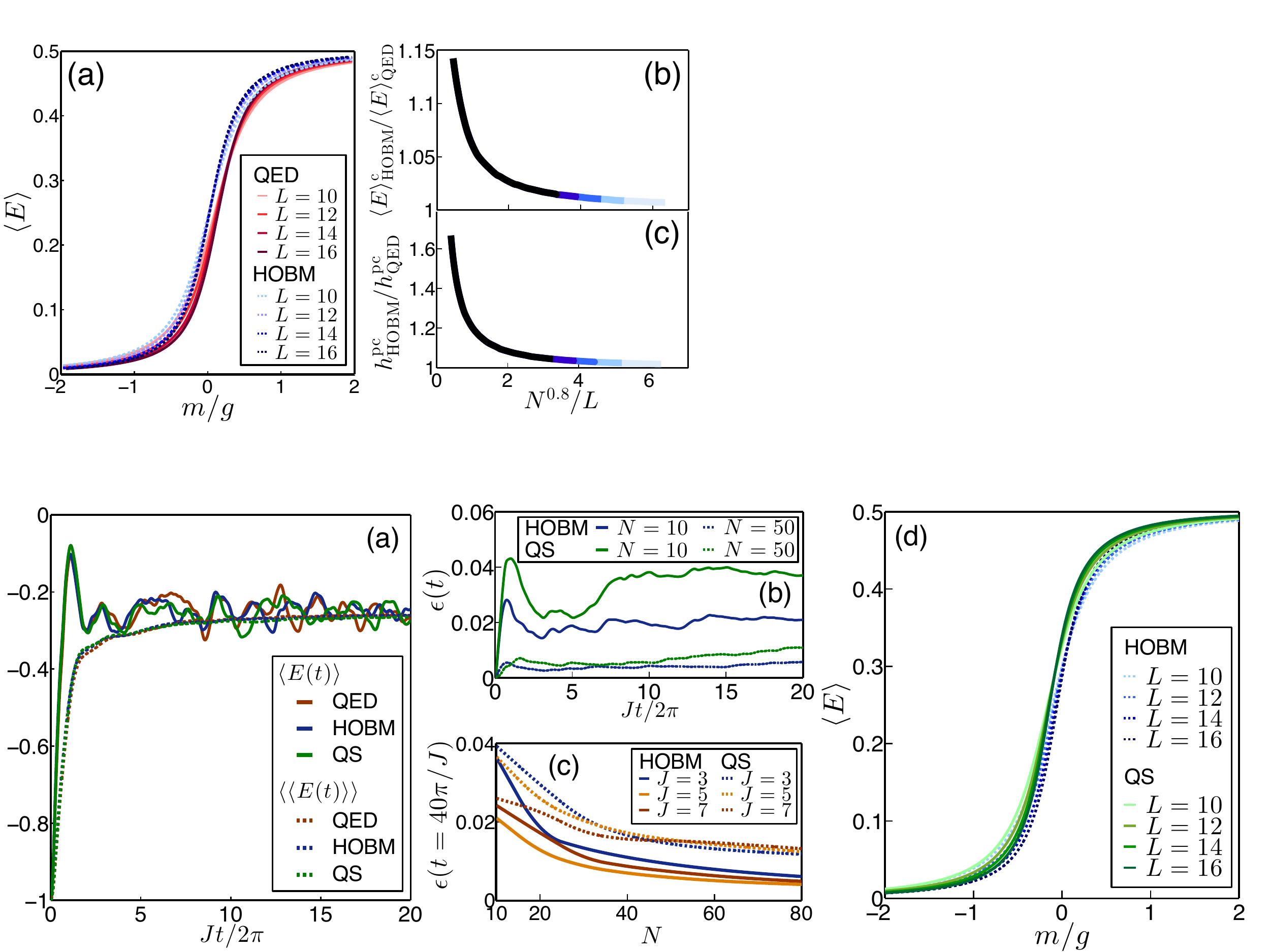}
\caption{
Quantitative analysis of Coleman's parity-symmetry breaking phase transition at vacuum angle $\pi$, for both lattice QED and the HOBM. 
The coupling parameter is chosen as $ga=0.3$, which locates the quantum critical point  of lattice QED at $(m/g)_\mathrm{c}=0.297$~\cite{Byrnes2002}.
(a) The phase transition is quantified by the order parameter $\langle E\rangle$, with a slight shift between the HOBM and lattice QED. 
(b,c) Finite-size scaling for (b) the order parameter at criticality and (c) the position of the pseudo-critical point. Boson occupation number $N\in[10,200]$, while curves from light to dark blue denote increasing lattice sites $L=10,12,14,16,18$.
The rescaled data collapse onto a single curve, demonstrating that the HOBM has the same quantum critical behavior as lattice QED, as long as $N\to\infty$ faster than $L\to\infty$.}
\label{ColemanQPT}
\end{center}
\end{figure}

As a second illustration, we study the parity-symmetry-breaking quantum phase transition of the lattice QED at vacuum angle $\pi$~\cite{Coleman1976,Hamer1982,Byrnes2002}. 
Fig.~\ref{ColemanQPT}(a) displays the order parameter $\langle E \rangle$ across the quantum critical region from exact diagonalization of systems up to $L=16$, for both lattice QED and the HOBM. 
The precise position of the critical point is shifted at finite $N$, but as we will show now the replacement of gauge fields by boson operators does not affect the critical scaling behavior. 

In the critical region, one expects a universal scaling for the order parameter in the lattice QED as~\cite{Byrnes2002}
\begin{equation}
	\label{eq:scalingQED}
	\langle E\rangle_{\mathrm{QED}}\sim L^{\Delta} \phi(L^{1/\nu}h)\,, 
\end{equation}
where $\phi$ is a universal function, $h=m/g-(m/g)_c$ is the distance from the quantum critical point, and we have assumed the finite-size cut-off $1/L$ to be the most relevant perturbation, i.e., $1/L\gg h^\nu$. 
The precise position of $(m/g)_c$ depends on the other dimensionless parameter $ga$; nevertheless the critical exponents $\nu=1$ and $\Delta=-\beta/\nu=-1/8$, are independent of the lattice discretization $ga$ \cite{Byrnes2002}. 
For the HOBM, we assume $1/N$ as another relevant perturbation, and write down a tentative scaling Ansatz 
\begin{equation}
	\label{eq:scalingHOBM}
	\langle E\rangle_{\mathrm{HOBM}}\sim L^{\Delta'}\phi'(L^{1/{\nu'}}h,L^{1/{\eta'}}N^{-1})\,,
\end{equation}
with a universal function $\phi'$, still assuming $1/L\gg h^{\nu'}, N^{-\eta'}$. Here, $\Delta, \nu$ and $\Delta'$,$\nu'$,$\eta'$ are two, a priori different, sets of critical exponents. 

We now demonstrate numerically that the critical exponents $\Delta'$ and $\nu'$ are in fact the same as those of $(1+1)$D QED. Following Ref.~\cite{Hamer1997}, we map $(1+1)$D QED and the HOBM to the equivalent long-range-interacting spin models that facilitate numerical investigation, for which we perform exact diagonalization on finite lattices of size $L\in [10,18]$ and with various boson occupation offsets $N\in [10,200]$, and study their finite-size scaling behavior across the quantum critical region.
At the critical point, i.e., $h=0$, we expect $\langle E\rangle_{\mathrm{HOBM}}^\mathrm{c}/\langle E\rangle_{\mathrm{QED}}^\mathrm{c}\sim L^{\Delta' - \Delta}\phi''(L^{1/{\eta'}}N^{-1})$, where $\phi''(L^{1/{\eta'}}N^{-1})=\phi'(0,L^{1/{\eta'}}N^{-1})$.
As displayed in Fig.~\ref{ColemanQPT}(b), we obtain a perfect scaling collapse for various choices of $L$ and $N$, which indicates that $\Delta=\Delta'$. 
To determine $\nu'$, we calculate the pseudo-critical point $h^{\textrm{pc}}$, defined as the value of $m/g$ where $\partial \langle E \rangle /\partial h$ reaches its maximum. 
From Eqs.~\eqref{eq:scalingQED} and \eqref{eq:scalingHOBM}, and using $\Delta'=\Delta$, we expect the scaling behavior $h^{\textrm{pc}}_{\textrm{HOBM}}/h^{\textrm{pc}}_{\textrm{QED}}\sim L^{1/\nu'-1/\nu}\phi''(L^{1/{\eta'}}N^{-1})$. 
The perfect scaling collapse in Fig.~\ref{ColemanQPT}(c), indicates that also $\nu=\nu'$. The results shown in Fig.~\ref{ColemanQPT}(a-c) are performed at $ga=0.3$, but we have checked that a similar scaling collapse happens also at other values of $ga$.
These results indicate that the proposed HOBM has the same quantum critical behavior as $(1+1)$D lattice QED. 
What is more, the two scaling collapses in Fig.~\ref{ColemanQPT}(b) and (c) show that the scaling Ansatz \eqref{eq:scalingHOBM} is indeed valid with $\eta'\simeq 0.8$. This means that, with decreasing $1/N$, the HOBM approaches lattice QED in a well-controlled and continuous manner. 

In Sec.~\ref{sec3}, we discuss a feasible experimental scheme to realize the HOBM with trapped ions. Importantly, however, the above general analysis is independent of experimental platforms and may also be fruitful, for example, for quantum simulators based on superconducting qubits.

\subsection{Quantum link model (QLM)}
\label{QLMformulation}

Before discussing a possible realization of the HOBM, we want to briefly remark on a related truncation scheme for LGTs, the quantum link model (QLM)~\cite{Horn1981,Orland1990,Chandrasekharan1997,Brower1999}. The QLM formalism preserves the local gauge invariance strictly by expressing the gauge field living on the link as spin operators  in a proper spin-$S$ representation [see Fig.~\ref{overallillustration}(c)]. 
For the $(1+1)$D $U(1)$ model considered here, the conversion from the standard Kogut--Susskind formalism to the QLM is straightforward, by replacing in Eqs.~\eqref{SchwingerModelHamiltonian} and \eqref{SchwingerModelGauseLaw} the parallel transporter $U_{i,i+1}(U_{i,i+1}^\dag)$  by the spin variable $s_{i,i+1}^+(s_{i,i+1}^-)$ and the electric field $E_{i,i+1}$ by $s_{i,i+1}^z$. Since $[s_{i,i+1}^+,s_{i,i+1}^-]=2 s_{i,i+1}^z$, the parallel transporter realized in this way is not unitary. 
The representation approaches the standard Kogut--Susskind formulation in the large $S$ limit~\cite{Wiese2013,Wiese2014}, which in turn approaches the continuum theory in the limit of infinitely small lattice constant. 
Remarkably, representations with modest $S$ already provide qualitatively similar physics to the standard LGT. For example, for $(1+1)$D QED considered here, the corresponding $S=1/2$ QLM manifests parity-symmetry-breaking quantum phase transition, the same as the lattice QED at vacuum angle $\pi$, and the $S=1$ QLM displays string breaking dynamics~\cite{Banerjee2012,Rico2014}.  

Since in QLMs the gauge variables on the links have finite dimensional Hilbert spaces [see Fig.~\ref{overallillustration}(e)], they are very appealing for quantum simulations~\cite{Wiese2013,Wiese2014}. 
For example, quantum simulators for Abelian and non-Abelian QLMs with cold atoms in optical lattices have been proposed in Refs.~\cite{Banerjee2012,Banerjee2013}, with an emphasis on observing the underlying physics in a small-$S$ setting, rather than taking the $S\to \infty$ limit. 
In Sec.~\ref{sec4: QLM}, we will describe a scheme to build an analog quantum simulator of the $S=1/2$ Abelian QLM in $(1+1)$ space-time dimensions in trapped ions with modest experimental demands. While this scheme obtains systematic deviations for increasing ion numbers, it strongly improves the working energy scale over existing proposal~\cite{Hauke2013b}, and may thus provide an alternative to the HOBM for small-scale proof-of-principle experiments. 

\section{A quantum simulator of $(1+1)$D lattice QED: the HOBM scheme}
\label{sec3}
In this section, we propose and discuss a scheme to simulate the HOBM, by encoding the spins in Eq.~\eqref{HOBMHamiltonian} in ionic internal states, while the bosonic gauge fields are represented by local vibrational quanta of the ions. The scheme can be scaled up to tens of ions trapped in an array of micro-traps.
The bosonic occupation offset $N$ is controlled by initial state preparation, allowing the quantum simulator to approach smoothly the standard $(1+1)$D lattice QED. Moreover, by exploiting phonons as active DOFs participating in the dynamics of the gauge theory, the proposed quantum simulator works on a favorable energy scale compared to typical decoherence rates.  We first discussing in detail the engineering of the HOBM Hamiltonian with the microscopic building blocks from the trapped-ion toolbox, after which we move on to discuss experimental issues, including initial-state preparation and numerical predictions for a modest-size quantum simulator.

\subsection{Trapped-ion implementation of the HOBM}
\label{sec3A: Trapped-ion implementation of the HOBM}
In this section, we introduce our envisioned experimental setup and show how the HOBM can be realized within it by designed sideband-addressing using lasers. 

\subsubsection{Employed degrees of freedom: pseudo-spins and localized phonon modes}
\label{Employed degrees of freedom: pseudo-spins and localized phonon modes}
Our envisioned setup consists of $L$ ions, each trapped in a local potential minimum, which form a string of micro-traps in space, as can be generated with designed surface-trap electrodes~\cite{Harlander2011,Wilson2014,Mehta2014,Mielenz2015}. The internal DOFs of the ions are restricted to two electronic levels and can be described by a collection of pseudo-spin operators. Their Hamiltonian is $H_{\textrm{s}} = \sum_l\omega_{eg}\sigma_l^z$, with $\sigma_l^z$ the Pauli operator associated with ion $l$ and $\omega_{eg}$ the corresponding electronic transition frequency. We use these pseudo-spin DOFs to represent the spin-matter $\tau_i$ appearing in Eq.~\eqref{HOBMHamiltonian}.

To encode the bosonic gauge fields $a_{i,i+1}$ in Eq.~\eqref{HOBMHamiltonian}, we consider a special arrangement of the transverse frequencies of individual micro-traps along the ion string, so that the collective phonon modes describing the transverse vibration of the ions are effectively localized between two nearest-neighbor ions. We will use these localized phonons to represent the bosonic gauge fields. In the following, we shall first briefly review the physics of the phonons in a micro-trap setting in as far as pertinent to the present proposal. Then, we move on to the engineering of these phonon modes by local adjustment of the trapping frequencies of individual micro-traps.

We assume that the micro-traps are nearly equally spaced along the string with the distance between neighboring trap centres being a constant $d$. Similar to the linear Paul traps~\cite{Wineland1998,James1998}, the balance between local trapping potential and mutual Coulomb interaction determines the equilibrium position of the ions, around which they vibrate. We assume the trapping frequencies along radial $x,y$ directions are far larger than those along the axial $z$ direction, so that the ions form a $1$D chain. Due to the symmetry of this $1$D geometry, in the harmonic regime of small amplitude vibrations, the motion of the ions in different directions decouple~\cite{James1998}, $H_{\textrm{ph}}=\sum_{\alpha}H_{\textrm{ph}}^\alpha$, with $\alpha=x,y,z$, and
\begin{equation}
\label{phononH}
H_{\textrm{ph}}^\alpha = \sum_{l}\frac{(p_{l}^\alpha)^2}{2M_{\mathrm{I}}} + \frac{1}{2} M_{\mathrm{I}}\sum_{lm} V_{lm}^\alpha r_l^\alpha r_m^\alpha,
\end{equation}
where $M_{\mathrm{I}}$ is the ion mass, $r_l^\alpha$ is the deviation of the $l$-th ion along the $\alpha$ direction from its equilibrium position, and $p_l^\alpha$ is the conjugate momentum. The strength of the quadratic potential induced by the Coulomb interaction, $V_{lm}^\alpha$, depends on the distance between the ions in their equilibrium configuration~\cite{Porras2004},
\begin{equation}
\label{CoulombCoupling}
V_{lm}^\alpha=\left\{ \begin{array}{ll}
(\omega_l^\alpha)^2 - \gamma_\alpha \sum_{n\neq l}\frac{e^2/M_{\mathrm{I}}}{4\pi\epsilon_0|z_l^0 - z_n^0|^3}, & l=m, \\
\gamma_\alpha \frac{e^2/M_{\mathrm{I}}}{4\pi\epsilon_0|z_l^0 - z_m^0|^3}, & l\neq m,
\end{array}\right.
\end{equation}
with $\gamma_{x,y}=1,\gamma_{z}=-2$. Here, $\omega_l^\alpha$ is the trapping frequency along the $\alpha$ direction of the $l$-th micro-trap, and $z_l^0$ is the equilibrium position of the $l$-th ion. The above Hamiltonian Eq.~\eqref{phononH} can be diagonalized by defining collective phonon modes shared by the whole string of ions, $r_l^\alpha=\sum_q M_{lq}^\alpha(c_q^{\alpha\dag}+c_q^\alpha)/\sqrt{2M_{\mathrm{I}}\epsilon_q^\alpha}$, yielding
\begin{equation}
\label{phononHamiltonian}
H_{\textrm{ph}}^\alpha = \sum_q \epsilon_q^\alpha c_q^{\alpha\dag} c_q^\alpha.
\end{equation}
The matrix $M^\alpha$ diagonalizes $V_{lm}^\alpha$, $\sum_{lm}M_{lq}^\alpha V_{lm}^\alpha M_{mq'}^\alpha=\delta_{qq'}(\epsilon_q^\alpha)^2$, and relates the local vibration of the ions to collective phonon modes.

\begin{figure*}[t]
\begin{center}
\includegraphics[width=0.9\textwidth]{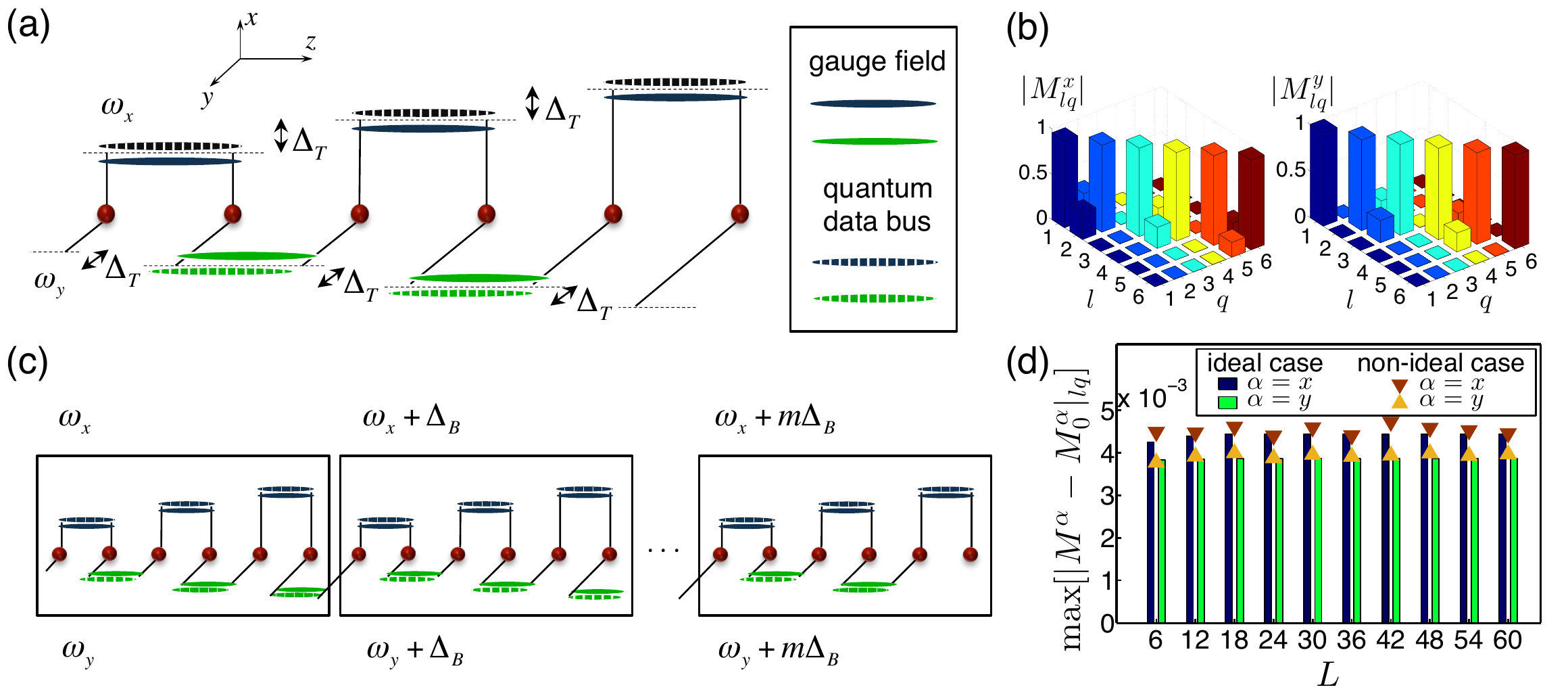}
\caption{Engineering of localized radial phonon modes in a string of micro-traps. 
(a) The radial trapping frequencies $\omega_{x,y}$ increase by $\Delta_T$ along the chain in a stepwise fashion, with an offset of one lattice site between $x$ (blue) and $y$ (green). 
The large trapping-frequency mismatch $\Delta_T$ effectively localizes hybridized phonon modes within pairs of ions. 
Phonon modes $c^\alpha_l$ and $c^\alpha_{l+1}$ are shared by ions $l$ and $l+1$ (with $\alpha=x$ for $l=\mathrm{odd}$ and $\alpha=y$ for $l=\mathrm{even}$). 
In each pair, the phonon mode $c_{l}^\alpha$ is chosen to encode the bosonic gauge field, while the residual phonon $c_{l+1}^\alpha$ serves as quantum data-bus to transmit interactions. 
(b) Calculation of the matrix $M_{lq}^{x(y)}$, which diagonalizes the vibrations, for realistic parameters ($^9$Be$^+$ ions, with the trap parameters $\Delta_T=2\pi\times 500 $kHz, $\delta_T=2\pi\times5$kHz, $\omega_x=\omega_y=2\pi\times 5$MHz, and distance between trap centers $d=30\mu$m). 
The leakage of $M_{lq}^{\alpha}$ out of pairs of ions is well below $0.01$ along both $x$ and $y$ directions, showing that the phonons are efficiently localized in pairs of ions. 
(c) Since the trapping frequency cannot be increased infinitely, scaling to long ion chains requires repetition of elementary blocks containing a few pairs of ions [such as depicted in panel (a)]. 
A global trapping-frequency offset $\Delta_B$ between each block eliminates long-range inter-block hopping of the localized phonons. 
(d) For realistic parameters, the leak-out of vibrational modes outside of the desired pair remains well below 0.01 even for a large number of blocks. Here, $\Delta_B=2\pi\times 50$kHz, with other parameters the same as in panel (b); block size is 6 ions, number of blocks is in the range of 1-10. The leak-out is measured by the maximum element of $|M^\alpha-M^\alpha_0|$, where $M^\alpha$ is the normal-mode distribution matrix in the $\alpha$-direction, and the block-diagonal $M^\alpha_0$ is its zeroth-order approximation. Two situations are considered: the ideal case, where the radial trapping frequencies are designated by Eqs.~\eqref{generalizedTrapFrequencyDesignX} and \eqref{generalizedTrapFrequencyDesignY}; the non-ideal case, where a local randomness of the trapping frequencies, uniformly distributed in $2\pi\times[-30,30]$kHz, is added on top of Eqs.~\eqref{generalizedTrapFrequencyDesignX} and \eqref{generalizedTrapFrequencyDesignY}, to reflect the uncertainties in the frequency control in experiments.
}
\label{fig4:setup}
\end{center}
\end{figure*}

In view of the experimental progress in local control of the trapping frequencies for individual ions~\cite{Schmied2009, Mielenz2015}, we consider in the following a suitable segmentation of the micro-trap frequencies, where the collective phonon modes become localized between pairs of ions (see Fig.~\ref{fig4:setup} as an illustration). To this end, we choose the frequency of the $l$-th micro-trap along the $x$ direction as
\begin{equation}
\label{microTrapFrequencyX}
\omega_l^x = \omega_x + \left \lfloor{\frac{l-1}{2}}\right \rfloor \Delta_T + \frac{1+(-1)^l}{2}\delta_T,
\end{equation}
while in the $y$ direction
\begin{equation}
\label{microTrapFrequencyY}
\omega_l^y = \omega_y +  \left \lfloor{\frac{l}{2}}\right \rfloor \Delta_T+ \frac{1-(-1)^{l}}{2}\delta_T.
\end{equation}
Here, $\omega_{x(y)}$ is a (large) global trapping frequency that we assume equal for each ion. 
On top of this, the trapping frequency is adjusted locally by two frequency-offsets, $\Delta_T$ and $\delta_T$. We will use $\delta_T$ further below to adjust the parameter $J$ of the model Eq.~\eqref{HOBMHamiltonian}, while $\Delta_T$ will serve to suitably engineer the vibrational modes. 
To this end, we choose the frequencies to satisfy the condition
\begin{equation}
\label{localizedModeCondition}
\Delta_T\gg \delta_T\sim \max[V_{l,l+1}^{x(y)}]/\omega_{x(y)},
\end{equation}
in which $V_{l,l+1}^\alpha/\omega_{\alpha}$ measures the hopping rate of the quanta of local vibrations between ion $l$ and $l+1$ along the $\alpha$ direction~\cite{Brown2011}.

With the choice of Eqs.~\eqref{microTrapFrequencyX} and \eqref{microTrapFrequencyY}, $\omega_{2n-1}^x$ is near-resonant to $\omega_{2n}^x$, in the sense that their frequency difference $\delta_T$ is comparable to the hopping rate of local vibrational quanta. In contrast, $\omega_{2n}^x$ and $\omega_{2n+1}^x$ have a frequency difference $\sim\Delta_T$, far larger than their mutual coupling strength. Under such a trap arrangement, the local vibration of the $(2n-1)$-th and the $2n$-th ion along the $x$ direction hybridize into two collective phonon modes, which contain little contribution from the vibration of other ions because of their large frequency mismatch. Similarly, along the $y$ direction, $\omega_{2n}^y$ is near-resonant to $\omega_{2n+1}^y$, while being far off-resonant to all the other trapping frequencies. The mutual Coulomb interaction hybridizes these vibrations into two collective phonon modes shared mainly by the $2n$-th and $(2n+1)$-th ion.

These localized radial phonon modes are exploited to build the bosonic gauge-fields $a_{i,i+1}$ in Eq.~\eqref{HOBMHamiltonian}. To be more concrete, $a_{2n-1,2n}$ is encoded by $c_{2n-1}^x$, the $(2n-1)$-th localized phonon along the $x$ direction, while $a_{2n,2n+1}$ is encoded by $c_{2n}^y$ [see Fig.~\ref{fig4:setup}(a)]. The remaining radial phonon DOFs, namely $c_{2n}^x$ and $c_{2n+1}^y$, will be exploited as quantum data-bus to transmit the interaction between nearest-neighbor ions (see Sec.~\ref{sec:SpinGaugeFieldCoupling} below). 
Here, the radial phonon modes are labeled so that $c_q^{x(y)}$ connects adiabatically to the local vibration of the $q$-th ion in the $x(y)$ direction in the $d\to \infty$ limit, where $d$ is the distance between neighboring trap centers.

As can be seen in Fig.~\ref{fig4:setup}(b), the desired hybridization of phonon modes is well achievable with realistic experimental parameters. 
It is possible to quantify the residual undesired coupling between phonon modes in perturbation theory in the small parameter $V_{lm}^\alpha/(\Delta_T\omega_\alpha)$. 
In zeroth order, the matrix $M^{x(y)}$ that diagonalizes the couplings of local vibrations becomes block-diagonal, $M^x\simeq M_0^x=\mathrm{diag}(T_{12},T_{34},...,T_{2n-1,2n},...)$ and similarly $M^y\simeq M^y_0=\mathrm{diag}(1, T_{23},T_{45},...,T_{2n,2n+1},...,1)$. The $2\times2$ matrix $T_{l,l+1}$ diagonalizes the near-resonant blocks consisting of ion $l$ and $l+1$, with $l=\mathrm{odd}$ in the $x$ direction and $l=\mathrm{even}$ in the $y$ direction,
\begin{equation}
T_{l,l+1}=\left( \begin{array}{cc}
\cos\theta_{l,l+1}&\sin\theta_{l,l+1}\\
-\sin\theta_{l,l+1}&\cos\theta_{l,l+1}\end{array} \right),
\end{equation}
where the angle $\theta_{l,l+1}$ characterizes the distribution of the two localized modes within the pair of ions $l$ and $l+1$,
\begin{equation}
\label{phononDistributionAngle}
\theta_{l,l+1}=\left\{ \begin{array}{lcr} \frac{1}{2}\arctan \left(\frac{2V_{l,l+1}^x}{V_{l+1,l+1}^x-V_{l,l}^x}\right),&& l=\mathrm{odd},\\\\
\frac{1}{2}\arctan\left(\frac{2V_{l,l+1}^y}{V_{l+1,l+1}^y-V_{l,l}^y}\right),&& l=\mathrm{even}.
\end{array}\right.
\end{equation}
By tuning the small frequency-offset $\delta_T$ within the near-resonant pairs, the relative strength between $V_{l,l}^{x(y)}$ and  $V_{l+1,l+1}^{x(y)}$ can be adjusted via Eq.~\eqref{CoulombCoupling}, thus allowing us to control the angle $\theta_{l,l+1}$. In Appendix~\ref{sec:perturbativeCalculationOfTheMMatrix}, we calculate perturbatively the elements of $M^{x(y)}$ beyond the lowest order, i.e., $[M^{x(y)}-M^{x(y)}_0]_{lq}$, and find they are bounded by $\max[V_{l,l+1}^{x(y)}]/(\Delta_T\omega_{x(y)})$. Under the condition~\eqref{localizedModeCondition}, the leak-out of these localized phonon modes from the correponding pair of ions is thus negligible. 

The micro-trap design in Eqs.~\eqref{microTrapFrequencyX} and~\eqref{microTrapFrequencyY} requires a stepwise increase of the trapping frequencies along the radial directions. Thus, its scalability is restricted by the strength of radial confinement achievable in experiments. Nevertheless, the rapid dipolar power-law decay of the coupling $V^\alpha_{lm}$ [see Eq.~\eqref{CoulombCoupling}] allows for further scaling-up of such a local-phonon setup. We envision a micro-trap array consisting of several blocks, each block containing a string of segmented micro-traps constructed as Eqs.~\eqref{microTrapFrequencyX} and~\eqref{microTrapFrequencyY} [see Fig.~\ref{fig4:setup}(a)], while a global trapping-frequency shift $\Delta_B$ between blocks avoids the cross-talk of local phonons in different blocks, as schematically illustrated in Fig.~\ref{fig4:setup}(c). In such a micro-trap array, the frequency selections for the $l$-th trap, $\omega_l^{x(y)}$, can be generalized from Eqs.~\eqref{microTrapFrequencyX} and~\eqref{microTrapFrequencyY} to 
\begin{equation}
\label{generalizedTrapFrequencyDesignX}
\omega^x_{mN_I+j} = \omega_x + m\Delta_B + \left \lfloor{\frac{j-1}{2}}\right \rfloor \Delta_T + \frac{1+(-1)^j}{2}\delta_T,
\end{equation}
\begin{equation}
\label{generalizedTrapFrequencyDesignY}
\omega^y_{mN_I+j+1} = \omega_y + m\Delta_B + \left \lfloor{\frac{j+1}{2}}\right \rfloor \Delta_T + \frac{1+(-1)^j}{2}\delta_T,
\end{equation}
where $N_I=\mathrm{even}$ is the number of micro-traps in each block, $m\geq 0$ is the block index and $1\leq j\leq N_I$ specifies individual traps in each block. The frequency hierarchy Eq.~\eqref{localizedModeCondition} is generalized to 
\begin{equation}
\label{GeneralLocalizedPhononCondition}
\Delta_T\gg \Delta_B \gg\delta_T\sim \max[V_{l,l+1}^{x(y)}]/\omega_{x(y)},
\end{equation}
which leads to an upper bound $ \max[V_{l,l+1}^{x(y)}]/[(N_I-1)^3\Delta_B\omega_{x(y)}]$ to the cross-talk between localized modes in different blocks (see Appendix~\ref{sec:perturbativeCalculationOfTheMMatrix} for a detailed derivation). 

The scalability of such a micro-trap array can be estimated as $L\lesssim N_I\Delta_T/\Delta_B$, where $L$ is the total number of ions. For ion arrays even longer, the local phonons in far-separated blocks become near-resonant due to the accumulation of block-dependent trapping energy offset $m\Delta_B$. Because of the finite accuracy of frequency control, they can become accidentally resonant, leading to significant leak-out of the phonon modes. Another practical limitation is the zig-zag transition~\cite{Birkl1992} in such a micro-trap array, which depends on the trapping frequencies in all three spatial directions as well as the spacing between individual traps in experiments~\cite{Szymanski2012}.

The conditions in Eq.~\eqref{GeneralLocalizedPhononCondition} can be experimentally realized in current microfabricated surface ion traps~\cite{Harlander2011,Wilson2014,Mehta2014,Mielenz2015}. For concreteness, we consider the surface-trap setup for $^9\textrm{Be}^+$ in the NIST group~\cite{Wilson2014}, in which the typical value of radial trapping frequencies is, $\omega_{x(y)} = 2\pi\times 5$MHz, while in the axial direction $\omega_z =  2\pi\times0.5$MHz. The separation of nearest-neighbor trap centers is $d = 30 \mu$m, leading to a Coulomb coupling $V_{l,l+1}^{x(y)} \sim (2\pi\times0.12\textrm{MHz})^2$. To satisfy Eq.~\eqref{GeneralLocalizedPhononCondition}, one can choose $\delta_T\simeq 2\pi\times 5$kHz, $\Delta_B\simeq 2\pi\times 50$kHz and $\Delta_T\simeq 2\pi\times 500$kHz, suppressing undesired mode leak-out within the same block to below $\max[V_{l,l+1}^{x(y)}]/\Delta_T\omega_{x(y)}<0.01$. We assume each block contains $N_I=6$ ions, leading to a mode leak-out between different blocks of $ \max[V_{l,l+1}^{x(y)}]/[(N_I-1)^3\Delta_B\omega_{x(y)}]<0.005$. This is confirmed by the numerical calculation presented in Fig.~\ref{fig4:setup}(b,d), where we calculate the maximum leak-out of localized phonon modes in ion arrays consisting of various number of ion blocks. The extremely small leakage of the local phonons out of each ion-pair shows that even at a length of 60 ions the engineered vibrational modes behave as desired. 

\subsubsection{Spin--gauge-field coupling: design of sideband transitions}
\label{sec:SpinGaugeFieldCoupling}

\begin{figure}
\begin{center}
\includegraphics[width=0.99\columnwidth]{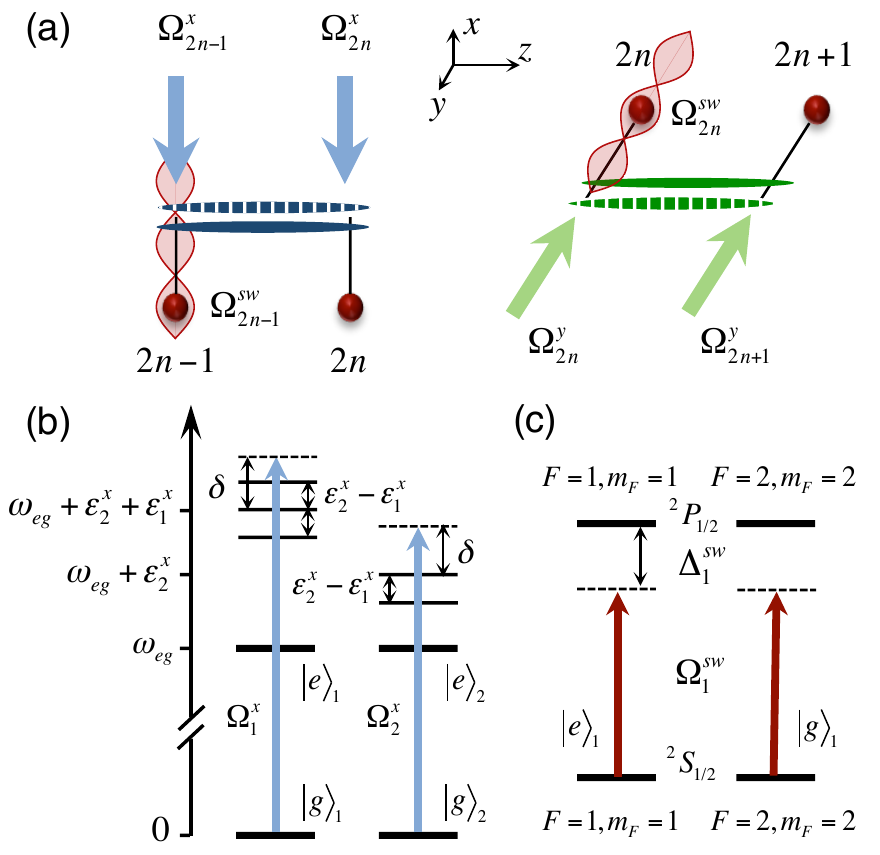}
\caption{
Design of the HOBM Hamiltonian Eq.~\eqref{HOBMHamiltonian}. 
(a) The engineering is independent in each basic element consisting of two neighboring ions $l$ and $l+1$. 
The design of the spin--gauge-field interaction $\propto J$ involves the pseudo-spins with states $\{\ket{g}_l,\ket{e}_l\}$ and $\{\ket{g}_{l+1},\ket{e}_{l+1}\}$, as well as the local phonon modes $c^\alpha_{l}$ and $c^\alpha_{l+1}$, and uses two local laser beams $\Omega_l^\alpha$ and $\Omega_{l+1}^\alpha$ (with $\alpha=x$ for $l=2n-1=\mathrm{odd}$ and $\alpha=y$ for $l=2n=\mathrm{even}$). The creation of the gauge-field energy term $\propto V$, relies on a strongly detuned standing-wave radiation (red sinusoid) to the first ion in the element, applied in the $x\,(y)$ direction depending on $l=\mathrm{odd}\,(\mathrm{even})$. 
(b) Generation of the spin--gauge-field coupling illustrated for the element consisting of ions 1 and 2. 
$\Omega_1^x$ drives the first ion on the second blue sideband consisting of the excitation of one gauge-field phonon $(c_1^x)$ and one data-bus phonon $(c_2^x)$, while $\Omega_2^x$ drives the second ion on the first blue sideband of the data-bus phonon $c_2^x$. 
The detuning $\delta$ is chosen far larger than the sideband-transition strengths, so the resonant process is a flip-flop transition between the two pseudo-spins. 
(c) Creation of the gauge-field energy term $\propto V$, exemplified for the first element, where a standing wave acts on ion $1$ along the $x$ direction. 
Here, we consider the example of $^9$Be$^+$ ions, where the pseudo-spin is encoded by the two hyperfine levels $(F,m_F)=(1,1)$ and $(F,m_F)=(2,2)$ of the $^2S_{1/2}$ manifold. The standing wave drives off-resonantly the optical transitions from both spin states to the $^2P_{1/2}$ manifold, with detuning $\Delta_1^{\mathrm{sw}}$ and Rabi frequency $\Omega_1^\mathrm{sw}$. For large enough $\Delta_1^{\mathrm{sw}}$, it creates nearly equal AC-Stark shifts to both spin-up and spin-down states, as discussed in detail in Sec.~\ref{phonon nonlinearity}.}
\label{fig5:levels}
\end{center}
\end{figure}
With the matter-field and gauge-field DOFs at hand, the next step is to design the desired spin--gauge-field coupling [the term $\propto J$ in Eq.~\eqref{HOBMHamiltonian}]. This is achieved by applying two local laser beams to each ion, one along the $x$ direction (with frequency $\omega_l^{\mathrm{L}x}$ and Rabi frequency $\Omega_l^x$) and the other along the $y$ direction (with frequency $\omega_l^{\mathrm{L}y}$ and Rabi frequency $\Omega_l^y$) respectively, as illustrated in Fig.~\ref{fig5:levels}(a). 
The corresponding light-matter coupling Hamiltonian is \cite{Wineland1998}
\begin{equation}
H_{\textrm{d}} = \sum_{l,\alpha}\frac{\Omega_{l}^\alpha}{2} \textrm{exp}(-i\omega_{l}^{\mathrm{L}\alpha}t+ik_{l}^\alpha r_{l}^\alpha) \sigma_l^+ + \textrm{h.c.}, 
\end{equation}
with $\alpha=x,y$. 
The total Hamiltonian of the system is then $H=H_\mathrm{s} +H^x_{\mathrm{ph}} + H^y_{\mathrm{ph}} + H_\mathrm{d}$, with $H_{\textrm{s}} = \sum_l\omega_{eg}\sigma_l^z/2$ describing the dynamics of the internal electronic DOFs and the phonon Hamiltonian $H_{\mathrm{ph}}^\alpha$ given by Eq.~\eqref{phononHamiltonian}. 

In the frame rotating with $H_{\mathrm{s}}+H_{\mathrm{ph}}^x+H_{\mathrm{ph}}^y$, and neglecting the leak-out of the localized phonons from each ion pair, the total Hamiltonian can be written as $H=\sum_l H^{l,l+1}$, where  $H^{l,l+1}$ describes the interaction within the $l$-th element, consisting of the ions $l$ and $l+1$ and the two local phonon modes, $c^\alpha_l$ and $c^\alpha_{l+1}$ (with $\alpha=x$ for $l=\mathrm{odd}$ while $\alpha=y$ for $l=\mathrm{even}$), coupled by the two local lasers $\Omega_l^\alpha$ and $\Omega_{l+1}^\alpha$ [see Fig.~\ref{fig5:levels}(a)].
The engineering of the spin--gauge-field interaction becomes independent in each element, with the two local lasers $\Omega_l^\alpha$ and $\Omega_{l+1}^\alpha$ driving designed transitions only to the sidebands involving the gauge-field phonon mode $c_{l}^\alpha$ and the data-bus phonon mode $c_{l+1}^\alpha$. In the following, we take as an example the first element, $(l,l+1)=(1,2)$, $\alpha=x$, of which the relevant internal transitions are illustrated in Fig.~\ref{fig5:levels}(b). 
The corresponding Hamiltonian in the rotating frame is
\begin{equation}
\label{rotatingH}
H^{1,2}= \sum_{l=1}^{2}\frac{\Omega_{l}^x}{2} \textrm{exp}[-i\delta_{l}^{\mathrm{L}x}t+ik_{l}^x x_{l}(t)] \sigma_l^+ + \textrm{h.c.},
\end{equation}
where $\delta_l^{\mathrm{L}x} = \omega_l^{\mathrm{L}x} - \omega_{eg}$ is the detuning between laser frequency and internal transition frequency. 
The momentum-kick from the photon provides the coupling to the vibrational modes. It can be expressed in terms of the two localized phonon modes by 
\begin{eqnarray}
\left( \begin{array}{c}
k_{1}^xx_{1}(t)\\
k_{2}^xx_{2}(t)
\end{array} \right)=&&\left( \begin{array}{cc}
\eta^x_{1,1}\cos\theta_{1,2}&\eta_{1,2}^x\sin\theta_{1,2}\\
-\eta^x_{2,1}\sin\theta_{1,2}&\eta^x_{2,2}\cos\theta_{1,2}\end{array} \right)
\left( \begin{array}{c} 
c_{1}^x e^{-i\epsilon_{1}^x t}\nonumber\\
c_{2}^x e^{-i\epsilon_{2}^x t}\end{array} \right) \\
&&+ \textrm{h.c.}\,,
\end{eqnarray}
where the angle $\theta_{1,2}$ is given in Eq.~\eqref{phononDistributionAngle}, and the Lamb--Dicke parameter is defined as $\eta_{lq}^x = k_{l}^x/\sqrt{2M_{\mathrm{I}}\epsilon_{q}^x}$. In the following, we shall neglect the small difference between the Lamb--Dicke parameter defined inside the same block, i.e., we take $\eta^x_{1(2),1(2)}\simeq\eta_{1,2}^x$.

We assume the near-resonant condition $\delta_{1}^{\textrm{L}x}=\epsilon_{1}^x+\epsilon_{2}^x+\delta+\mu$ and $\delta_{2}^{\mathrm{L}x}=\epsilon_{2}^x+\delta-\mu$, with $\delta\ll\epsilon^x_{1(2)}$. Here, $\mu\ll\delta$ is a small offset, which will contribute to the correct alternating mass term of the spins. In the Lamb--Dicke regime $\eta_{1,2}^x\sqrt{N}\ll 1$, we can expand the photon kick into a series of phonon-sideband transitions. Here, we consider only the near-resonant terms, as shown schematically in Fig.~\ref{fig5:levels}(b), and defer the analysis of the impact of far off-resonant transitions to other sidebands to Appendix~\ref{AC-Stark shifts and their compensations in the HOBM scheme}.  The near-resonant transitions of ion 1 consist of the three blue second-sidebands of the two local phonons, of which the transition frequencies are $\omega_{eg} + 2\epsilon_1^x,\omega_{eg}+\epsilon_1^x+\epsilon_2^x,\omega_{eg}+2\epsilon_2^x$ respectively. The near-resonant transitions of ion 2 include the two blue first-sidebands, with transition frequencies $\omega_{eg}+
\epsilon_1^x$ and $\omega_{eg}+\epsilon_2^x$. Keeping only these near-resonant sidebands, we thus have
\begin{eqnarray}
\label{sidebandTransitionsList}
H^{1,2} =-\frac{1}{2}\Big[&&f_{1,2}\cos\theta_{1,2}\sin\theta_{1,2}\sigma_{1}^+ c_{1}^{x\dag} c_{2}^{x\dag}e^{-i(\delta+\mu) t}\nonumber\\
&&+f_{1,2}\cos^2\theta_{1,2}\sigma_1^+ (c_1^{x\dag})^2 e^{-i(\delta+\epsilon_2^x-\epsilon_1^x+\mu)t}\nonumber\\
&&+f_{1,2}\sin^2\theta_{1,2}\sigma_1^+ (c_2^{x\dag})^2 e^{-i(\delta-\epsilon_2^x+\epsilon_1^x+\mu)t}\nonumber\\
&&+g_{1,2}\sin\theta_{1,2}\sigma_2^+c_{1}^{x\dag}e^{-i(\delta+\epsilon_2^x-\epsilon_1^x-\mu)t}\nonumber\\
&&- g_{1,2}\cos\theta_{1,2}\sigma_{2}^+ c_{2}^{x\dag}e^{-i(\delta-\mu) t}\Big]+ \textrm{h.c.},
\end{eqnarray}
with $f_{1,2}=\Omega_{1}^x(\eta_{1,2}^x)^2$ and $g_{1,2}=i\Omega_{2}^x\eta^x_{1,2}$, which measure the strengths of the corresponding sideband transitions of each ion. Further, assuming the conditions $|f_{1,2}|\sqrt{N}\sin\theta_{1,2},|g_{1,2}|\cos\theta_{1,2}\ll 2|\delta|$ and $|f_{1,2}|N, |g_{1,2}|\sqrt{N}\sin\theta_{1,2}\ll 2|\delta \pm (\epsilon_2^x-\epsilon_1^x)|$, the terms in Eq.~\eqref{sidebandTransitionsList} are off-resonant due to their relatively small sideband-transition strengths. In these conditions, we assumed that the operators $c_1^x$ describe the gauge-field mode, which has occupation $\sim N$, while the operators $c_2^x$ describe the quantum data-bus with occupation $\simeq 0$.

The system dynamics is governed by the effective Hamiltonian $H_{\mathrm{eff}}^{1,2}$ obtained through second-order perturbation theory on top of Eq.~\eqref{sidebandTransitionsList}, taking all resonant processes into account. $H_{\textrm{eff}}^{1,2}$ can be decomposed into three contributions, $H_{\textrm{eff}}^{1,2}=H_{\textrm{sg}}^{1,2}+H_{\textrm{sm}}^{1,2}+H_{\textrm{ls,n}}^{1,2}$, in which $H_{\textrm{sg}}^{1,2}$ and $H_{\textrm{sm}}^{1,2}$ are the desired spin--gauge-field coupling and spin-mass term in the HOBM respectively, while $H_{\textrm{ls,n}}^{1,2}$ describes undesired AC-Stark shifts (light shifts) induced by the excitation--de-excitation cycle involving the near-resonant sideband transitions of individual ions. Explicitly, in the frame where the pseudo-spins rotate with frequency $\mu$, we have 
\begin{equation}
\label{eq:Hsm}
H_{\textrm{sm}}^{1,2}=\frac{\mu}{2}(\sigma^z_{2}-\sigma^z_{1})\,, 
\end{equation}
whereas the effective spin--gauge-field coupling comes from two resonant second-order processes on top of Eq.~\eqref{sidebandTransitionsList}, one involving the first and the fifth line therein, and the other involving the second and the fourth line,
\begin{equation}
\label{eq:Hvert}
H_{\textrm{sg}}^{1,2} = -\frac{1}{\sqrt{N}} J_{1,2} \sigma_{1}^+ c_{1}^{x\dag}\sigma_{2}^- +\textrm{h.c.},
\end{equation}
with the effective tunnelling strength $J_{1,2}=\sqrt{N}f_{1,2}g_{1,2}^{*}\cos^2\theta_{1,2}\sin\theta_{1,2}[1/(4\delta)-1/2(\delta+\epsilon_2^x-\epsilon_1^x)]$. 
Other second-order processes generate undesired additional AC-Stark shifts $H_{\mathrm{ls,n}}^{1,2}$. In Appendix~\ref{AC-Stark shifts and their compensations in the HOBM scheme}, we make a thorough analysis of these terms, together with the AC-Stark shifts from far off-resonant sideband transitions. We show they are \emph{gauge-invariant}, and can be highly suppressed, thus inducing negligible detriment to the reliability of the proposed quantum simulator.

The above analysis applies equally to any other element involving ions $l$ and $l+1$, as long as the labels are replaced correspondingly, i.e., $1\to l$, $2\to l+1$, and $x\to\alpha$ with $\alpha=x$ for $l=\mathrm{odd}$, while $\alpha=y$ for $l=\mathrm{even}$. To contribute correctly to the staggered mass term, the small detuning offset $\mu$ should be replaced by $\mu\to(-1)^{l+1}\mu$ for the ion pair $(l,l+1)$. The tunneling strength in each element can be made constant, i.e., $J_{l,l+1}=J$, by tuning the Rabi frequency $\Omega_l^{x(y)}$ of the local lasers. The effective Hamiltonian for the whole system is thus $H_{\mathrm{eff}}=\sum_l H_{\mathrm{eff}}^{l,l+1}$. 
This correctly reproduces the spin--gauge-field couplings and spin mass terms in Eq.~\eqref{HOBMHamiltonian} via Eqs.~\eqref{eq:Hsm} and~\eqref{eq:Hvert}.

\subsubsection{Gauge-field energy: phonon nonlinearity}
\label{phonon nonlinearity}
To complete the HOBM Hamiltonian, Eq.~(\ref{HOBMHamiltonian}), we require the gauge-field energy term $\propto V$. It can be realized in a similar manner as proposed in Ref.~\cite{Porras2004b}. For the element consisting of the ion pair $(l,l+1)$, we apply a standing-wave laser beam to the $l$-th ion, of which the wavevector is along the $x(y)$ direction depending on $l=\mathrm{even(odd)}$ [see Fig.~\ref{fig5:levels}(a)]. The applied standing-wave field is far off-resonant to the internal electronic transitions of the ions. Nevertheless, it induces an appropriate amount of nonlinearity to the local phonon modes through the position-dependent AC-Stark shift. 

We assume the pair of levels forming the pseudo-spin, $\ket{g}_{l}$ and $\ket{e}_{l}$, receive the same AC-Stark shift from the standing-wave light field. This can be realized, e.g., in the hyperfine-qubit configuration for $^9$Be$^+$ ions, where the pseudo-spin is encoded by the two hyperfine levels $(F,m_F)=(1,1)$ and $(F,m_F)=(2,2)$ of the $^2S_{1/2}$ manifold, as depicted in Fig.~\ref{fig5:levels}(c). We take the element $(l,l+1)=(1,2)$ as an example. 
Two phase-locked counter-propagating lasers are applied to ion 1 along the $x$ direction, which are linearly polarized along the quantization axis of the electronic DOFs of the ions thus form a standing wave. 
It drives off-resonantly the optical transitions between the $^2S_{1/2}$ and the $^2P_{1/2}$ manifold. The two off-resonant transitions shown in (c) have the same Rabi frequency, with a typical value of $\Omega_1^\mathrm{sw}\sim2\pi\times 1$GHz at the anti-node of the standing wave. The detuning of the standing wave has a typical value of $\Delta_1^\mathrm{sw}\sim 2\pi\times 1$THz, far larger than the hyperfine splitting in the ground/excited state manifold (on the order of $10$GHz~\cite{Bollinger1985}). As a result, the standing-wave laser creates nearly equal AC-Stark shifts to both spin-up and spin-down states.

We assume the equilibrium position of ion 1 is at one of the anti-nodes of the applied standing wave laser. The position-dependent AC-Stark shift of ion 1 can be written as 
\begin{eqnarray}
H_{\textrm{sw}}^{1,2}&=&\frac{|\Omega^{\mathrm{sw}}_1|^2}{4\Delta^{\mathrm{sw}}_1}\textrm{cos}^2(k^{\mathrm{sw}}_1 x_{1}) \\
&=&  \frac{|\Omega^\mathrm{sw}_1|^2}{4\Delta^\mathrm{sw}_1}\bigg(\alpha + \beta c_{1}^{x\dag} c_{1}^x+\gamma (c_{1}^{x\dag} c_{1}^x)^2\bigg) + \mathcal{O}\left[(\eta_1^{\mathrm{sw}})^4\right]\nonumber\,
\end{eqnarray}
where $2\pi/k^{\mathrm{sw}}_1$ is the spatial periodicity of the standing wave, $\alpha=1-(\eta^{\mathrm{sw}}_1)^2+(\eta^{\mathrm{sw}}_1)^4$, $\beta=-2(\eta^{\mathrm{sw}}_2)^2(1+(\eta^{\mathrm{sw}}_1)^2)\cos^2\theta_{1,2}$, $\gamma=2(\eta^{\mathrm{sw}}_{1})^4\cos^4\theta_{1,2}$, and $\eta^{\mathrm{sw}}_1=k^{\mathrm{sw}}_1/\sqrt{2M_{\mathrm{I}}\epsilon_{1}^x}$, and non-resonant terms have been neglected in the rotating-wave approximation. After reabsorbing the frequency correction $\propto \beta$ into the local vibrational frequency $\epsilon_{1}^x$, we arrive at the desired gauge-field energy term with effective coupling strength $V=\gamma|\Omega^{\mathrm{sw}}_1|^2/{4\Delta^{\mathrm{sw}}_1}$. By replacing the labels $(1,2)\to(l,l+1)$ and $c_1^x\to c_{l}^{x(y)}$ depending on $l=\mathrm{odd(even)}$, the above analysis applies equally to any other element $(l,l+1)$. By tuning the Rabi frequency of the standing-wave lasers in each element, we can adjust their phonon nonlinearity to the same value $V$, which gives the desired gauge-field energy term in Eq.~(\ref{HOBMHamiltonian}).

\subsection{Experimental feasibility}
\label{sec3: Experimental feasibility}
Having outlined the construction of a trapped-ion quantum simulator for the HOBM in Sec.~\ref{sec3A: Trapped-ion implementation of the HOBM}, in this section we discuss its experimental feasibility. In Sec.~\ref{sec3: Experimental feasibility}, we first analyze the scalability and error sources of the proposed scheme. Then, we discuss in Sec.~\ref{surfaceTrapRealization} the possibility to realize it in an array of individual traps. Based on these analyses,  in Sec.~\ref{sec3B2: numericsHOBMsimumator} we perform a numerical study of the predicted performance of such a trapped-ion quantum simulator.

\subsubsection{Practical limitations and imperfections}
\label{sec3: Experimental feasibility}
As discussed in Sec.~\ref{sec3A: Trapped-ion implementation of the HOBM}, the segmentation scheme of the micro-trap frequencies allows for scaling up to several tens of ions, as long as the radial phonon modes are effectively localized. Nevertheless, there exists another technical restriction for the number of ions. The HOBM scheme requires initial preparation of the gauge-field phonons into a Fock state with phonon number $N$, and quantitatively approaches $(1+1)$D lattice QED only in the regime $N\geq L$, where $L$ is the number of lattice sites. The useful number of ions $L$ in the quantum simulator is thus in practice limited by the maximum achievable phonon Fock-state in the initial-state preparation. Already twenty years ago, experiments using repeated sideband-pulses have created Fock states with $N=16$ with good accuracy~\cite{Meekhof1996}, and there seem to be no fundamental roadblocks for the preparation of higher Fock states. Our envisioned quantum simulator thus has the scalability to a few tens of ions, when implemented with current trapped-ion technology.

Additionally, to simplify the analyses of the sideband selections in Sec.~\ref{sec:SpinGaugeFieldCoupling}, we have assumed the perfect localization of the radial phonon modes within each ion pair. In practice, the small but non-zero leak-out of these local phonons communicates with the ions far away. Thus, the addressing lasers for each ion pair also stimulate sideband transitions involving phonon modes outside the pair. They correspond to an unphysical, gauge-symmetry violating process where a fermion is created but the gauge field changes at a different link. Fortunately, such sideband transitions are extremely weak, as their strength is proportional to the phonon leakage (which is extremely small, see Fig.~\ref{fig4:setup}(d) as a realistic example), and are highly off-resonant, thanks to the large frequency difference between the phonons localized in different ion pairs. Thus, they induce negligible detriments to the quantum simulation, and the ideal localization of the phonons remains an excellent approximation.

Finally, in the design of the spin--gauge-field interaction, the laser configuration in Fig.~\ref{fig5:levels}(b) inevitably induces AC-Stark shifts due to off-resonant virtual population of the phonon sidebands of individual ions. As explained in detail in Appendix~\ref{AC-Stark shifts and their compensations in the HOBM scheme}, upon appropriate compensation by additional laser beams in each element, all the relevant AC-Stark shifts can be summed up to a compact form
\begin{equation}
\sum_{l=1}^{L-1}\sum_{m=l}^{l+1}[E_m^{l,l+1}+F_{m}^{l,l+1}(c_{l}^{\alpha\dag}c^\alpha_{l}-N)]\sigma_m^z,
\end{equation}
with $\alpha=x$ for $l=\mathrm{odd}$, and $\alpha=y$ for $l=\mathrm{even}$. The energy scale $E_l^{l,l+1}$ for the phonon-independent AC-Stark shift is typically larger than the working energy scale of the quantum simulator. However, it can be compensated nearly completely by adjusting the detuning of the local laser beams, $\delta_{l}^{\textrm{L}x}\to\delta_{l}^{\textrm{L}x}+2E_l^{l-1,l}+2E_l^{l,l+1}$ and $\delta_{l}^{\textrm{L}y}\to\delta_{l}^{\textrm{L}y}+2E_l^{l-1,l}+2E_l^{l,l+1}$. The phonon dependent AC-Stark shifts cannot be compensated by adjusting the laser frequency, but their energy scale $F_m^{l,l+1}$ can be made far smaller than the working energy scale of the quantum simulator, as quantified in Appendix~\ref{AC-Stark shifts and their compensations in the HOBM scheme}, using typical experimental parameters. Thus, once compensated properly, the AC-Stark shifts induce negligible errors to the performance of the quantum simulator.

\subsubsection{Experimental parameters in a surface-trap realization}
\label{surfaceTrapRealization}
The envisioned array of micro-traps is best realized by current microfabricated surface traps~\cite{Harlander2011,Wilson2014,Mehta2014,Mielenz2015}, where the ions are trapped above the plane of electrodes in electromagnetic potential landscapes.  
These offer the necessary control to implement the segmented trapping potential described in Sec.~\ref{Employed degrees of freedom: pseudo-spins and localized phonon modes}. 
In this section, we consider such a surface-trap realization, by presenting typical experimental parameters and analyzing possible experimental imperfections. 

We take the microfabricated surface traps for $^9$Be$^+$ ions in the NIST group~\cite{Wilson2014} as a concrete illustration. We consider a trap array consisting of 2 blocks of segmented traps, i.e., 12 ions in total, built according to the scheme described in Sec.~\ref{Employed degrees of freedom: pseudo-spins and localized phonon modes}, with the same trap parameters as therein. The gauge-field phonons are prepared initially in the Fock state with $N=10$. Since the engineering of the HOBM Hamiltonian is independent in each element consisting of two nearest-neighbor ions $(l,l+1)$, we once again take the first element $(l,l+1)=(1,2)$ as an example. The trapping frequencies, as selected in Sec.~\ref{Employed degrees of freedom: pseudo-spins and localized phonon modes}, lead to $\theta_{1,2}=0.25$ for the local-phonon distribution, and $\epsilon_2^x-\epsilon_1^x\simeq2\pi\times 10$kHz for the mode splitting. We assume $\eta_{1,2}^x=0.08$, thus fulfilling the Lamb--Dicke condition $\eta_{1,2}^x\sqrt{N}\ll1$. By choosing $\delta=-2\pi\times 50$kHz, the two lasers are near-resonant to the desired sideband transitions [see Fig.~\ref{fig5:levels}(b)], while other sidebands are detuned at least $\sim\omega_{x(y)}=2\pi\times 5 $MHz. We further assume the Rabi frequencies $\Omega_1^x=2\pi\times 180$kHz and $\Omega_2^x=2\pi\times 210$kHz, which leads to the corresponding sideband-transition strengths $f_{1,2}=2\pi\times 1.2$kHz and $g_{1,2}=2\pi\times 17$kHz. As a result, the effective spin--gauge-field coupling strength becomes $J=2\pi\times 120$Hz, on the order of energy scales of current experiments with effective spin models~\cite{Johanning2009,Blatt2012,Schneider2012}. By adjusting the small frequency offset $\mu$, one can create easily the spin mass term at the same energy scale as $J$. To generate the desired gauge-field nonlinearity, we choose $\Omega^{\mathrm{sw}}_1\simeq 2\pi\times 1 $GHz, $\Delta^{\mathrm{sw}}_1\simeq 2\pi\times 1$THz, and $\eta^{\mathrm{sw}}_1\simeq 0.08$, yielding $V\simeq 2\pi\times 20$Hz.

Realistic surface ion traps contain uncertainties in the control of the trap frequencies as aimed at by Eqs.~\eqref{generalizedTrapFrequencyDesignX} and \eqref{generalizedTrapFrequencyDesignY}, due to the complexity in engineering the potential landscapes above the electrodes~\cite{Schmied2009, Mielenz2015}. Nevertheless, the effectiveness of the frequency-segmentation only relies on the clear separation of the energy scales, Eq.~\eqref{GeneralLocalizedPhononCondition}. As long as the radial frequencies of individual traps are controlled to an accuracy significantly smaller than $\Delta_B$, the phonon modes are still localized efficiently. To reflect the limited accuracy of frequency-control we add a local randomness, uniformly distributed in the interval $2\pi\times [-30,30]$kHz, to the radial frequency of each trap. As shown in Fig.~\ref{fig4:setup}(d), the calculated normal phonon modes are still highly concentrated in pairs of ions. Thus, the full scheme presented in Sec.~\ref{sec3A: Trapped-ion implementation of the HOBM} applies well, even with such experimental imperfections.

Among the sources of decoherence in trapped-ion systems, phonon heating due to the electromagnetic field noise is the most detrimental one. Operating at cryogenic temperature can reduce phonon heating significantly. For example, the phonon heating rate for axial phonons at $\omega_z\simeq 2\pi\times 2.3$MHz can be reduced as low as $70/$s for ion-spacing $d\sim 30\mu$m in the cryogenic surface traps in the NIST group~\cite{Brown2011}. Even lower phonon heating rates are being actively pursued~\cite{McConnell2015}. We expect similar low heating rates for the radial phonon modes considered here. Additionally, fluctuations of the global magnetic field (which determines the quantization axis of the pseudo-spins) result in dephasing of the internal DOFs of the ions. Nevertheless, for Hamiltonians that preserve the total polarization $\sum_l\sigma_l^z$, e.g., for the HOBM Hamiltonian Eq.~\eqref{HOBMHamiltonian} considered here, the dynamics in decoherence-free subspaces possesses coherent evolution for as long as 10ms, as shown in Refs.~\cite{Jurcevic2014,Jurcevic2015}. Since these decoherence rates are smaller by nearly one order of magnitude than the working energy scale of the proposed quantum simulator, we expect that they have a small effect on the performance of the proposed quantum simulator.

\subsubsection{Numerical studies of a modest-size quantum simulator}
\label{sec3B2: numericsHOBMsimumator}
\begin{figure*}[t]
\begin{center}
\includegraphics[width=0.99\textwidth]{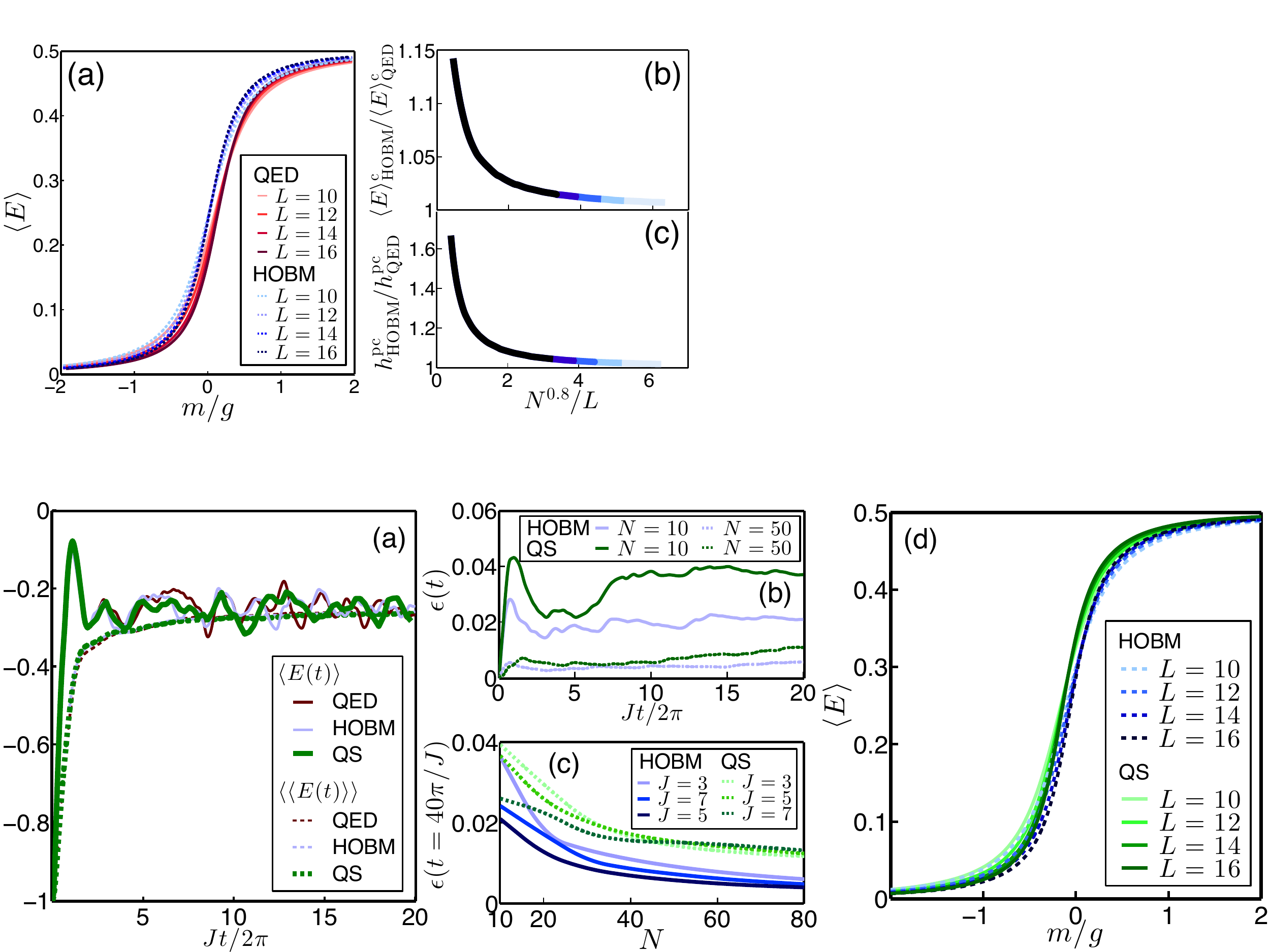}
\caption{Numerical predictions for string breaking and Coleman's quantum phase transition on a modest-size quantum simulator, built according to the HOBM scheme proposed in Sec.~\ref{sec3}, compared to that of the ideal HOBM and $(1+1)$D QED. (a) String breaking dynamics in a system of $L=12$ ions, over a relatively long evolution period $t\in [0,40\pi/J]$, to benchmark the imperfections induced by the AC-Stark shifts. The experimental parameters are chosen so that $J=2\pi\times 120$Hz, $V=\mu=2\pi\times 25$Hz, and the local radial phonon modes which encode the gauge-field DOFs are prepared into an initial Fock state with occupation number $N=10$. (b,c) Similar to the case of the ideal HOBM, Fig.~\ref{stringbreaking}, the time-averaged error for the quantum simulator, $\epsilon(t)=\int_0^t dt' |\langle E(t')\rangle_{\textrm{QS}} - \langle E(t')\rangle_{\textrm{QED}}|/t$, is bounded during time evolution and is suppressed with increasing boson number $N$. The error is dominated by the AC-Stark shifts, which however only induce small deviations from the ideal results and do not break the Gauss law. (d) Comparison of the critical behavior of the order parameter $\langle E\rangle$, between the quantum simulator and the ideal HOBM, in the Coleman's quantum phase transition at vacuum angle $\pi$. We choose $J=2\pi\times 120$Hz and $V=2\pi\times 25$Hz, yielding $g=2\sqrt{JV}\simeq 2\pi\times 55$Hz. By changing the small detuning $\mu=m$, we scan across the quantum critical point, and find that the quantum simulator represents the critical behavior of the HOBM faithfully.}
\label{fig5QSperformance}
\end{center}
\end{figure*}

Based on the experimental parameters in Sec.~\ref{surfaceTrapRealization}, in this section we study numerically the string-breaking dynamics and Coleman's quantum phase transition on a modest-size quantum simulator consisting of $L=12$ ions, as a quantitative measure of the reliability of the present proposal. To estimate the potential errors in the performance of the proposed quantum simulator, we compare the behaviour of the ideal model, Eq.~\eqref{HOBMHamiltonian}, to the dynamics taking the systematic imperfections due to AC-Stark shifts into account, choosing experimentally feasible parameters $J=2\pi\times 120$Hz, and $V,\mu$ on the order of $2\pi\times 20$Hz. The initial boson occupation offset is $N=10$, realistic for current trapped-ion technology.

Figure~\ref{fig5QSperformance}(a-c) presents the expected string-breaking dynamics. The exact evolution of the space-averaged electric field $\langle E(t) \rangle$ and the space- and time-averaged electric field $\langle\langle E(t) \rangle\rangle$ are compared to the behavior of the ideal HOBM as well as lattice QED. 
The systematic AC-Stark shifts only slightly alter $\langle E(t) \rangle$, while the difference in $\langle\langle E(t) \rangle\rangle$ is hard to discern. Moreover, the resulting error is bounded during time evolution [Fig.~\ref{fig5QSperformance}(b)], and decreases with increasing $N$ [Fig.~\ref{fig5QSperformance}(c)]. With other experimental imperfections, especially the decoherence rates discussed in Sec.~\ref{surfaceTrapRealization}, it will be possible to observe the dynamics of string breaking over several steps of $1/J$.

In Fig.~\ref{fig5QSperformance}(d), we show the behavior of the space-averaged electric field across the parity-symmetry-breaking phase transition at vacuum angle $\pi$, and compare it to that of the HOBM. Again, the error due to the AC-Stark effect only slightly shifts the critical point, and the quantum simulator is expected to represent the ground-state phase diagram well.

\section{a quantum simulator of the $S=1/2$ QLM: the energy lattice scheme}
\label{sec4: QLM}
The quantum simulator for the HOBM proposed above allows for the exploration of rich physics and has good scalability, but requires an advanced experimental apparatus with high tunability of vibrational frequencies. In this section, we propose and discuss an alternative scheme based on the $S=1/2$ QLM (see Sec.~\ref{QLMformulation}), which is feasible with current trapped-ion technology such as available in the setup described in Ref.~\cite{M.Johanning2009,Kim2009}. This scheme requires only phase-coherent second sideband transitions, and even improves on the energy scale compared to the HOBM scheme discussed in the preceding sections.  
These advantages come at the price of certain intrinsic (though \emph{gauge-invariant}) errors, which reduce the quantitative agreement with the ideal QLM for increasing ion number, making the quantum simulator proposed here ideal for small-scale proof-of-principle experiments.  

To benchmark the proposed quantum simulator, we study the quench dynamics across the parity-symmetry-breaking quantum phase transition present in the $S=1/2$ QLM, as a simple example of false vacuum decay, a non-perturbative phenomenon that exists commonly also in more complicated gauge theories~\cite{Hooft2005}. 
Below, we first briefly introduce the model we exploit to design the quantum simulator, a bosonic realization of the $S=1/2$ QLM. Then, we move on to discuss in detail a possible implementation in a trapped-ion setup, as well as potential sources of errors.

\subsection{Bosonic realization of the $S=1/2$ QLM}
\label{Bosonic realization of the $S=1/2$ QLM}
As introduced in Sec.~\ref{QLMformulation}, QLMs can be viewed as a special class of spin models with three-body interaction terms plus gauge constraints. Thus, in principle they can be realized by exploiting the internal pseudo-spin DOFs of trapped ions, as discussed in Ref.~\cite{Hauke2013b}. However, the energy scale of such higher-oder spin--spin interactions is suppressed by the small Lamb--Dicke parameter $\eta$. 
An interesting feature of the $S=1/2$ QLM can help overcome this difficulty: the spin matter $\sigma_i$ can be replaced by bosonic DOFs $c_i$. As we shall prove in a moment, in the physical gauge sector prescribed by the Gauss law, double occupation of bosons is forbidden, thus making the bosonic theory equivalent to the original $S=1/2$ QLM. This fact allows us to exploit the phonon DOFs in the ion trap to design a quantum simulation scheme that works on an energy scale containing less powers of Lamb--Dicke parameters.

To arrive at the bosonic realization of the $S=1/2$ QLM, we start from its spin version, which is obtained straightforwardly by the substitutions $U_{i,i+1}\to s_{i,i+1}^+$, $E_{i,i+1}\to s_{i,i+1}^z$ in Eqs.~\eqref{SchwingerHamiltonianJordanWigner} and \eqref{GaussLawJordanWigner}. Note that the matter spin $\tau_i$ in Eq.~\eqref{SchwingerHamiltonianJordanWigner} and \eqref{GaussLawJordanWigner} are Pauli matrices, while the gauge spin $s_{i,i+1}$ here are $S=1/2$ representation of the angular-momentum algebra (which contain a prefactor $1/2$ compared to Pauli operators).
Further, it proves convenient to remove the alternating signs in these equations by a staggered rotation~\cite{Hauke2013b}, $U=\prod_{i=\mathrm{odd}}\mathrm{exp}[-i\pi(\tau_i^x/2+s_{i-1,i}^x)]$, to both the matter ($\tau_i$) and the gauge ($s_{i,i+1}$) spins, resulting in the transformation $U\tau_i^{z(y)}U^\dag = (-1)^i \tau_i^{z(y)}$, $Us_{i,i+1}^{z(y)}U^\dag = (-1)^{i+1} s_{i,i+1}^{z(y)}$. The spin Hamiltonian is then transformed to
\begin{equation}
\label{staggeredrotatedQLM}
H = -J \sum_i (\tau_i^- s_{i,i+1}^+ \tau_{i+1}^- + \textrm{h.c.}) + \frac{\mu}{2} \sum_i \tau_i^z,
\end{equation}
while the local gauge generator is converted to 
\begin{equation}
\label{staggeredrotatedGaugeConstraint}
G_i = (-1)^i [\frac{1}{2}(\tau_i^z+1)+ (s_{i,i+1}^z + s_{i-1,i}^z)].
\end{equation}
The prefactor $(-1)^i$ in Eq.~\eqref{staggeredrotatedGaugeConstraint} is unimportant as we stay in the physical Coulomb sector $\mathcal{G}$ prescribed by $G_i\mathcal{G}=0$, and will be omitted in the following.

The bosonic realization of the $S=1/2$ QLM is obtained simply by replacing the spin matter by bosonic DOFs in Eq.~\eqref{staggeredrotatedQLM},
\begin{equation}
\label{staggeredrotatedBosonicQLM}
H = -J \sum_i (c_i s_{i,i+1}^+ c_{i+1} + \textrm{h.c.}) + \mu \sum_i c_i^\dag c_i,
\end{equation}
and similarly for the local gauge generator
\begin{equation}
\label{staggeredrotatedBosonicGaugeConstraint}
G_i = c_i^\dag c_i + (s_{i,i+1}^z + s_{i-1,i}^z).
\end{equation}
Compared to the HOBM, Eqs.~\eqref{HOBMHamiltonian} and \eqref{BosonAppGauss}, the role of bosons and pseudo-spins is interchanged. 

The proof that Eqs.~\eqref{staggeredrotatedBosonicQLM} and \eqref{staggeredrotatedBosonicGaugeConstraint} yield a valid gauge theory in the QLM formalism is straightforward: The quantum link operators $s_{i,i+1}$ are $S=1/2$ spins with eigenvalues $\pm1/2$. When enforcing the Gauss law $G_i\mathcal{G}=0$, the boson number $c_i^\dag c_i$ at a single site can only be $0$ or $1$. This means in the physical sector $\mathcal{G}$ the bosons are equivalent to $S=1/2$ spins. We note, however, such a feature is unique only to the $S=1/2$ QLM. For $S\geq1$ the Gauss law does not forbid higher occupancy of the bosonic DOFs, and the representation of matter field by bosons is no longer valid. 

\begin{figure}[t]
\begin{center}
\includegraphics[width=0.999\columnwidth]{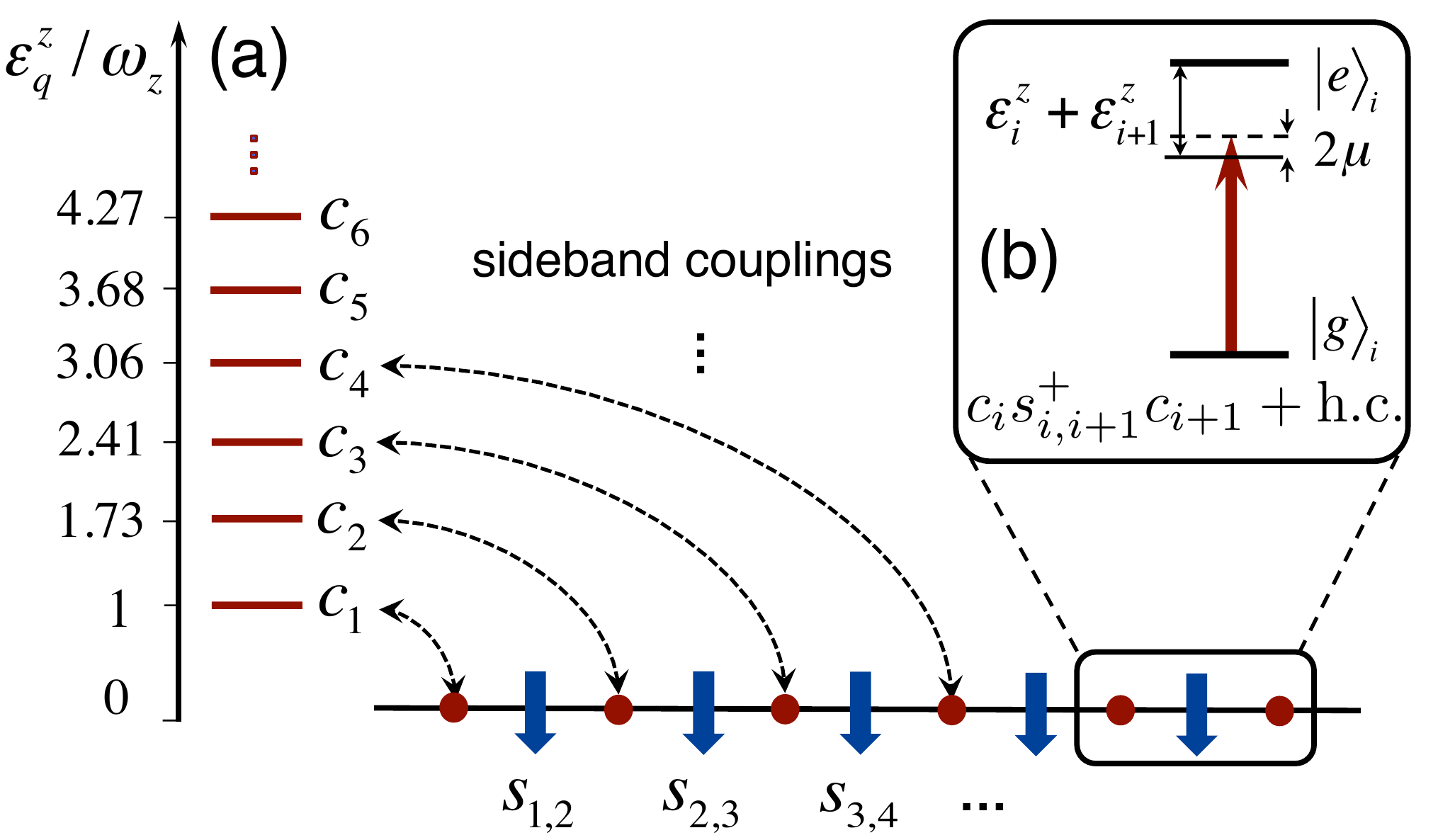}
\caption{The energy lattice scheme for the quantum simulation of a $S=1/2$ $U(1)$ quantum link model in $(1+1)$ space-time dimensions. 
(a) In this scheme, the $S=1/2$ gauge link variable $s_{i,i+1}$ is encoded in the internal pseudo-spin states of the $i$-th ion, $\ket{g}_i$ and $\ket{e}_i$. The matter field $c_i$ is realized by the collective phonon mode $c_q^z$ with same index number, ordered by energy $\epsilon_q^z$. 
The Gauss law, implemented by initial state preparation, forbids double excitation of the phonons, thus rendering the bosonic model a valid $S=1/2$ QLM. 
(b) The gauge-field--matter interaction can be realized by near-resonant second red-sideband transitions on individual ions, which can be implemented straightforwardly with single-ion addressability.
}
\label{fig5:energyLatticeScheme}
\end{center}
\end{figure}

\subsection{Implementation with trapped ions}
\label{subsec:Implementation with trapped ions}
To implement the QLM Hamiltonian, Eq.~\eqref{staggeredrotatedBosonicQLM}, 
we consider an array of $L$ ions confined in a standard linear Paul trap, with axial collective phonon modes $c_q^z$, $q=1\dots L$, distributed among all ions. 
As sketched in Fig.~\ref{fig5:energyLatticeScheme}, we order these by increasing energy (i.e., $\epsilon_q^z<\epsilon_{q+1}^z$ with $q=1$ being the axial COM mode) and identify each bosonic matter field $c_i$ in Eq.~\eqref{staggeredrotatedBosonicQLM}, $i=1\dots L$, with the phonon mode $c_q^z$ with the same index number, $q=i$. 
To implement periodic boundary conditions, we further make the identification $c_{L+1}=c_1$. 
This construction effectively yields a lattice in energy with sites labeled by $q$. 
The $S=1/2$ link variables $s_{i,i+1}$, on the other hand, are encoded in the internal states of the ions. 

The matter--gauge-field interaction term, $c_i s_{i,i+1}^+ c_{i+1}$, can then be realized as a set of properly designed second sideband transitions, which can be achieved straightforwardly with the capability of single-ion addressing. 
For  concreteness, we consider a microwave driving scheme as in Ref.~\cite{Piltz2015}, although the idea also applies to laser driving schemes with coherent single-site addressability (see, e.g., Refs.~\cite{Kim2009}). In this scheme, each ion is driven independently by a plane-wave microwave field (with frequency $\omega^{\mathrm{M}}_l$ and Rabi frequency $\Omega_l$). 
Assuming that the microwave fields are applied along the axial $z$ direction (with wavevector $k_l$), the total Hamiltonian for such radiation-ion interaction can be written in the frame rotating with the spin and phonon frequencies [similar to Eq.~(\ref{rotatingH}), but coupling to axial collective phonon modes] as
\begin{equation}
\label{LaserIonIntQLM}
H_{\textrm{d}}=\frac{1}{2}\sum_{l=1}^L \Omega_l e^{-i\Delta_l t}\textrm{exp}\bigg(i\sum_{q}\eta_{lq}^z M_{lq}^z z_q(t)\bigg)\sigma_l^++\textrm{h.c.},
\end{equation}
with $z_q(t)=c_q^z \textrm{exp}(-i\epsilon_q^zt)+(c_q^z)^\dag \textrm{exp}(i\epsilon_q^zt)$. The detunings are defined as $\Delta_l=\omega^{\mathrm{M}}_l-\omega_{eg}$, while the Lamb-Dicke parameters are $\eta_{lq}^z=k_l/\sqrt{2M_{\mathrm{I}}\epsilon_q^z}$ with $M_{\mathrm{I}}$ the ion mass.

For each ion $l$, the microwave detuning $\Delta_l$ is chosen individually so as to drive the internal transition to the second red sideband, with simultaneous absorption of one phonon with frequency $\epsilon_l^z$ and another one with frequency $\epsilon_{l+1}^z$. For this, we require the near-resonant condition $\Delta_l = -\epsilon_l^z -\epsilon_{l+1}^z+2\mu$, where the small detuning offset $2\mu$ accounts for the nonzero mass of the matter fields. 
In the microwave setup of Ref.~\cite{M.Johanning2009}, the required single-ion addressing is realized in a linear Paul trap, by $(\mathrm{i})$ applying a linear static magnetic gradient to modify the hyperfine transition frequency of the ions along the string and $(\mathrm{ii})$ simultaneously applying microwave fields to drive the transitions of specific ions. Since adjacent ions have a transition frequency difference much larger than the radiation linewidth, individual driving is highly accurate in the sense that crosstalk between different ions and microwave fields is negligible~\cite{Piltz2014}. 

Keeping only the near-resonant terms, and moving into a frame where the axial phonons rotate with the frequency $\mu$, the Hamiltonian Eq.~\eqref{LaserIonIntQLM} takes a form analogous to the QLM Hamiltonian Eq.~\eqref{staggeredrotatedQLM}, 
\begin{equation}
\label{QSHamiltonianQLM}
H = -\sum_l ( J_l c_l \sigma_{l}^+ c_{l+1}+\textrm{h.c.}) + \mu\sum_l c_l^\dag c_l,
\end{equation}
with $J_l=\Omega_{l}\eta_{l,l}^z \eta_{l,l+1}^z M_{l,l}^z M_{l,l+1}^z/2$. By the identification $\sigma_l^+\to s_{l,l+1}^+$, Eq.~\eqref{QSHamiltonianQLM} recovers the QLM Hamiltonian. The tunnelling strength can be made homogeneous, $J_l=J$, by properly arranging the Rabi frequencies $\Omega_l$  along the ion string. 
By exploiting the phonon DOFs to encode the matter field, the present energy lattice scheme improves on the energy scale $O(\eta^4)$ of the proposal~\cite{Hauke2013b} by two order of $\eta$. Moreover, it is on the same order of $\eta$'s as existing quantum simulations of spin models \cite{Johanning2009,Blatt2012,Schneider2012}, with the additional advantage of being a resonant instead of a perturbative interaction. 

Finally, the Gauss law, $G_i\mathcal{G}=0$, with $G_i$ as given in Eq.~\eqref{staggeredrotatedBosonicGaugeConstraint}, is enforced by initial state preparation (in contrast to the proposal~\cite{Hauke2013b} where it is protected by a large energy constraint). 
In the subsequent section, we show that the interaction term realized through Eq.~\eqref{QSHamiltonianQLM} works on an energy scale much higher than the typical decoherence rates in trapped ion experiments, so the Gauss law remains intact during a typical experimental period. 
The employed set of designed second-sideband transitions thus realizes the desired $S=1/2$ QLM. 

\subsection{Performance under realistic imperfections}
\label{sec3c: numericsOnQLM}

\begin{figure}[t]
\begin{center}
\includegraphics[width=0.99\columnwidth]{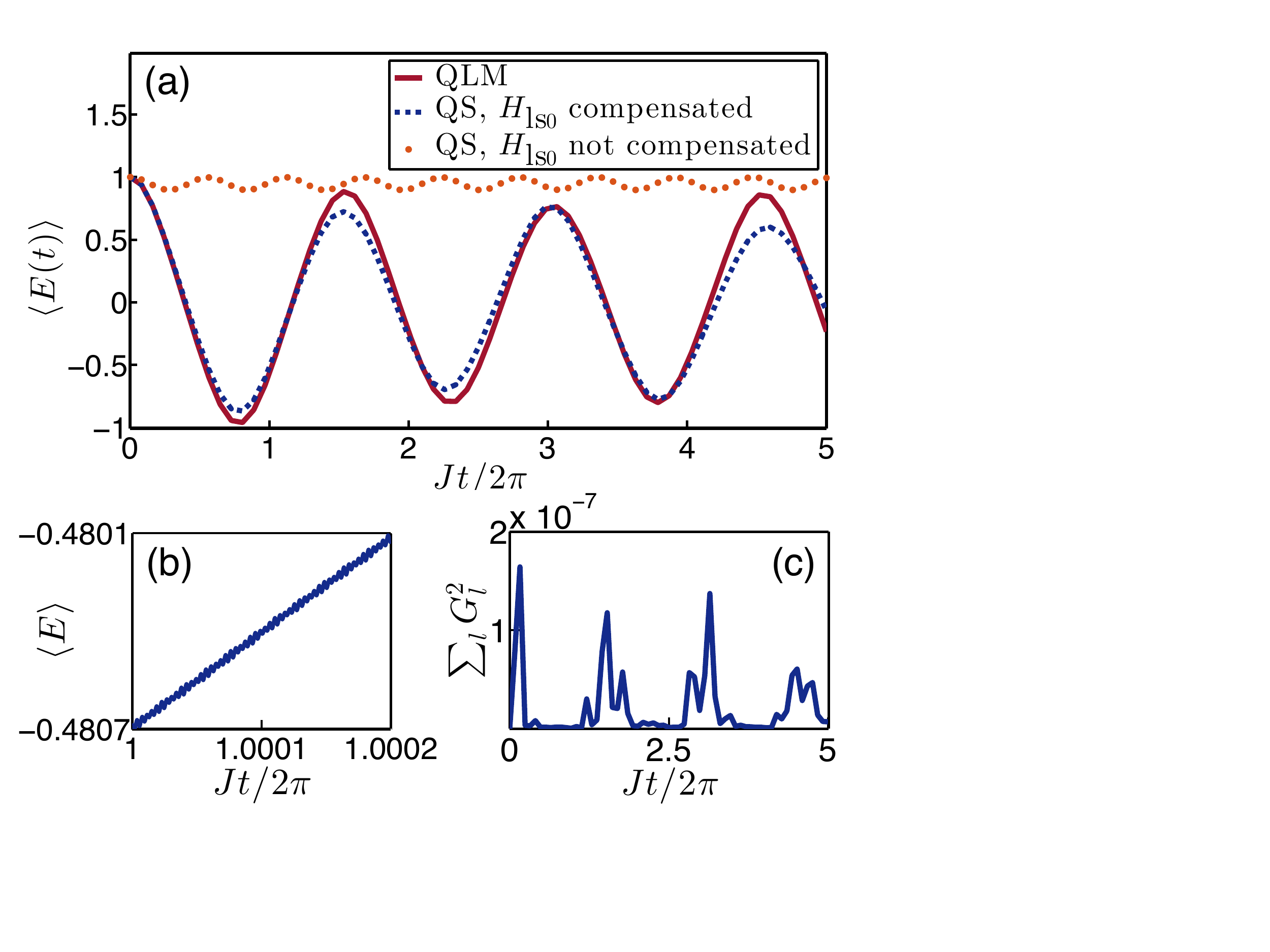}
\caption{Numerical comparison of the dynamics of false vacuum decay, between the ideal $S=1/2$ QLM and a small-scale quantum simulation according to the energy lattice scheme proposed in Sec.~\ref{sec4: QLM}. The lattice size is chosen as $L=6$, and the matter--gauge-field coupling strength $J=2\pi\times 500$Hz. The system is initially prepared in a parity-symmetry-broken ground state at large $\mu$. For $t>0$ it evolves under the Hamiltonian at $\mu=0$, which favours a parity-symmetry-restored ground state. (a) Coherent oscillations of the order parameter $\langle E \rangle = \sum_l (-1)^l\sigma_l^z/L$. The carrier-transition light shift, $H_{\mathrm{ls}0}$, if not compensated, induces significant errors (orange dotted curve). If $H_{\mathrm{ls}0}$ is compensated by a local adjustment of the detuning of the driving frequencies, the quantum simulator (blue dashed curve) simulates the quantum link model (red solid curve) well, with small gauge-invariant errors induced by light shifts from first-sideband transitions. 
(b) Evolution of the order parameter for the quantum simulator, taken from (a) but with amplified resolution. The fast, small-amplitude oscillations are due to off-resonant transitions to other sidebands. They have negligible influence on the quantum simulation. 
(c) The Gauss law remains intact to an extremely high degree during the entire quench dynamics, since the main errors due to AC-Stark shifts are gauge invariant. 
The insignificant violation of the Gauss law is due to small off-resonant transitions to other sidebands, as seen in panel (b).
} 
\label{falsevacuumdecayQLM}
\end{center}
\end{figure}

While the quantum simulator proposed in this section is characterized by excellent energy scales and straightforward experimental implementation, there are nevertheless some possible sources of errors. Most important are off-resonant transitions to sidebands that we neglected in Sec.~\ref{subsec:Implementation with trapped ions}. As we will demonstrate now, these will not break the Gauss law and do not significantly modify the dynamics for small systems. 

For illustration purposes, we study numerically the dynamics of false vacuum decay, a key phenomenon of the $S=1/2$ $U(1)$ QLM, with various error sources taken into account. As mentioned in Sec.~\ref{QLMformulation}, the $S=1/2$ QLM displays a parity-symmetry-breaking quantum phase transition: for small $\mu$ the ground state (the so-called false vacuum) of the system is parity-invariant, indicated by the order parameter $\langle E \rangle=0$, while for sufficiently large $\mu$ the system favors one of the parity-symmetry-broken states as its ground state (the true vacuum), with $\langle E\rangle\neq 0$. For the bosonic version of the QLM realized here, the order parameter is $\langle E\rangle=\sum_l(-1)^l\sigma_l^z/L$, where the alternating minus-sign is due to the staggered rotation leading to Eq.~\eqref{staggeredrotatedQLM}. 
We quench across this transition by initially preparing the system in $\ket{\psi_{\rm spins}}\otimes\ket{\psi_{\rm phonons}}=\ket{gegege}\otimes\ket{000000}$, a parity-symmetry-broken ground state at large $\mu$, and then reducing the mass of the matter field $\mu$ abruptly at time $t=0$ to $\mu=0$, thus realizing an inverse situation to false-vacuum decay. This leads to coherent oscillation of $\langle E\rangle$ in a finite-size system. To calibrate the dynamics of a real system in experiment, we numerically propagate the system under the full Hamiltonian Eq.~\eqref{LaserIonIntQLM} (in a frame rotating with the frequency off-set $\mu$), rather than the near-resonant one Eq.~\eqref{QSHamiltonianQLM}. The phonon operators are truncated to a maximum occupation $N_{\mathrm{max}}^{\mathrm{b}}=4$, sufficiently large for the study here.
In Fig.~\ref{falsevacuumdecayQLM}(a), we show the resulting dynamics in a system with lattice size $L=6$, and compare the ideal evolution to the one with the imperfections due to off-resonant transitions and AC-Stark shifts, which we quantify now. 

As mentioned in the previous section, the idea of the current simulation scheme is to be near-resonant to desired second sidebands while keeping away from any other unwanted transitions. A strong undesired contribution, which however can be compensated, comes from AC-Stark shifts from the carrier transition of each ion, 
\begin{equation}
H_{\mathrm{ls}0} = - \sum_l \frac{ |\Omega_l|^2}{4\Delta_l}\sigma_l^z.
\end{equation}
Its energy scale is significantly higher than the working energy scale of the proposed quantum simulator (which is $\propto\eta^2$). 
However, by controlling the detuning $\Delta_l\to\Delta_l-|\Omega_l|^2/(2\Delta_l)$ of the microwave fields, these phonon-number independent AC-Stark shifts can be compensated nearly completely. 
This is demonstrated in Fig.~\ref{falsevacuumdecayQLM}(a): the uncompensated evolution (orange dotted line) deviates significantly from the ideal evolution, but the compensated dynamics (blue dashed) remains close to the desired one (red solid). Thus, in practice these phonon-number independent shifts prove insignificant. 

The residual deviation from the ideal dynamics seen in Fig.~\ref{falsevacuumdecayQLM}(a) arises mainly from phonon-number dependent AC-Stark shifts due to first sideband transitions, which cannot be compensated in this  manner. They can be written compactly as
\begin{equation}
H_{\mathrm{ls}1} =- \sum_{lq}(E_{lq}^- + E_{lq}^+) c_q^{z\dag} c_q^z \sigma_l^z,
\end{equation}
with $E_{lq}^{\pm}=|\Omega_{l}|^2(M_{lq}\eta_{lq})^2/4(\Delta_{l}\pm\epsilon_q)$, where we omitted the phonon-independent part of $H_{\mathrm{ls}1}$, as it can be compensated along with $H_{\mathrm{ls}0}$.
These AC-Stark shifts describe an undesired interaction between spin and phonon populations. Their energy scale is suppressed by a factor $|\Omega_l|/(\Delta_l\pm\epsilon_q)$ compared to the energy scale of the desired QLM, Eq.~\eqref{QSHamiltonianQLM}. However, their weight grows faster as the number of ions increases, due to the double summation $\sum_{lq}$. This increase limits the scalability of the present scheme to a small number of ions. 
Through numerical studies, we find that good quantitative agreement can be achieved in a standard linear Paul trap for $L\lesssim7$ ions. 
Importantly, however, all the above AC-Stark shifts are gauge invariant errors, as $[H_{\mathrm{ls}0(1)},G_i]=0$. Consequently, they only modify the dynamics within the physical Hilbert space, but do not break the Gauss law, and the model, though modified, remains a valid gauge theory. 

Additionally, there are off-resonant transitions that may break the Gauss law, but these are strongly suppressed by their large detuning. Figure~\ref{falsevacuumdecayQLM}(b) displays a zoom on the time evolution, where all relevant sideband transitions are taken into account. In this strong magnification, one can identify fast but extremely small-amplitude oscillations of the order parameter $\langle E\rangle$. These fast oscillations do not affect the slow dynamics of the false vacuum decay, and they break the Gauss law only extremely weakly, as seen in Fig.~\ref{falsevacuumdecayQLM}(c).

Finally, a realistic quantum simulation has intrinsic sources of decoherence, in particular phonon heating as well as dephasing of the pseudo-spin DOF~\cite{Wineland1998}. 
Nevertheless, these detriments are small compared to the excellent energy scale of the proposed scheme. To be more concrete, we consider $L=6$ ions, with $\Omega_1=2\pi\times 150$kHz, $\eta_{1,1}=\sqrt[4]{3}\eta_{1,2}\simeq 0.15$, and $\Omega_l$ arranged accordingly by requiring $J_l=J_1=J$. The resulting gauge-field--matter coupling strength is $J\simeq 2\pi\times 500$Hz. This energy scale is by one order of magnitude larger than the typical dephasing rate of the spins (about a few hundred Hz~\cite{Monz2011}, since, contrary to Sec.~\ref{sec3}, the dynamics here is not restricted to decoherence-free subspaces that conserve total spin), as well as heating rates of the axial phonons (which in a linear Paul trap can be as low as a few quanta/s, see Ref.~\cite{Lucas2007,Benhelm2008}). The energy lattice scheme thus provides a practical platform for proof-of-principle experiments with current technology---for small systems the physics of the ideal $U(1)$ QLM is reliably reproduced, qualitatively and quantitatively.

\section{Conclusion}
\label{sec6}
In summary, we have demonstrated that the combination of internal pseudo-spins and localized or collective vibrational modes provides a rich toolbox for implementing lattice gauge theories in trapped-ion chains, relying on current technology. 
To illustrate our ideas, we have proposed two approaches to quantum simulate $(1+1)$D QED. 
By a careful analysis of the most relevant error sources, as well as numerical calculations with realistic parameters, we have demonstrated the experimental feasibility of both schemes. 
Besides trapped ions, the introduced models may be of interest also for other quantum simulation platforms such as superconducting qubits. 

With the first scheme, we have introduced a lattice gauge theory that approximates the $U(1)$ link variables by bosons at high occupation number, while the fermionic matter is encoded in pseudo-spin DOFs. Via a scaling analysis in the ground state as well as numerical calculations of quench dynamics, we have demonstrated that this model approaches $(1+1)$D QED in a well-controlled manner, with good agreement already at moderate boson occupation numbers. 
The implementation of this scheme relies on purpose-engineered local trapping frequencies and off-resonant sideband couplings with single-site addressing. 
Considering realistic parameters, the scheme works on energy scales of current experiments on effective spin models, and is scalable to dozens of ions thus could reach the interesting regime where real-time dynamics becomes inaccessible on classical computers. 

The second scheme inverts the use of bosons and spins: the gauge fields are represented by spins $1/2$ in a QLM formalism; this allows us to replace the fermions by bosons, enabling an efficient implementation in trapped ions.   
The realization is rather straightforward in setups with coherent multicolor laser or microwave radiation, the main ingredient being a set of resonant second-sideband couplings.  
It displays energy scales even better than the first scheme. Though the discussed implementation has gauge-invariant errors that scale unfavorably with increasing system size, we have demonstrated numerically by studying the phenomenon of false-vacuum decay that it provides quantitively reliable predictions for systems of up to about 6 matter sites. This scheme is therefore ideally suited for small-scale proof-of-principle experiments, where the most relevant dynamics already becomes visible. 

The present proposals open an avenue towards analog quantum simulation of a simple but relevant lattice gauge theory in state-of-the-art experiments. 
In the future, it will be interesting to analyze how the proposed schemes can be generalized to more complex lattice gauge theories. 
For example, by exploiting two-dimensional ion crystals \cite{Britton2012,Bohnet2015} one could design bosonic lattice gauge theories in higher dimensions, and it is conceivable that the use of a larger number of vibrational modes in different spatial directions allows for the realization of non-Abelian gauge theories.  

\section*{Acknowledgments} We thank M. Dalmonte and C. Muschik for helpful discussions. D.Y., P.Z. and P.H. acknowledge support from the ERC Synergy Grant UQUAM, the EU grant SIQS and the Austrian Science Fund through SFB FOQUS (FWF Project No.~F4016-N23). G.S.G., M.J. and Ch.W. acknowledge funding from the Bundesministerium f\"ur Bildung und Forschung (FK 01BQ1012), and from Deutsche Forschungsgemeinschaft. G.S.G. also acknowledges support from the European Commission's Horizon 2020 research and innovation program Marie Sk\l{}odowska-Curie Action under Grant Agreement No. 657261.

\begin{appendix}
\section{perturbative calculation of the $M$ matrix}
\label{sec:perturbativeCalculationOfTheMMatrix}
The calculation of the matrix $M_{lq}^{x(y)}$ in Sec.~\ref{Employed degrees of freedom: pseudo-spins and localized phonon modes} is done to the lowest order, i.e., considering only the hybridization of phonon vibrations within each near-resonant ion pair. It is straightforward to calculate corrections to $M_{lq}^{x(y)}$ perturbatively, which helps bound the maximum leakage of the localized phonon modes out of each ion pair. Thanks to the frequency hierarchy,~Eq.\eqref{GeneralLocalizedPhononCondition} in the main text, we can choose the small expansion parameters as $V_{ll'}^{x(y)}/[\Delta_T\omega_{x(y)}]$ and $V_{ll'}^{x(y)}/[\Delta_B\omega_{x(y)}]$, where $V_{ll'}^{x(y)}$ is defined in~Eq.~\eqref{CoulombCoupling} whereas $\Delta_T$ and $\Delta_B$ are the frequency offsets of the segmented trap array, defined through Eqs.~\eqref{generalizedTrapFrequencyDesignX} and~\eqref{generalizedTrapFrequencyDesignY}.

We consider a micro-trap array consisting of several blocks, each containing $N_I$ micro-traps, as described in Sec.~\ref{Employed degrees of freedom: pseudo-spins and localized phonon modes} and shown schematically in Fig.~\ref{fig4:setup}(c). For such a segmented configuration, 
the $l$-th micro-trap can be alternatively specified by two indices $(m_x,j_x)$ using the unique decomposition $l=m_xN_I+j_x$, where $m_x\geq 0$ is the block index while $1\leq j_x\leq N_I$ specifies individual traps inside the same block. Similarly, through the decomposition $l=m_yN_I+j_y+1$, where $m_y\geq 0$ and $1\leq j_y\leq N_I$, the same trap can be labeled by a set of two indices $(m_y,j_y)$. Notice $(m_x,j_x)$ and $(m_y,j_y)$ can be different for the same $l$, due to the different segmentation of the trapping potential along the $x$ and the $y$ direction (see Fig.~\ref{fig4:setup}). The radial confinement strength of the $l$-th trap, $\omega_l^{x(y)}$, can thus be determined using these two sets of indices along with Eqs.~\eqref{generalizedTrapFrequencyDesignX} and~\eqref{generalizedTrapFrequencyDesignY}.

For this general setting, we still label the radial phonon modes so that $c_l^{x(y)}$ connects adiabatically to the local vibration of the $l$-th ion in the $x(y)$ direction in the $d\to \infty$ limit, where $d$ is the distance between neighboring trap centers. Under this convention, the phonon distribution matrix element $M_{lq}^{x}$ with first order corrections can be written as
\begin{eqnarray}
M_{ll'}^x=
&\pm& \sin\theta_{l,l\pm 1}\left[\delta_{l\pm1,l'}+\frac{1-\delta_{l\pm1,l'}}{(\omega_{l'}^x)^2-(\omega_{l}^x)^2}V_{l\pm1,l'}^x\right]\nonumber\\
&+&\cos\theta_{l,l\pm 1}\left[\delta_{ll'}+\frac{1-\delta_{ll'}}{(\omega_{l'}^x)^2-(\omega_{l}^x)^2}V_{ll'}^x\right]\nonumber\\
&+&\mathcal{O}\left[\frac{(V_{ll'}^{x})^2}{\omega_x^2\Delta_T^2},\frac{(V_{ll'}^{x})^2}{\omega_x^2\Delta_B^2}\right],
\end{eqnarray}
where $\pm$ is chosen depending on $l=\mathrm{odd(even)}$, $\delta_{ll'}$ is the Kronecker delta, $\theta_{l,l+1}$ is defined in Eq.~\eqref{phononDistributionAngle} and we have made the convention $\theta_{l,l-1}=\theta_{l-1,l}$. Similarly, along the $y$ direction we have
\begin{eqnarray}
M_{ll'}^y=&\mp& \sin\theta_{l\mp1,l}\left[\delta_{l\mp1,l'}+\frac{1-\delta_{l\mp1,l'}}{(\omega_{l'}^y)^2-(\omega_{l}^y)^2}V_{l\mp1,l'}^y\right]\nonumber\\
&+&\cos\theta_{l\mp1,l}\left[\delta_{ll'}+\frac{1-\delta_{ll'}}{(\omega_{l'}^y)^2-(\omega_{l}^y)^2}V_{ll'}^y\right]\nonumber\\
&+&\mathcal{O}\left[\frac{(V_{ll'}^{y})^2}{\omega_y^2\Delta_T^2},\frac{(V_{ll'}^{y})^2}{\omega_y^2\Delta_B^2}\right],
\end{eqnarray}
where $\mp$ is chosen depending on $l=\mathrm{odd(even)}$.

With the help of Eqs.~\eqref{generalizedTrapFrequencyDesignX} and~\eqref{generalizedTrapFrequencyDesignY}, it is then straightforward to verify that for two ions $l$ and $l'$ that do not form a near-resonant pair, the cross-talk between their individual vibrations along the $\alpha$ direction $(\alpha=x,y)$ is bounded by
\begin{equation}
\label{boundprove}
\begin{split}
|M_{ll'}^\alpha|\leq&\frac{\max[V_{ll'}^\alpha,V_{l\pm1,l'}^\alpha]}{|(\omega_{l'}^\alpha)^2-(\omega_{l}^\alpha)^2|}\\
\leq& \max\left[\frac{\max[V_{l,l+1}^{\alpha}]}{\Delta_T\omega_{\alpha}},\frac{\max[V_{l,l+1}^{\alpha}]}{(N_I-1)^3\Delta_B\omega_{\alpha}}\right],
\end{split}
\end{equation}
where $\max[V_{l,l+1}^\alpha]$, $\alpha=x,y$, is the maximum value of the dipolar coupling strength between adjacent ions in the whole ion array. The two quantities in the last line of Eq.~\eqref{boundprove} have a clear physical meaning: while the former bounds the leakage of localized phonons to an off-resonant micro-trap within the same block, the latter quantifies the maximum cross-talk between two micro-traps lying in different blocks.

\section{AC-Stark shifts and their compensations in the HOBM scheme }
\label{AC-Stark shifts and their compensations in the HOBM scheme}
In the engineering of the spin--gauge-field coupling and spin-mass terms as discussed in Sec.~\ref{sec:SpinGaugeFieldCoupling}, the two applied laser beams $\Omega_l^\alpha,\Omega_{l+1}^{\alpha}$ in each element $(l,l+1)$ inevitably induce AC-Stark shifts to individual ions. In the present appendix we make a detailed analysis of all relevant AC-Stark shifts, and present a concrete scheme to compensate them. Since the laser-matter coupling is essentially independent in each element, here we shall again make use of the first element $(l,l+1)=(1,2)$ as a concrete example to illustrate our ideas.

We first quantify $H_{\mathrm{ls,n}}^{1,2}$, the AC-Stark shifts from near-resonant sideband transitions as shown schematically in Fig.~\ref{fig5:levels}. We assume that the radial phonon modes are initially ground-state cooled, and neglect phonon-heating during the dynamical evolution  (which is a good approximation as long as the phonon heating is far slower than the system dynamics, as fulfilled by the cryogenic surface-traps discussed in Sec.~\ref{surfaceTrapRealization}), thus having $c_{2}^{x\dag} c_{2}^x\simeq 0$. $H_{\mathrm{ls,n}}^{1,2}$ can be written as the sum of the AC-Stark shifts of individual ions, $H_{\mathrm{ls,n}}^{1,2}=H_{\mathrm{ls,n1}}^{1,2}+H_{\mathrm{ls,n2}}^{1,2}$, where
\begin{eqnarray}
\label{nearResonanceLightShift1}
H_{\textrm{ls,n1}}^{1,2} = &&-\frac{|f_{1,2}|^2}{8\delta}\cos\theta_{1,2}^2\sin^2\theta_{1,2}\left[c_{1}^{x\dag}c_{1}^x(\sigma_1^z-1)+\sigma_1^z\right]\nonumber\\
&&-\frac{|f_{1,2}|^2\cos^4\theta_{1,2}}{4(\delta+\epsilon_2^x-\epsilon_1^x)}\left[(c_1^{x}c_1^{x\dag})^2-c_1^{x\dag}c_1^x(2\sigma_1^z+1)\right]\sigma_1^z\nonumber\\
&&-\frac{|f_{1,2}|^2\sin^4\theta_{1,2}}{8(\delta-\epsilon_2^x+\epsilon_1^x)}\sigma_1^z
\end{eqnarray}
describes the AC-Stark shifts due to the three near-resonant second-sideband transitions of the first ion, while for ion 2 we have
\begin{eqnarray}
\label{nearResonanceLightShift2}
H_{\textrm{ls,n2}}^{1,2}=
&&-\frac{|g_{1,2}|^2\sin^2\theta_{1,2}}{4(\delta+\epsilon_2^x-\epsilon_1^x)}\left(c_1^{x\dag}c_1^x+\frac{1}{2}\right)\sigma_2^z\nonumber\\
&&-\frac{|g_{1,2}|^2}{8\delta}\cos^2\theta_{1,2}\sigma^z_{2}.
\end{eqnarray}
The sideband transition strengths are $f_{1,2}=\Omega_{1}^x(\eta_{1,2}^x)^2$ and $g_{1,2}=i\Omega_{2}^x\eta^x_{1,2}$, the same as in Sec.~\ref{sec:SpinGaugeFieldCoupling}.

Besides these near-resonant sidebands, virtual transitions to far off-resonant sidebands also give rise to AC-Stark shifts, which we denote as $H_{\mathrm{ls,o}}^{1,2}$. Nevertheless, thanks to the small Lamb--Dicke parameter, the AC-Stark shifts from higher sideband transitions will decrease exponentially with increasing the phonon-number difference in such a transition. For our case, essentially only two far off-resonant transitions contribute significant AC-Stark shifts, namely the carrier transitions and the (far off-resonant) first sideband transitions. Similar to $H_{\mathrm{ls,n}}^{1,2}$, we can decompose $H_{\mathrm{ls,o}}^{1,2}$ into the contribution from each ion in pair (1,2), $H_{\mathrm{ls,o}}^{1,2}=H_{\mathrm{ls,o1}}^{1,2}+H_{\mathrm{ls,o2}}^{1,2}$. For ion 1 we have\begin{eqnarray}
\label{offResonanceLightShift1}
H_{\mathrm{ls,o1}}^{1,2}=&&-\frac{\Omega_1^{x*}f_{1,2}\cos^2\theta_{1,2}(\delta+\epsilon_2^x+\epsilon_1^x)}{4(\delta+\epsilon_2^x)(\delta+\epsilon_2^x+2\epsilon_1^x)}\left(c_1^{x\dag}c_1^x+\frac{1}{2}\right)\sigma_1^z\nonumber\\
&&-\frac{\Omega_1^{x*}f_{1,2}\sin^2\theta_{1,2}(\delta+\epsilon_1^x+\epsilon_2^x)}{4(\delta+\epsilon_1^x)(\delta+\epsilon_1^x+2\epsilon_2^x)}\sigma_1^z\nonumber\\
&&-\frac{|\Omega_1^x|^2}{4(\delta+\epsilon_1^x+\epsilon_2^x)}\sigma_1^z,
\end{eqnarray}
in which the first two (third) lines come from the first-sideband (carrier) transition respectively. The contribution from ion 2 reads
\begin{eqnarray}
\label{offResonanceLightShift2}
H_{\mathrm{ls,o2}}^{1,2}=&&-\frac{|g_{1,2}|^2\sin^2\theta_{1,2}}{4(\delta+\epsilon_2^x+\epsilon_1^x)}
\left(c_1^{x\dag}c_1^x+\frac{1}{2}\right)\sigma_2^z\nonumber\\
&&-\frac{|g_{1,2}|^2\cos^2\theta_{1,2}}{8(\delta+2\epsilon_2^x)}\sigma_2^z-\frac{|\Omega_2^x|^2}{4(\delta+\epsilon_1^x)}\sigma_2^z.
\end{eqnarray} 
Other far off-resonant sidebands are detuned with the laser frequency by at least $\omega^{x(y)}$, while containing at least two powers of $\eta_{1,2}^x\sqrt{N}$ in their sideband transition strength. Taking the typical experimental parameters as discussed in Sec.~\ref{surfaceTrapRealization}, we find that they are below or on the order of $2\pi\times 1$Hz, and thus can be safely neglected.

The relevant AC-Stark shifts in the element $(1,2)$ are thus only the above, $H_{\mathrm{ls}}^{1,2}=H_{\mathrm{ls,n}}^{1,2}+H_{\mathrm{ls,o}}^{1,2}$. Notice that they are gauge-invariant, $[H_{\mathrm{ls}}^{1,2},G_i]=0$, where the local gauge generator $G_i$ is given by Eq.~\eqref{BosonAppGauss} in the main text. The AC-Stark shifts do not break the Gauss law, but only induce errors to the system dynamics within the physical Hilbert space. In the following we show that with appropriate compensation these errors can be highly suppressed compared to the working energy scale of the proposed quantum simulator.

\begin{figure}[t]
\begin{center}
\includegraphics[width=0.7\columnwidth]{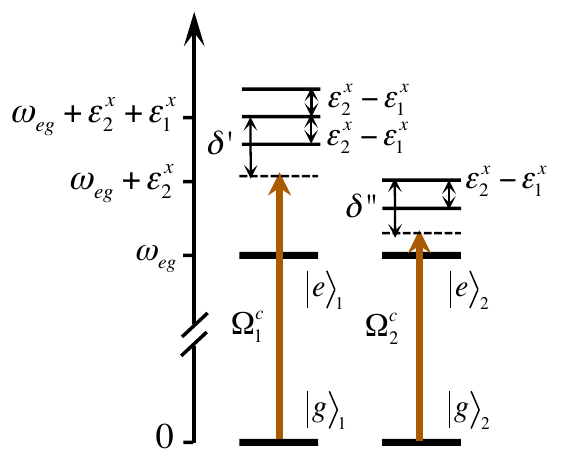}
\caption{
Compensation scheme of the AC-Stark shifts, here shown for the element $(1,2)$ as an example. An additional laser beam, with parameters $(\Omega_{l}^{\mathrm{c}}, \omega_l^{\mathrm{Lc}})$ is applied to the first ion. It is off-resonant to any transition inside the ion pair $(1,2)$, nevertheless it induces an appropriate amount of AC-Stark shift, which compensates the part of the light shift quadratic on the gauge-field phonon number induced by $\Omega_{1}^{x}$. Similarly, a laser beam with parameter $(\Omega_{2}^{\mathrm{c}}, \omega_2^{\mathrm{Lc}})$, is applied to ion 2, to compensate part of the light shift linearly dependent on the gauge-field phonon number. The residual AC-Stark shifts can be compensated largely by local adjustment of the frequencies of the two sideband laser beams $\omega_{1}^{\mathrm{L}x}$ and $\omega_{2}^{\mathrm{L}x}$.
}
\label{appBlevels}
\end{center}
\end{figure}

Our compensation scheme consists of two ingredients. First, we apply an additional laser beam (see Fig.~\ref{appBlevels}) to each ion, to compensate the most serious AC-Stark shift errors, i.e., the second line in Eq.~\eqref{nearResonanceLightShift1} and the first line in Eq.~\eqref{nearResonanceLightShift2}. These AC-Stark shifts are dependent on the gauge-field phonon thus cannot be eliminated in simple ways, and their energy scale is comparable to the working energy scale of the quantum simulator. For ion 1, we assume the additional laser has Rabi frequency $\Omega_1^\mathrm{c}$, Lamb--Dicke parameter $\eta_{1,2}^{\mathrm{c}1}\simeq\eta_{1,2}^x$, while its frequency $\omega_1^{\mathrm{Lc}}=\omega_{eg}+\epsilon_1^x+\epsilon_2^x+\delta'$ is far from any sideband transitions. Moreover, we choose $\delta'$ so that the transition is off-resonant to any transition driven by the two original lasers $\Omega_{1(2)}^x$. Thus, all its effects are the AC-Stark shifts from virtual sideband transitions, $H_\mathrm{{ls,c1}}^{1,2}=H_{\mathrm{ls,nc1}}^{1,2}+H_{\mathrm{ls,oc1}}^{1,2}$, which can be obtained straightforwardly from Eqs.~\eqref{nearResonanceLightShift1} and \eqref{offResonanceLightShift1} by the replacement $\delta\to\delta'$ and $\Omega_1^x\to\Omega_1^\mathrm{c}$. Importantly, we assume the condition $|\Omega_1^x|^2/(\delta+\epsilon_2^x-\epsilon_1^x)\simeq-|\Omega_1^c|^2/(\delta'+\epsilon_2^x-\epsilon_1^x)$, by which we eliminate AC-Stark shifts quadratic in phonon number. 
Similarly, the other compensation laser, with Rabi frequency $\Omega_2^\mathrm{c}$, Lamb--Dicke parameter $\eta_{1,2}^{\mathrm{c}2}=\eta_{1,2}^x$, and frequency $\omega_2^\mathrm{Lc}=\omega_{eg}+\epsilon_2^x+\delta''$ (off-resonant to any sideband transitions), induces an approriate amount of AC-Stark shift to ion 2, described by Eq.~\eqref{nearResonanceLightShift2} and \eqref{offResonanceLightShift2} with the replacement $\delta\to\delta''$ and $\Omega_{2}^x\to\Omega_{2}^\mathrm{c}$. The requirement $|\Omega_2^x|^2/(\delta+\epsilon_2^x-\epsilon_1^x)\simeq-|\Omega_2^\mathrm{c}|^2/(\delta''+\epsilon_2^x-\epsilon_1^x)$ then compensates the first term in Eq.~\eqref{nearResonanceLightShift2}.

Second, the residual AC-Stark shifts in the element $(1,2)$, together with that induced by the additional lasers $\Omega_{1(2)}^\mathrm{c}$, can be compensated to a large extent by local adjustment of $\omega_{1}^{\mathrm{L}x}$ and $\omega_{2}^{\mathrm{L}x}$, the detunings of the two sideband lasers (see Fig.~\ref{fig5:levels}). Neglecting the tiny correction to the phonon vibrational frequency ($\leq 1$Hz) contained in the first line of Eq.~\eqref{nearResonanceLightShift1}, these AC-Stark shifts are independent of or depend linearly on the occupation number of the gauge-field mode, and can be summed up to a compact expression as $H_{\mathrm{ls}}^{1,2}+H_{\mathrm{ls,c}}^{1,2}=\sum_{l=1}^2 [E^{1,2}_l + F^{1,2}_l (c_{1}^{x\dag}c_{1}^x -N)]\sigma_l^z$. 
This can be generalized to the whole array of ions, of which the total AC-Stark shifts can be written as
\begin{equation}
\sum_{l=1}^{L-1}\sum_{m=l}^{l+1}[E_m^{l,l+1}+F_{m}^{l,l+1}(c_{l}^{\alpha\dag}c^\alpha_{l}-N)]\sigma_m^z,
\end{equation}
with $\alpha=x$ for $l=\mathrm{odd}$ while $\alpha=y$ for $l=\mathrm{even}$. The phonon-independent AC-Stark shift $\propto E_m^{l,l+1}$ can be compensated by adjustment of the frequencies of the local laser beams, $\delta_{l}^{\textrm{L}x}\to\delta_{l}^{\textrm{L}x}+2E_l^{l-1,l}+2E_l^{l,l+1}$ and $\delta_{l}^{\textrm{L}y}\to\delta_{l}^{\textrm{L}y}+2E_l^{l-1,l}+2E_l^{l,l+1}$. The residual terms mainly come from far off-resonant first sideband transitions, as exemplified by the first line in Eqs.~\eqref{offResonanceLightShift1} and \eqref{offResonanceLightShift2}. Their energy scale $F_{m}^{l,l+1}$ is far smaller than the energy scale of system dynamics, thus have tiny impact. This is clearly illustrated in Fig.~\ref{fig5QSperformance}, where we compare the imperfect dynamics with these residual AC-Stark shifts $\propto F_{m}^{l,l+1}$, to the behavior of the ideal HOBM. 

As a concrete example, we consider the experimental parameters for a surface-trap realization as presented in Sec.~\ref{surfaceTrapRealization}. For the compensation beams in element $(1,2)$, we can chose $\delta'\simeq 2\pi\times 80$kHz and $|\Omega_1^\mathrm{c}|\simeq2\pi\times 270$kHz, while $\delta''\simeq 2\pi\times 120$kHz and $|\Omega_2^\mathrm{c}|\simeq 2\pi\times 380$kHz. Without the requirement of fine tuning, compensation of $90\%$ of such an AC-Stark shifts will suffice to suppress the error to be smaller than the working energy of the quantum simulator by one order of magnitude. The residual AC-Stark shifts are on the order of $E_{1(2)}^{1,2}\sim 2\pi\times1$kHz while $F_{1(2)}^{1,2}\sim 2\pi\times 10$Hz. Since the former can be compensated by adjustment of the laser frequency, while the latter is far smaller than $J$, these imperfections bring in only tiny detriment to the performance of the proposed quantum simulator.

\end{appendix}

\bibliography{LGTIons_bib}

\begin{thebibliography}{80}%
\makeatletter
\providecommand \@ifxundefined [1]{%
 \@ifx{#1\undefined}
}%
\providecommand \@ifnum [1]{%
 \ifnum #1\expandafter \@firstoftwo
 \else \expandafter \@secondoftwo
 \fi
}%
\providecommand \@ifx [1]{%
 \ifx #1\expandafter \@firstoftwo
 \else \expandafter \@secondoftwo
 \fi
}%
\providecommand \natexlab [1]{#1}%
\providecommand \enquote  [1]{``#1''}%
\providecommand \bibnamefont  [1]{#1}%
\providecommand \bibfnamefont [1]{#1}%
\providecommand \citenamefont [1]{#1}%
\providecommand \href@noop [0]{\@secondoftwo}%
\providecommand \href [0]{\begingroup \@sanitize@url \@href}%
\providecommand \@href[1]{\@@startlink{#1}\@@href}%
\providecommand \@@href[1]{\endgroup#1\@@endlink}%
\providecommand \@sanitize@url [0]{\catcode `\\12\catcode `\$12\catcode
  `\&12\catcode `\#12\catcode `\^12\catcode `\_12\catcode `\%12\relax}%
\providecommand \@@startlink[1]{}%
\providecommand \@@endlink[0]{}%
\providecommand \url  [0]{\begingroup\@sanitize@url \@url }%
\providecommand \@url [1]{\endgroup\@href {#1}{\urlprefix }}%
\providecommand \urlprefix  [0]{URL }%
\providecommand \Eprint [0]{\href }%
\providecommand \doibase [0]{http://dx.doi.org/}%
\providecommand \selectlanguage [0]{\@gobble}%
\providecommand \bibinfo  [0]{\@secondoftwo}%
\providecommand \bibfield  [0]{\@secondoftwo}%
\providecommand \translation [1]{[#1]}%
\providecommand \BibitemOpen [0]{}%
\providecommand \bibitemStop [0]{}%
\providecommand \bibitemNoStop [0]{.\EOS\space}%
\providecommand \EOS [0]{\spacefactor3000\relax}%
\providecommand \BibitemShut  [1]{\csname bibitem#1\endcsname}%
\let\auto@bib@innerbib\@empty
\bibitem [{\citenamefont {Cirac}\ and\ \citenamefont
  {Zoller}(2012)}]{Cirac2012}%
  \BibitemOpen
  \bibfield  {author} {\bibinfo {author} {\bibfnamefont {J.~I.}\ \bibnamefont
  {Cirac}}\ and\ \bibinfo {author} {\bibfnamefont {P.}~\bibnamefont {Zoller}},\
  }\href {\doibase doi:10.1038/nphys2275} {\bibfield  {journal} {\bibinfo
  {journal} {Nat. Phys.}\ }\textbf {\bibinfo {volume} {8}},\ \bibinfo {pages}
  {264} (\bibinfo {year} {2012})}\BibitemShut {NoStop}%
\bibitem [{\citenamefont {Montvay}\ and\ \citenamefont
  {Muenster}(1994)}]{Montvay1994}%
  \BibitemOpen
  \bibfield  {author} {\bibinfo {author} {\bibfnamefont {I.}~\bibnamefont
  {Montvay}}\ and\ \bibinfo {author} {\bibfnamefont {G.}~\bibnamefont
  {Muenster}},\ }\href@noop {} {\emph {\bibinfo {title} {Quantum Fields on a
  lattice}}}\ (\bibinfo  {publisher} {Cambridge Univ. Press, Cambridge},\
  \bibinfo {year} {1994})\BibitemShut {NoStop}%
\bibitem [{\citenamefont {Gattringer}\ and\ \citenamefont
  {Lang}(2010)}]{Gattringer2010}%
  \BibitemOpen
  \bibfield  {author} {\bibinfo {author} {\bibfnamefont {C.}~\bibnamefont
  {Gattringer}}\ and\ \bibinfo {author} {\bibfnamefont {C.~B.}\ \bibnamefont
  {Lang}},\ }\href@noop {} {\emph {\bibinfo {title} {Quantum Chromodynamics on
  the Lattice}}}\ (\bibinfo  {publisher} {Springer-Verlag},\ \bibinfo {year}
  {2010})\BibitemShut {NoStop}%
\bibitem [{\citenamefont {Kogut}(1979)}]{Kogut1979}%
  \BibitemOpen
  \bibfield  {author} {\bibinfo {author} {\bibfnamefont {J.~B.}\ \bibnamefont
  {Kogut}},\ }\href {\doibase 10.1103/RevModPhys.51.659} {\bibfield  {journal}
  {\bibinfo  {journal} {Rev. Mod. Phys.}\ }\textbf {\bibinfo {volume} {51}},\
  \bibinfo {pages} {659} (\bibinfo {year} {1979})}\BibitemShut {NoStop}%
\bibitem [{\citenamefont {Lee}\ \emph {et~al.}(2006)\citenamefont {Lee},
  \citenamefont {Nagaosa},\ and\ \citenamefont {Wen}}]{Lee2006}%
  \BibitemOpen
  \bibfield  {author} {\bibinfo {author} {\bibfnamefont {P.~A.}\ \bibnamefont
  {Lee}}, \bibinfo {author} {\bibfnamefont {N.}~\bibnamefont {Nagaosa}}, \ and\
  \bibinfo {author} {\bibfnamefont {X.-G.}\ \bibnamefont {Wen}},\ }\href
  {\doibase 10.1103/RevModPhys.78.17} {\bibfield  {journal} {\bibinfo
  {journal} {Rev. Mod. Phys.}\ }\textbf {\bibinfo {volume} {78}},\ \bibinfo
  {eid} {17} (\bibinfo {year} {2006})}\BibitemShut {NoStop}%
\bibitem [{\citenamefont {Lacroix}\ \emph {et~al.}(2010)\citenamefont
  {Lacroix}, \citenamefont {Mendels},\ and\ \citenamefont
  {Mila}}]{Lacroix2010}%
  \BibitemOpen
  \bibinfo {editor} {\bibfnamefont {C.}~\bibnamefont {Lacroix}}, \bibinfo
  {editor} {\bibfnamefont {P.}~\bibnamefont {Mendels}}, \ and\ \bibinfo
  {editor} {\bibfnamefont {F.}~\bibnamefont {Mila}},\ eds.,\ \href@noop {}
  {\emph {\bibinfo {title} {Introduction to Frustrated Magnetism}}}\ (\bibinfo
  {publisher} {Springer Series in Solid-State Sciences Vol. 164},\ \bibinfo
  {year} {2010})\BibitemShut {NoStop}%
\bibitem [{\citenamefont {Wiese}(2013)}]{Wiese2013}%
  \BibitemOpen
  \bibfield  {author} {\bibinfo {author} {\bibfnamefont {U.-J.}\ \bibnamefont
  {Wiese}},\ }\href {\doibase 10.1002/andp.201300104} {\bibfield  {journal}
  {\bibinfo  {journal} {Ann. Phys.}\ }\textbf {\bibinfo {volume} {525}},\
  \bibinfo {pages} {777} (\bibinfo {year} {2013})}\BibitemShut {NoStop}%
\bibitem [{\citenamefont {Wiese}(2014)}]{Wiese2014}%
  \BibitemOpen
  \bibfield  {author} {\bibinfo {author} {\bibfnamefont {U.-J.}\ \bibnamefont
  {Wiese}},\ }\href
  {http://www.sciencedirect.com/science/article/pii/S0375947414004849}
  {\bibfield  {journal} {\bibinfo  {journal} {Nucl. Phys. A}\ }\textbf
  {\bibinfo {volume} {931}},\ \bibinfo {pages} {246} (\bibinfo {year}
  {2014})}\BibitemShut {NoStop}%
\bibitem [{\citenamefont {Erez~Zohar}(2016)}]{Zohar2016}%
  \BibitemOpen
  \bibfield  {author} {\bibinfo {author} {\bibfnamefont {B.~R.}\ \bibnamefont
  {Erez~Zohar}, \bibfnamefont {J.~Ignacio~Cirac}},\ }\href {\doibase
  10.1088/0034-4885/79/1/014401} {\bibfield  {journal} {\bibinfo  {journal}
  {Rep. Prog. Phys.}\ }\textbf {\bibinfo {volume} {79}},\ \bibinfo {pages}
  {014401} (\bibinfo {year} {2016})}\BibitemShut {NoStop}%
\bibitem [{\citenamefont {Dalmonte}\ and\ \citenamefont
  {Montangero}()}]{Dalmonte2016}%
  \BibitemOpen
  \bibfield  {author} {\bibinfo {author} {\bibfnamefont {M.}~\bibnamefont
  {Dalmonte}}\ and\ \bibinfo {author} {\bibfnamefont {S.}~\bibnamefont
  {Montangero}},\ }\href {http://arxiv.org/abs/1602.03776} {\bibinfo  {journal}
  {arXiv:1602.03776}\ }\BibitemShut {NoStop}%
\bibitem [{\citenamefont {Dou{\c c}ot}\ \emph {et~al.}(2004)\citenamefont
  {Dou{\c c}ot}, \citenamefont {Ioffe},\ and\ \citenamefont
  {Vidal}}]{Doucot2004}%
  \BibitemOpen
\bibfield  {journal} {  }\bibfield  {author} {\bibinfo {author} {\bibfnamefont
  {B.}~\bibnamefont {Dou{\c c}ot}}, \bibinfo {author} {\bibfnamefont {L.~B.}\
  \bibnamefont {Ioffe}}, \ and\ \bibinfo {author} {\bibfnamefont
  {J.}~\bibnamefont {Vidal}},\ }\href
  {http://journals.aps.org/prb/abstract/10.1103/PhysRevB.69.214501} {\bibfield
  {journal} {\bibinfo  {journal} {Phys. Rev. B}\ }\textbf {\bibinfo {volume}
  {69}},\ \bibinfo {pages} {214501} (\bibinfo {year} {2004})}\BibitemShut
  {NoStop}%
\bibitem [{\citenamefont {Marcos}\ \emph {et~al.}(2013)\citenamefont {Marcos},
  \citenamefont {Rabl}, \citenamefont {Rico},\ and\ \citenamefont
  {Zoller}}]{Marcos2013}%
  \BibitemOpen
  \bibfield  {author} {\bibinfo {author} {\bibfnamefont {D.}~\bibnamefont
  {Marcos}}, \bibinfo {author} {\bibfnamefont {P.}~\bibnamefont {Rabl}},
  \bibinfo {author} {\bibfnamefont {E.}~\bibnamefont {Rico}}, \ and\ \bibinfo
  {author} {\bibfnamefont {P.}~\bibnamefont {Zoller}},\ }\href {\doibase
  10.1103/PhysRevLett.111.110504} {\bibfield  {journal} {\bibinfo  {journal}
  {Phys. Rev. Lett.}\ }\textbf {\bibinfo {volume} {111}},\ \bibinfo {pages}
  {110504} (\bibinfo {year} {2013})}\BibitemShut {NoStop}%
\bibitem [{\citenamefont {Marcos}\ \emph {et~al.}(2014)\citenamefont {Marcos},
  \citenamefont {Widmer}, \citenamefont {Rico}, \citenamefont {Hafezi},
  \citenamefont {Rabl}, \citenamefont {Wiese},\ and\ \citenamefont
  {Zoller}}]{Marcos2014}%
  \BibitemOpen
  \bibfield  {author} {\bibinfo {author} {\bibfnamefont {D.}~\bibnamefont
  {Marcos}}, \bibinfo {author} {\bibfnamefont {P.}~\bibnamefont {Widmer}},
  \bibinfo {author} {\bibfnamefont {E.}~\bibnamefont {Rico}}, \bibinfo {author}
  {\bibfnamefont {M.}~\bibnamefont {Hafezi}}, \bibinfo {author} {\bibfnamefont
  {P.}~\bibnamefont {Rabl}}, \bibinfo {author} {\bibfnamefont {U.}~\bibnamefont
  {Wiese}}, \ and\ \bibinfo {author} {\bibfnamefont {P.}~\bibnamefont
  {Zoller}},\ }\href
  {http://linkinghub.elsevier.com/retrieve/pii/S0003491614002711} {\bibfield
  {journal} {\bibinfo  {journal} {Annals of physics}\ }\textbf {\bibinfo
  {volume} {351}},\ \bibinfo {pages} {634} (\bibinfo {year}
  {2014})}\BibitemShut {NoStop}%
\bibitem [{\citenamefont {Mezzacapo}\ \emph {et~al.}(2015)\citenamefont
  {Mezzacapo}, \citenamefont {Rico}, \citenamefont {Sab{\'\i}n}, \citenamefont
  {Egusquiza}, \citenamefont {Lamata},\ and\ \citenamefont
  {Solano}}]{Mezzacapo2015}%
  \BibitemOpen
  \bibfield  {author} {\bibinfo {author} {\bibfnamefont {A.}~\bibnamefont
  {Mezzacapo}}, \bibinfo {author} {\bibfnamefont {E.}~\bibnamefont {Rico}},
  \bibinfo {author} {\bibfnamefont {C.}~\bibnamefont {Sab{\'\i}n}}, \bibinfo
  {author} {\bibfnamefont {I.}~\bibnamefont {Egusquiza}}, \bibinfo {author}
  {\bibfnamefont {L.}~\bibnamefont {Lamata}}, \ and\ \bibinfo {author}
  {\bibfnamefont {E.}~\bibnamefont {Solano}},\ }\href
  {http://journals.aps.org/prl/abstract/10.1103/PhysRevLett.115.240502}
  {\bibfield  {journal} {\bibinfo  {journal} {Phys. Rev. Lett.}\ }\textbf
  {\bibinfo {volume} {115}},\ \bibinfo {pages} {240502} (\bibinfo {year}
  {2015})}\BibitemShut {NoStop}%
\bibitem [{\citenamefont {Garc{\'\i}a-{\'A}lvarez}\ \emph
  {et~al.}(2015)\citenamefont {Garc{\'\i}a-{\'A}lvarez}, \citenamefont
  {Casanova}, \citenamefont {Mezzacapo}, \citenamefont {Egusquiza},
  \citenamefont {Lamata}, \citenamefont {Romero},\ and\ \citenamefont
  {Solano}}]{GarciaAlvarez2015}%
  \BibitemOpen
  \bibfield  {author} {\bibinfo {author} {\bibfnamefont {L.}~\bibnamefont
  {Garc{\'\i}a-{\'A}lvarez}}, \bibinfo {author} {\bibfnamefont
  {J.}~\bibnamefont {Casanova}}, \bibinfo {author} {\bibfnamefont
  {A.}~\bibnamefont {Mezzacapo}}, \bibinfo {author} {\bibfnamefont
  {I.}~\bibnamefont {Egusquiza}}, \bibinfo {author} {\bibfnamefont
  {L.}~\bibnamefont {Lamata}}, \bibinfo {author} {\bibfnamefont
  {G.}~\bibnamefont {Romero}}, \ and\ \bibinfo {author} {\bibfnamefont
  {E.}~\bibnamefont {Solano}},\ }\href
  {http://journals.aps.org/prl/abstract/10.1103/PhysRevLett.114.070502}
  {\bibfield  {journal} {\bibinfo  {journal} {Phys. Rev. Lett.}\ }\textbf
  {\bibinfo {volume} {114}},\ \bibinfo {pages} {070502} (\bibinfo {year}
  {2015})}\BibitemShut {NoStop}%
\bibitem [{\citenamefont {Hauke}\ \emph {et~al.}(2013)\citenamefont {Hauke},
  \citenamefont {Marcos}, \citenamefont {Dalmonte},\ and\ \citenamefont
  {Zoller}}]{Hauke2013b}%
  \BibitemOpen
  \bibfield  {author} {\bibinfo {author} {\bibfnamefont {P.}~\bibnamefont
  {Hauke}}, \bibinfo {author} {\bibfnamefont {D.}~\bibnamefont {Marcos}},
  \bibinfo {author} {\bibfnamefont {M.}~\bibnamefont {Dalmonte}}, \ and\
  \bibinfo {author} {\bibfnamefont {P.}~\bibnamefont {Zoller}},\ }\href
  {http://link.aps.org/doi/10.1103/PhysRevX.3.041018} {\bibfield  {journal}
  {\bibinfo  {journal} {Phys. Rev. X}\ }\textbf {\bibinfo {volume} {3}},\
  \bibinfo {pages} {041018} (\bibinfo {year} {2013})}\BibitemShut {NoStop}%
\bibitem [{\citenamefont {Hamer}\ \emph {et~al.}(1997)\citenamefont {Hamer},
  \citenamefont {Weihong},\ and\ \citenamefont {Oitmaa}}]{Hamer1997}%
  \BibitemOpen
  \bibfield  {author} {\bibinfo {author} {\bibfnamefont {C.~J.}\ \bibnamefont
  {Hamer}}, \bibinfo {author} {\bibfnamefont {Z.}~\bibnamefont {Weihong}}, \
  and\ \bibinfo {author} {\bibfnamefont {J.}~\bibnamefont {Oitmaa}},\ }\href
  {http://link.aps.org/doi/10.1103/PhysRevD.56.55} {\bibfield  {journal}
  {\bibinfo  {journal} {Phys. Rev. D}\ }\textbf {\bibinfo {volume} {56}},\
  \bibinfo {pages} {55} (\bibinfo {year} {1997})}\BibitemShut {NoStop}%
\bibitem [{\citenamefont {{Esteban A. Martinez \emph{et
  al.}}}()}]{MartinezInPreparation}%
  \BibitemOpen
  \bibfield  {author} {\bibinfo {author} {\bibnamefont {{Esteban A. Martinez
  \emph{et al.}}}},\ }\href@noop {} {\bibinfo  {journal} {in preparation}\
  }\BibitemShut {NoStop}%
\bibitem [{\citenamefont {{Christine Muschik \emph{et
  al.}}}()}]{MuschikInPreparation}%
  \BibitemOpen
\bibfield  {journal} {  }\bibfield  {author} {\bibinfo {author} {\bibnamefont
  {{Christine Muschik \emph{et al.}}}},\ }\href@noop {} {\bibinfo  {journal}
  {in preparation}\ }\BibitemShut {NoStop}%
\bibitem [{\citenamefont {Ba{\~{n}}uls}\ \emph {et~al.}(2013)\citenamefont
  {Ba{\~{n}}uls}, \citenamefont {Cichy}, \citenamefont {Cirac},\ and\
  \citenamefont {Jansen}}]{Banuls2013}%
  \BibitemOpen
\bibfield  {journal} {  }\bibfield  {author} {\bibinfo {author} {\bibfnamefont
  {M.}~\bibnamefont {Ba{\~{n}}uls}}, \bibinfo {author} {\bibfnamefont
  {K.}~\bibnamefont {Cichy}}, \bibinfo {author} {\bibfnamefont
  {J.}~\bibnamefont {Cirac}}, \ and\ \bibinfo {author} {\bibfnamefont
  {K.}~\bibnamefont {Jansen}},\ }\href {\doibase 10.1007/JHEP11(2013)158}
  {\bibfield  {journal} {\bibinfo  {journal} {Journal of High Energy Physics}\
  }\textbf {\bibinfo {volume} {2013}},\ \bibinfo {pages} {158} (\bibinfo {year}
  {2013})}\BibitemShut {NoStop}%
\bibitem [{\citenamefont {Rico}\ \emph {et~al.}(2014)\citenamefont {Rico},
  \citenamefont {Pichler}, \citenamefont {Dalmonte}, \citenamefont {Zoller},\
  and\ \citenamefont {Montangero}}]{Rico2014}%
  \BibitemOpen
  \bibfield  {author} {\bibinfo {author} {\bibfnamefont {E.}~\bibnamefont
  {Rico}}, \bibinfo {author} {\bibfnamefont {T.}~\bibnamefont {Pichler}},
  \bibinfo {author} {\bibfnamefont {M.}~\bibnamefont {Dalmonte}}, \bibinfo
  {author} {\bibfnamefont {P.}~\bibnamefont {Zoller}}, \ and\ \bibinfo {author}
  {\bibfnamefont {S.}~\bibnamefont {Montangero}},\ }\href {\doibase
  10.1103/PhysRevLett.112.201601} {\bibfield  {journal} {\bibinfo  {journal}
  {Phys. Rev. Lett.}\ }\textbf {\bibinfo {volume} {112}},\ \bibinfo {pages}
  {201601} (\bibinfo {year} {2014})}\BibitemShut {NoStop}%
\bibitem [{\citenamefont {Buyens}\ \emph {et~al.}(2014)\citenamefont {Buyens},
  \citenamefont {Haegeman}, \citenamefont {Van~Acoleyen}, \citenamefont
  {Verschelde},\ and\ \citenamefont {Verstraete}}]{Buyens2014}%
  \BibitemOpen
  \bibfield  {author} {\bibinfo {author} {\bibfnamefont {B.}~\bibnamefont
  {Buyens}}, \bibinfo {author} {\bibfnamefont {J.}~\bibnamefont {Haegeman}},
  \bibinfo {author} {\bibfnamefont {K.}~\bibnamefont {Van~Acoleyen}}, \bibinfo
  {author} {\bibfnamefont {H.}~\bibnamefont {Verschelde}}, \ and\ \bibinfo
  {author} {\bibfnamefont {F.}~\bibnamefont {Verstraete}},\ }\href {\doibase
  10.1103/PhysRevLett.113.091601} {\bibfield  {journal} {\bibinfo  {journal}
  {Phys. Rev. Lett.}\ }\textbf {\bibinfo {volume} {113}},\ \bibinfo {pages}
  {091601} (\bibinfo {year} {2014})}\BibitemShut {NoStop}%
\bibitem [{\citenamefont {Silvi}\ \emph {et~al.}(2014)\citenamefont {Silvi},
  \citenamefont {Rico}, \citenamefont {Calarco},\ and\ \citenamefont
  {Montangero}}]{1367-2630-16-10-103015}%
  \BibitemOpen
  \bibfield  {author} {\bibinfo {author} {\bibfnamefont {P.}~\bibnamefont
  {Silvi}}, \bibinfo {author} {\bibfnamefont {E.}~\bibnamefont {Rico}},
  \bibinfo {author} {\bibfnamefont {T.}~\bibnamefont {Calarco}}, \ and\
  \bibinfo {author} {\bibfnamefont {S.}~\bibnamefont {Montangero}},\ }\href
  {http://stacks.iop.org/1367-2630/16/i=10/a=103015} {\bibfield  {journal}
  {\bibinfo  {journal} {New Journal of Physics}\ }\textbf {\bibinfo {volume}
  {16}},\ \bibinfo {pages} {103015} (\bibinfo {year} {2014})}\BibitemShut
  {NoStop}%
\bibitem [{\citenamefont {Tagliacozzo}\ \emph {et~al.}(2014)\citenamefont
  {Tagliacozzo}, \citenamefont {Celi},\ and\ \citenamefont
  {Lewenstein}}]{Tagliacozzo2014}%
  \BibitemOpen
  \bibfield  {author} {\bibinfo {author} {\bibfnamefont {L.}~\bibnamefont
  {Tagliacozzo}}, \bibinfo {author} {\bibfnamefont {A.}~\bibnamefont {Celi}}, \
  and\ \bibinfo {author} {\bibfnamefont {M.}~\bibnamefont {Lewenstein}},\
  }\href {\doibase 10.1103/PhysRevX.4.041024} {\bibfield  {journal} {\bibinfo
  {journal} {Phys. Rev. X}\ }\textbf {\bibinfo {volume} {4}},\ \bibinfo {pages}
  {1} (\bibinfo {year} {2014})}\BibitemShut {NoStop}%
\bibitem [{\citenamefont {Haegeman}\ \emph {et~al.}(2015)\citenamefont
  {Haegeman}, \citenamefont {{Van Acoleyen}}, \citenamefont {Schuch},
  \citenamefont {Cirac},\ and\ \citenamefont {Verstraete}}]{Haegeman2015}%
  \BibitemOpen
  \bibfield  {author} {\bibinfo {author} {\bibfnamefont {J.}~\bibnamefont
  {Haegeman}}, \bibinfo {author} {\bibfnamefont {K.}~\bibnamefont {{Van
  Acoleyen}}}, \bibinfo {author} {\bibfnamefont {N.}~\bibnamefont {Schuch}},
  \bibinfo {author} {\bibfnamefont {J.~I.}\ \bibnamefont {Cirac}}, \ and\
  \bibinfo {author} {\bibfnamefont {F.}~\bibnamefont {Verstraete}},\ }\href
  {\doibase 10.1103/PhysRevX.5.011024} {\bibfield  {journal} {\bibinfo
  {journal} {Physical Review X}\ }\textbf {\bibinfo {volume} {5}},\ \bibinfo
  {pages} {1} (\bibinfo {year} {2015})}\BibitemShut {NoStop}%
\bibitem [{\citenamefont {Pichler}\ \emph {et~al.}()\citenamefont {Pichler},
  \citenamefont {Dalmonte}, \citenamefont {Rico}, \citenamefont {Zoller},\ and\
  \citenamefont {Montangero}}]{Pichler2015}%
  \BibitemOpen
  \bibfield  {author} {\bibinfo {author} {\bibfnamefont {T.}~\bibnamefont
  {Pichler}}, \bibinfo {author} {\bibfnamefont {M.}~\bibnamefont {Dalmonte}},
  \bibinfo {author} {\bibfnamefont {E.}~\bibnamefont {Rico}}, \bibinfo {author}
  {\bibfnamefont {P.}~\bibnamefont {Zoller}}, \ and\ \bibinfo {author}
  {\bibfnamefont {S.}~\bibnamefont {Montangero}},\ }\href
  {http://arxiv.org/abs/1505.04440} {\ }\Eprint
  {http://arxiv.org/abs/1505.04440} {arXiv:1505.04440} \BibitemShut {NoStop}%
\bibitem [{\citenamefont {Schwinger}(1962)}]{Schwinger1962}%
  \BibitemOpen
  \bibfield  {author} {\bibinfo {author} {\bibfnamefont {J.}~\bibnamefont
  {Schwinger}},\ }\href
  {http://journals.aps.org/pr/abstract/10.1103/PhysRev.128.2425} {\bibfield
  {journal} {\bibinfo  {journal} {Phys. Rev.}\ }\textbf {\bibinfo {volume}
  {128}},\ \bibinfo {pages} {2425} (\bibinfo {year} {1962})}\BibitemShut
  {NoStop}%
\bibitem [{\citenamefont {Coleman}\ \emph {et~al.}(1975)\citenamefont
  {Coleman}, \citenamefont {Jackiw},\ and\ \citenamefont
  {Susskind}}]{Coleman1975}%
  \BibitemOpen
  \bibfield  {author} {\bibinfo {author} {\bibfnamefont {S.}~\bibnamefont
  {Coleman}}, \bibinfo {author} {\bibfnamefont {R.}~\bibnamefont {Jackiw}}, \
  and\ \bibinfo {author} {\bibfnamefont {L.}~\bibnamefont {Susskind}},\ }\href
  {\doibase 10.1016/0003-4916(75)90212-2} {\bibfield  {journal} {\bibinfo
  {journal} {Annals of Physics}\ }\textbf {\bibinfo {volume} {93}},\ \bibinfo
  {pages} {267} (\bibinfo {year} {1975})}\BibitemShut {NoStop}%
\bibitem [{\citenamefont {Coleman}(1976)}]{Coleman1976}%
  \BibitemOpen
  \bibfield  {author} {\bibinfo {author} {\bibfnamefont {S.}~\bibnamefont
  {Coleman}},\ }\href {\doibase 10.1016/0003-4916(76)90280-3} {\bibfield
  {journal} {\bibinfo  {journal} {Annals of Physics}\ }\textbf {\bibinfo
  {volume} {101}},\ \bibinfo {pages} {239} (\bibinfo {year}
  {1976})}\BibitemShut {NoStop}%
\bibitem [{\citenamefont {Horn}(1981)}]{Horn1981}%
  \BibitemOpen
  \bibfield  {author} {\bibinfo {author} {\bibfnamefont {D.}~\bibnamefont
  {Horn}},\ }\href
  {http://www.sciencedirect.com/science/article/pii/0370269381907632}
  {\bibfield  {journal} {\bibinfo  {journal} {Physics Letters B}\ }\textbf
  {\bibinfo {volume} {100}},\ \bibinfo {pages} {149} (\bibinfo {year}
  {1981})}\BibitemShut {NoStop}%
\bibitem [{\citenamefont {Orland}\ and\ \citenamefont
  {Rohrlich}(1990)}]{Orland1990}%
  \BibitemOpen
  \bibfield  {author} {\bibinfo {author} {\bibfnamefont {P.}~\bibnamefont
  {Orland}}\ and\ \bibinfo {author} {\bibfnamefont {D.}~\bibnamefont
  {Rohrlich}},\ }\href
  {http://www.sciencedirect.com/science/article/pii/055032139090646U}
  {\bibfield  {journal} {\bibinfo  {journal} {Nucl. Phys. B}\ }\textbf
  {\bibinfo {volume} {338}},\ \bibinfo {pages} {647} (\bibinfo {year}
  {1990})}\BibitemShut {NoStop}%
\bibitem [{\citenamefont {Chandrasekharan}\ and\ \citenamefont
  {Wiese}(1997)}]{Chandrasekharan1997}%
  \BibitemOpen
  \bibfield  {author} {\bibinfo {author} {\bibfnamefont {S.}~\bibnamefont
  {Chandrasekharan}}\ and\ \bibinfo {author} {\bibfnamefont {U.-J.}\
  \bibnamefont {Wiese}},\ }\href
  {http://www.sciencedirect.com/science/article/pii/S0550321397800417}
  {\bibfield  {journal} {\bibinfo  {journal} {Nucl. Phys. B}\ }\textbf
  {\bibinfo {volume} {492}},\ \bibinfo {pages} {455} (\bibinfo {year}
  {1997})}\BibitemShut {NoStop}%
\bibitem [{\citenamefont {Brower}\ \emph {et~al.}(1999)\citenamefont {Brower},
  \citenamefont {Chandrasekharan},\ and\ \citenamefont {Wiese}}]{Brower1999}%
  \BibitemOpen
  \bibfield  {author} {\bibinfo {author} {\bibfnamefont {R.}~\bibnamefont
  {Brower}}, \bibinfo {author} {\bibfnamefont {S.}~\bibnamefont
  {Chandrasekharan}}, \ and\ \bibinfo {author} {\bibfnamefont {U.-J.}\
  \bibnamefont {Wiese}},\ }\href
  {http://journals.aps.org/prd/abstract/10.1103/PhysRevD.60.094502} {\bibfield
  {journal} {\bibinfo  {journal} {Phys. Rev. D}\ }\textbf {\bibinfo {volume}
  {60}},\ \bibinfo {pages} {094502} (\bibinfo {year} {1999})}\BibitemShut
  {NoStop}%
\bibitem [{\citenamefont {Porras}\ \emph {et~al.}(2008)\citenamefont {Porras},
  \citenamefont {Marquardt}, \citenamefont {von Delft},\ and\ \citenamefont
  {Cirac}}]{Porras2008}%
  \BibitemOpen
  \bibfield  {author} {\bibinfo {author} {\bibfnamefont {D.}~\bibnamefont
  {Porras}}, \bibinfo {author} {\bibfnamefont {F.}~\bibnamefont {Marquardt}},
  \bibinfo {author} {\bibfnamefont {J.}~\bibnamefont {von Delft}}, \ and\
  \bibinfo {author} {\bibfnamefont {J.~I.}\ \bibnamefont {Cirac}},\ }\href
  {\doibase 10.1103/PhysRevA.78.010101} {\bibfield  {journal} {\bibinfo
  {journal} {Phys. Rev. A}\ }\textbf {\bibinfo {volume} {78}},\ \bibinfo
  {pages} {010101} (\bibinfo {year} {2008})}\BibitemShut {NoStop}%
\bibitem [{\citenamefont {Ivanov}\ \emph {et~al.}(2009)\citenamefont {Ivanov},
  \citenamefont {Ivanov}, \citenamefont {Vitanov}, \citenamefont {Mering},
  \citenamefont {Fleischhauer},\ and\ \citenamefont {Singer}}]{Ivanov2009}%
  \BibitemOpen
  \bibfield  {author} {\bibinfo {author} {\bibfnamefont {P.~A.}\ \bibnamefont
  {Ivanov}}, \bibinfo {author} {\bibfnamefont {S.~S.}\ \bibnamefont {Ivanov}},
  \bibinfo {author} {\bibfnamefont {N.~V.}\ \bibnamefont {Vitanov}}, \bibinfo
  {author} {\bibfnamefont {A.}~\bibnamefont {Mering}}, \bibinfo {author}
  {\bibfnamefont {M.}~\bibnamefont {Fleischhauer}}, \ and\ \bibinfo {author}
  {\bibfnamefont {K.}~\bibnamefont {Singer}},\ }\href
  {http://journals.aps.org/pra/abstract/10.1103/PhysRevA.80.060301} {\bibfield
  {journal} {\bibinfo  {journal} {Phys. Rev. A}\ }\textbf {\bibinfo {volume}
  {80}},\ \bibinfo {pages} {060301(R)} (\bibinfo {year} {2009})}\BibitemShut
  {NoStop}%
\bibitem [{\citenamefont {Casanova}\ \emph {et~al.}(2011)\citenamefont
  {Casanova}, \citenamefont {Lamata}, \citenamefont {Egusquiza}, \citenamefont
  {Gerritsma}, \citenamefont {Roos}, \citenamefont {Garc\'ia-Ripoll},\ and\
  \citenamefont {Solano}}]{Casanova2011}%
  \BibitemOpen
  \bibfield  {author} {\bibinfo {author} {\bibfnamefont {J.}~\bibnamefont
  {Casanova}}, \bibinfo {author} {\bibfnamefont {L.}~\bibnamefont {Lamata}},
  \bibinfo {author} {\bibfnamefont {I.~L.}\ \bibnamefont {Egusquiza}}, \bibinfo
  {author} {\bibfnamefont {R.}~\bibnamefont {Gerritsma}}, \bibinfo {author}
  {\bibfnamefont {C.~F.}\ \bibnamefont {Roos}}, \bibinfo {author}
  {\bibfnamefont {J.~J.}\ \bibnamefont {Garc\'ia-Ripoll}}, \ and\ \bibinfo
  {author} {\bibfnamefont {E.}~\bibnamefont {Solano}},\ }\href
  {http://journals.aps.org/prl/abstract/10.1103/PhysRevLett.107.260501}
  {\bibfield  {journal} {\bibinfo  {journal} {Phys. Rev. Lett.}\ }\textbf
  {\bibinfo {volume} {107}},\ \bibinfo {pages} {260501} (\bibinfo {year}
  {2011})}\BibitemShut {NoStop}%
\bibitem [{\citenamefont {Porras}\ \emph {et~al.}(2012)\citenamefont {Porras},
  \citenamefont {Ivanov},\ and\ \citenamefont {Schmidt-Kaler}}]{Porras2012}%
  \BibitemOpen
  \bibfield  {author} {\bibinfo {author} {\bibfnamefont {D.}~\bibnamefont
  {Porras}}, \bibinfo {author} {\bibfnamefont {P.~A.}\ \bibnamefont {Ivanov}},
  \ and\ \bibinfo {author} {\bibfnamefont {F.}~\bibnamefont {Schmidt-Kaler}},\
  }\href {\doibase 10.1103/PhysRevLett.108.235701} {\bibfield  {journal}
  {\bibinfo  {journal} {Phys. Rev. Lett.}\ }\textbf {\bibinfo {volume} {108}},\
  \bibinfo {pages} {235701} (\bibinfo {year} {2012})}\BibitemShut {NoStop}%
\bibitem [{\citenamefont {Mezzacapo}\ \emph {et~al.}(2012)\citenamefont
  {Mezzacapo}, \citenamefont {Casanova}, \citenamefont {Lamata},\ and\
  \citenamefont {Solano}}]{Mezzacapo2012b}%
  \BibitemOpen
  \bibfield  {author} {\bibinfo {author} {\bibfnamefont {A.}~\bibnamefont
  {Mezzacapo}}, \bibinfo {author} {\bibfnamefont {J.}~\bibnamefont {Casanova}},
  \bibinfo {author} {\bibfnamefont {L.}~\bibnamefont {Lamata}}, \ and\ \bibinfo
  {author} {\bibfnamefont {E.}~\bibnamefont {Solano}},\ }\href
  {http://journals.aps.org/prl/abstract/10.1103/PhysRevLett.109.200501}
  {\bibfield  {journal} {\bibinfo  {journal} {Phys. Rev. Lett.}\ }\textbf
  {\bibinfo {volume} {109}},\ \bibinfo {pages} {200501} (\bibinfo {year}
  {2012})}\BibitemShut {NoStop}%
\bibitem [{\citenamefont {Nevado}\ and\ \citenamefont
  {Porras}(2013)}]{Nevado2013}%
  \BibitemOpen
  \bibfield  {author} {\bibinfo {author} {\bibfnamefont {P.}~\bibnamefont
  {Nevado}}\ and\ \bibinfo {author} {\bibfnamefont {D.}~\bibnamefont
  {Porras}},\ }\href {http://dx.doi.org/10.1140/epjst/e2013-01751-1} {\bibfield
   {journal} {\bibinfo  {journal} {Eur. Phys. J. Special Topics}\ }\textbf
  {\bibinfo {volume} {217}},\ \bibinfo {pages} {29} (\bibinfo {year}
  {2013})}\BibitemShut {NoStop}%
\bibitem [{\citenamefont {Ivanov}\ \emph {et~al.}(2013)\citenamefont {Ivanov},
  \citenamefont {Porras}, \citenamefont {Ivanov},\ and\ \citenamefont
  {Schmidt-Kaler}}]{Ivanov2013}%
  \BibitemOpen
  \bibfield  {author} {\bibinfo {author} {\bibfnamefont {P.~A.}\ \bibnamefont
  {Ivanov}}, \bibinfo {author} {\bibfnamefont {D.}~\bibnamefont {Porras}},
  \bibinfo {author} {\bibfnamefont {S.~S.}\ \bibnamefont {Ivanov}}, \ and\
  \bibinfo {author} {\bibfnamefont {F.}~\bibnamefont {Schmidt-Kaler}},\ }\href
  {http://iopscience.iop.org/article/10.1088/0953-4075/46/10/104003/meta}
  {\bibfield  {journal} {\bibinfo  {journal} {J. Phys. B}\ }\textbf {\bibinfo
  {volume} {46}},\ \bibinfo {pages} {104003} (\bibinfo {year}
  {2013})}\BibitemShut {NoStop}%
\bibitem [{\citenamefont {Kurcz}\ \emph {et~al.}(2015)\citenamefont {Kurcz},
  \citenamefont {Garc{\'\i}a-Ripoll},\ and\ \citenamefont
  {Bermudez}}]{Kurcz2015}%
  \BibitemOpen
  \bibfield  {author} {\bibinfo {author} {\bibfnamefont {A.}~\bibnamefont
  {Kurcz}}, \bibinfo {author} {\bibfnamefont {J.~J.}\ \bibnamefont
  {Garc{\'\i}a-Ripoll}}, \ and\ \bibinfo {author} {\bibfnamefont
  {A.}~\bibnamefont {Bermudez}},\ }\href
  {http://iopscience.iop.org/article/10.1088/1367-2630/17/11/115011/meta}
  {\bibfield  {journal} {\bibinfo  {journal} {New J. Phys.}\ }\textbf {\bibinfo
  {volume} {17}},\ \bibinfo {pages} {115011} (\bibinfo {year}
  {2015})}\BibitemShut {NoStop}%
\bibitem [{\citenamefont {Nevado}\ and\ \citenamefont {Porras}()}]{Nevado2015}%
  \BibitemOpen
  \bibfield  {author} {\bibinfo {author} {\bibfnamefont {P.}~\bibnamefont
  {Nevado}}\ and\ \bibinfo {author} {\bibfnamefont {D.}~\bibnamefont
  {Porras}},\ }\href {http://arxiv.org/abs/1503.04614} {\bibinfo  {journal}
  {arXiv:1503.04614}\ }\BibitemShut {NoStop}%
\bibitem [{\citenamefont {Gerritsma}\ \emph {et~al.}(2010)\citenamefont
  {Gerritsma}, \citenamefont {Kirchmair}, \citenamefont {Z{\"a}hringer},
  \citenamefont {Solano}, \citenamefont {Blatt},\ and\ \citenamefont
  {Roos}}]{Gerritsma2010}%
  \BibitemOpen
\bibfield  {journal} {  }\bibfield  {author} {\bibinfo {author} {\bibfnamefont
  {R.}~\bibnamefont {Gerritsma}}, \bibinfo {author} {\bibfnamefont
  {G.}~\bibnamefont {Kirchmair}}, \bibinfo {author} {\bibfnamefont
  {F.}~\bibnamefont {Z{\"a}hringer}}, \bibinfo {author} {\bibfnamefont
  {E.}~\bibnamefont {Solano}}, \bibinfo {author} {\bibfnamefont
  {R.}~\bibnamefont {Blatt}}, \ and\ \bibinfo {author} {\bibfnamefont {C.~F.}\
  \bibnamefont {Roos}},\ }\href
  {http://www.nature.com/nature/journal/v463/n7277/abs/nature08688.html}
  {\bibfield  {journal} {\bibinfo  {journal} {Nature}\ }\textbf {\bibinfo
  {volume} {463}},\ \bibinfo {pages} {68} (\bibinfo {year} {2010})}\BibitemShut
  {NoStop}%
\bibitem [{\citenamefont {Gerritsma}\ \emph {et~al.}(2011)\citenamefont
  {Gerritsma}, \citenamefont {Lanyon}, \citenamefont {Kirchmair}, \citenamefont
  {Z{\"a}hringer}, \citenamefont {Hempel}, \citenamefont {Casanova},
  \citenamefont {Garc{\'\i}a-Ripoll}, \citenamefont {Solano}, \citenamefont
  {Blatt},\ and\ \citenamefont {Roos}}]{Gerritsma2011}%
  \BibitemOpen
  \bibfield  {author} {\bibinfo {author} {\bibfnamefont {R.}~\bibnamefont
  {Gerritsma}}, \bibinfo {author} {\bibfnamefont {B.}~\bibnamefont {Lanyon}},
  \bibinfo {author} {\bibfnamefont {G.}~\bibnamefont {Kirchmair}}, \bibinfo
  {author} {\bibfnamefont {F.}~\bibnamefont {Z{\"a}hringer}}, \bibinfo {author}
  {\bibfnamefont {C.}~\bibnamefont {Hempel}}, \bibinfo {author} {\bibfnamefont
  {J.}~\bibnamefont {Casanova}}, \bibinfo {author} {\bibfnamefont {J.~J.}\
  \bibnamefont {Garc{\'\i}a-Ripoll}}, \bibinfo {author} {\bibfnamefont
  {E.}~\bibnamefont {Solano}}, \bibinfo {author} {\bibfnamefont
  {R.}~\bibnamefont {Blatt}}, \ and\ \bibinfo {author} {\bibfnamefont {C.~F.}\
  \bibnamefont {Roos}},\ }\href
  {http://journals.aps.org/prl/abstract/10.1103/PhysRevLett.106.060503}
  {\bibfield  {journal} {\bibinfo  {journal} {Phys. Rev. Lett.}\ }\textbf
  {\bibinfo {volume} {106}},\ \bibinfo {pages} {060503} (\bibinfo {year}
  {2011})}\BibitemShut {NoStop}%
\bibitem [{\citenamefont {Banks}\ \emph {et~al.}(1976)\citenamefont {Banks},
  \citenamefont {Susskind},\ and\ \citenamefont {Kogut}}]{Banks1976}%
  \BibitemOpen
  \bibfield  {author} {\bibinfo {author} {\bibfnamefont {T.}~\bibnamefont
  {Banks}}, \bibinfo {author} {\bibfnamefont {L.}~\bibnamefont {Susskind}}, \
  and\ \bibinfo {author} {\bibfnamefont {J.}~\bibnamefont {Kogut}},\ }\href
  {http://link.aps.org/doi/10.1103/PhysRevD.13.1043} {\bibfield  {journal}
  {\bibinfo  {journal} {Phys. Rev. D}\ }\textbf {\bibinfo {volume} {13}},\
  \bibinfo {pages} {1043} (\bibinfo {year} {1976})}\BibitemShut {NoStop}%
\bibitem [{\citenamefont {Hamer}\ \emph {et~al.}(1982)\citenamefont {Hamer},
  \citenamefont {Kogut}, \citenamefont {Crewther},\ and\ \citenamefont
  {Mazzolini}}]{Hamer1982}%
  \BibitemOpen
  \bibfield  {author} {\bibinfo {author} {\bibfnamefont {C.}~\bibnamefont
  {Hamer}}, \bibinfo {author} {\bibfnamefont {J.}~\bibnamefont {Kogut}},
  \bibinfo {author} {\bibfnamefont {D.}~\bibnamefont {Crewther}}, \ and\
  \bibinfo {author} {\bibfnamefont {M.}~\bibnamefont {Mazzolini}},\ }\href
  {\doibase 10.1016/0550-3213(82)90229-2} {\bibfield  {journal} {\bibinfo
  {journal} {Nuclear Physics B}\ }\textbf {\bibinfo {volume} {208}},\ \bibinfo
  {pages} {413} (\bibinfo {year} {1982})}\BibitemShut {NoStop}%
\bibitem [{\citenamefont {Byrnes}\ \emph {et~al.}(2002)\citenamefont {Byrnes},
  \citenamefont {Sriganesh}, \citenamefont {Bursill},\ and\ \citenamefont
  {Hamer}}]{Byrnes2002}%
  \BibitemOpen
  \bibfield  {author} {\bibinfo {author} {\bibfnamefont {T.}~\bibnamefont
  {Byrnes}}, \bibinfo {author} {\bibfnamefont {P.}~\bibnamefont {Sriganesh}},
  \bibinfo {author} {\bibfnamefont {R.~J.}\ \bibnamefont {Bursill}}, \ and\
  \bibinfo {author} {\bibfnamefont {C.~J.}\ \bibnamefont {Hamer}},\ }\href
  {\doibase 10.1103/PhysRevD.66.013002} {\bibfield  {journal} {\bibinfo
  {journal} {Physical Review D}\ }\textbf {\bibinfo {volume} {66}},\ \bibinfo
  {pages} {38} (\bibinfo {year} {2002})}\BibitemShut {NoStop}%
\bibitem [{\citenamefont {Kogut}\ and\ \citenamefont
  {Susskind}(1975)}]{Kogut1975}%
  \BibitemOpen
  \bibfield  {author} {\bibinfo {author} {\bibfnamefont {J.}~\bibnamefont
  {Kogut}}\ and\ \bibinfo {author} {\bibfnamefont {L.}~\bibnamefont
  {Susskind}},\ }\href {\doibase 10.1103/PhysRevD.11.395} {\bibfield  {journal}
  {\bibinfo  {journal} {Phys. Rev. D}\ }\textbf {\bibinfo {volume} {11}},\
  \bibinfo {pages} {395} (\bibinfo {year} {1975})}\BibitemShut {NoStop}%
\bibitem [{\citenamefont {Banerjee}\ \emph {et~al.}(2012)\citenamefont
  {Banerjee}, \citenamefont {Dalmonte}, \citenamefont {M\"uller}, \citenamefont
  {Rico}, \citenamefont {Stebler}, \citenamefont {Wiese},\ and\ \citenamefont
  {Zoller}}]{Banerjee2012}%
  \BibitemOpen
  \bibfield  {author} {\bibinfo {author} {\bibfnamefont {D.}~\bibnamefont
  {Banerjee}}, \bibinfo {author} {\bibfnamefont {M.}~\bibnamefont {Dalmonte}},
  \bibinfo {author} {\bibfnamefont {M.}~\bibnamefont {M\"uller}}, \bibinfo
  {author} {\bibfnamefont {E.}~\bibnamefont {Rico}}, \bibinfo {author}
  {\bibfnamefont {P.}~\bibnamefont {Stebler}}, \bibinfo {author} {\bibfnamefont
  {U.-J.}\ \bibnamefont {Wiese}}, \ and\ \bibinfo {author} {\bibfnamefont
  {P.}~\bibnamefont {Zoller}},\ }\href {\doibase
  http://journals.aps.org/prl/abstract/10.1103/PhysRevLett.109.175302}
  {\bibfield  {journal} {\bibinfo  {journal} {Phys. Rev. Lett.}\ }\textbf
  {\bibinfo {volume} {109}},\ \bibinfo {pages} {175302} (\bibinfo {year}
  {2012})}\BibitemShut {NoStop}%
\bibitem [{\citenamefont {Banerjee}\ \emph {et~al.}(2013)\citenamefont
  {Banerjee}, \citenamefont {B\"ogli}, \citenamefont {Dalmonte}, \citenamefont
  {Rico}, \citenamefont {Stebler}, \citenamefont {Wiese},\ and\ \citenamefont
  {Zoller}}]{Banerjee2013}%
  \BibitemOpen
  \bibfield  {author} {\bibinfo {author} {\bibfnamefont {D.}~\bibnamefont
  {Banerjee}}, \bibinfo {author} {\bibfnamefont {M.}~\bibnamefont {B\"ogli}},
  \bibinfo {author} {\bibfnamefont {M.}~\bibnamefont {Dalmonte}}, \bibinfo
  {author} {\bibfnamefont {E.}~\bibnamefont {Rico}}, \bibinfo {author}
  {\bibfnamefont {P.}~\bibnamefont {Stebler}}, \bibinfo {author} {\bibfnamefont
  {U.-J.}\ \bibnamefont {Wiese}}, \ and\ \bibinfo {author} {\bibfnamefont
  {P.}~\bibnamefont {Zoller}},\ }\href
  {http://journals.aps.org/prl/abstract/10.1103/PhysRevLett.110.125303}
  {\bibfield  {journal} {\bibinfo  {journal} {Phys. Rev. Lett.}\ }\textbf
  {\bibinfo {volume} {110}},\ \bibinfo {pages} {125303} (\bibinfo {year}
  {2013})}\BibitemShut {NoStop}%
\bibitem [{\citenamefont {Harlander}\ \emph {et~al.}(2011)\citenamefont
  {Harlander}, \citenamefont {Lechner}, \citenamefont {Brownnutt},
  \citenamefont {Blatt},\ and\ \citenamefont {H{\"{a}}nsel}}]{Harlander2011}%
  \BibitemOpen
  \bibfield  {author} {\bibinfo {author} {\bibfnamefont {M.}~\bibnamefont
  {Harlander}}, \bibinfo {author} {\bibfnamefont {R.}~\bibnamefont {Lechner}},
  \bibinfo {author} {\bibfnamefont {M.}~\bibnamefont {Brownnutt}}, \bibinfo
  {author} {\bibfnamefont {R.}~\bibnamefont {Blatt}}, \ and\ \bibinfo {author}
  {\bibfnamefont {W.}~\bibnamefont {H{\"{a}}nsel}},\ }\href {\doibase
  10.1038/nature09800} {\bibfield  {journal} {\bibinfo  {journal} {Nature}\
  }\textbf {\bibinfo {volume} {471}},\ \bibinfo {pages} {200} (\bibinfo {year}
  {2011})}\BibitemShut {NoStop}%
\bibitem [{\citenamefont {Wilson}\ \emph {et~al.}(2014)\citenamefont {Wilson},
  \citenamefont {Colombe}, \citenamefont {Brown}, \citenamefont {Knill},
  \citenamefont {Leibfried},\ and\ \citenamefont {Wineland}}]{Wilson2014}%
  \BibitemOpen
  \bibfield  {author} {\bibinfo {author} {\bibfnamefont {A.~C.}\ \bibnamefont
  {Wilson}}, \bibinfo {author} {\bibfnamefont {Y.}~\bibnamefont {Colombe}},
  \bibinfo {author} {\bibfnamefont {K.~R.}\ \bibnamefont {Brown}}, \bibinfo
  {author} {\bibfnamefont {E.}~\bibnamefont {Knill}}, \bibinfo {author}
  {\bibfnamefont {D.}~\bibnamefont {Leibfried}}, \ and\ \bibinfo {author}
  {\bibfnamefont {D.~J.}\ \bibnamefont {Wineland}},\ }\href {\doibase
  10.1038/nature13565} {\bibfield  {journal} {\bibinfo  {journal} {Nature}\
  }\textbf {\bibinfo {volume} {512}},\ \bibinfo {pages} {57} (\bibinfo {year}
  {2014})}\BibitemShut {NoStop}%
\bibitem [{\citenamefont {Mehta}\ \emph {et~al.}(2014)\citenamefont {Mehta},
  \citenamefont {Eltony}, \citenamefont {Bruzewicz}, \citenamefont {Chuang},
  \citenamefont {Ram}, \citenamefont {Sage},\ and\ \citenamefont
  {Chiaverini}}]{Mehta2014}%
  \BibitemOpen
  \bibfield  {author} {\bibinfo {author} {\bibfnamefont {K.~K.}\ \bibnamefont
  {Mehta}}, \bibinfo {author} {\bibfnamefont {A.~M.}\ \bibnamefont {Eltony}},
  \bibinfo {author} {\bibfnamefont {C.~D.}\ \bibnamefont {Bruzewicz}}, \bibinfo
  {author} {\bibfnamefont {I.~L.}\ \bibnamefont {Chuang}}, \bibinfo {author}
  {\bibfnamefont {R.~J.}\ \bibnamefont {Ram}}, \bibinfo {author} {\bibfnamefont
  {J.~M.}\ \bibnamefont {Sage}}, \ and\ \bibinfo {author} {\bibfnamefont
  {J.}~\bibnamefont {Chiaverini}},\ }\href
  {http://scitation.aip.org/content/aip/journal/apl/105/4/10.1063/1.4892061}
  {\bibfield  {journal} {\bibinfo  {journal} {Applied Physics Letters}\
  }\textbf {\bibinfo {volume} {105}},\ \bibinfo {eid} {044103} (\bibinfo {year}
  {2014})}\BibitemShut {NoStop}%
\bibitem [{\citenamefont {Mielenz}\ \emph {et~al.}(2015)\citenamefont
  {Mielenz}, \citenamefont {Kalis}, \citenamefont {Wittemer}, \citenamefont
  {Hakelberg}, \citenamefont {Schmied}, \citenamefont {Blain}, \citenamefont
  {Maunz}, \citenamefont {Leibfried}, \citenamefont {Warring},\ and\
  \citenamefont {Schaetz}}]{Mielenz2015}%
  \BibitemOpen
  \bibfield  {author} {\bibinfo {author} {\bibfnamefont {M.}~\bibnamefont
  {Mielenz}}, \bibinfo {author} {\bibfnamefont {H.}~\bibnamefont {Kalis}},
  \bibinfo {author} {\bibfnamefont {M.}~\bibnamefont {Wittemer}}, \bibinfo
  {author} {\bibfnamefont {F.}~\bibnamefont {Hakelberg}}, \bibinfo {author}
  {\bibfnamefont {R.}~\bibnamefont {Schmied}}, \bibinfo {author} {\bibfnamefont
  {M.}~\bibnamefont {Blain}}, \bibinfo {author} {\bibfnamefont
  {P.}~\bibnamefont {Maunz}}, \bibinfo {author} {\bibfnamefont
  {D.}~\bibnamefont {Leibfried}}, \bibinfo {author} {\bibfnamefont
  {U.}~\bibnamefont {Warring}}, \ and\ \bibinfo {author} {\bibfnamefont
  {T.}~\bibnamefont {Schaetz}},\ }\href {http://arxiv.org/abs/1512.03559}
  {\bibfield  {journal} {\bibinfo  {journal} {arXiv:1512.03559}\ } (\bibinfo
  {year} {2015})}\BibitemShut {NoStop}%
\bibitem [{\citenamefont {Wineland}\ \emph {et~al.}(1998)\citenamefont
  {Wineland}, \citenamefont {Monroe}, \citenamefont {Itano}, \citenamefont
  {Leibfried}, \citenamefont {King},\ and\ \citenamefont
  {Meekhof}}]{Wineland1998}%
  \BibitemOpen
  \bibfield  {author} {\bibinfo {author} {\bibfnamefont {D.}~\bibnamefont
  {Wineland}}, \bibinfo {author} {\bibfnamefont {C.}~\bibnamefont {Monroe}},
  \bibinfo {author} {\bibfnamefont {W.}~\bibnamefont {Itano}}, \bibinfo
  {author} {\bibfnamefont {D.}~\bibnamefont {Leibfried}}, \bibinfo {author}
  {\bibfnamefont {B.}~\bibnamefont {King}}, \ and\ \bibinfo {author}
  {\bibfnamefont {D.}~\bibnamefont {Meekhof}},\ }\href {\doibase
  10.6028/jres.103.019} {\bibfield  {journal} {\bibinfo  {journal} {Journal of
  Research of the National Institute of Standards and Technology}\ }\textbf
  {\bibinfo {volume} {103}},\ \bibinfo {pages} {259} (\bibinfo {year}
  {1998})}\BibitemShut {NoStop}%
\bibitem [{\citenamefont {James}(1998)}]{James1998}%
  \BibitemOpen
  \bibfield  {author} {\bibinfo {author} {\bibfnamefont {D.}~\bibnamefont
  {James}},\ }\href {\doibase 10.1007/s003400050373} {\bibfield  {journal}
  {\bibinfo  {journal} {Applied Physics B}\ }\textbf {\bibinfo {volume} {66}},\
  \bibinfo {pages} {181} (\bibinfo {year} {1998})}\BibitemShut {NoStop}%
\bibitem [{\citenamefont {Porras}\ and\ \citenamefont
  {Cirac}(2004{\natexlab{a}})}]{Porras2004}%
  \BibitemOpen
  \bibfield  {author} {\bibinfo {author} {\bibfnamefont {D.}~\bibnamefont
  {Porras}}\ and\ \bibinfo {author} {\bibfnamefont {J.~I.}\ \bibnamefont
  {Cirac}},\ }\href {\doibase 10.1103/PhysRevLett.92.207901} {\bibfield
  {journal} {\bibinfo  {journal} {Phys. Rev. Lett.}\ }\textbf {\bibinfo
  {volume} {92}},\ \bibinfo {pages} {207901} (\bibinfo {year}
  {2004}{\natexlab{a}})}\BibitemShut {NoStop}%
\bibitem [{\citenamefont {Schmied}\ \emph {et~al.}(2009)\citenamefont
  {Schmied}, \citenamefont {Wesenberg},\ and\ \citenamefont
  {Leibfried}}]{Schmied2009}%
  \BibitemOpen
  \bibfield  {author} {\bibinfo {author} {\bibfnamefont {R.}~\bibnamefont
  {Schmied}}, \bibinfo {author} {\bibfnamefont {J.~H.}\ \bibnamefont
  {Wesenberg}}, \ and\ \bibinfo {author} {\bibfnamefont {D.}~\bibnamefont
  {Leibfried}},\ }\href {\doibase 10.1103/PhysRevLett.102.233002} {\bibfield
  {journal} {\bibinfo  {journal} {Phys. Rev. Lett.}\ }\textbf {\bibinfo
  {volume} {102}},\ \bibinfo {pages} {2} (\bibinfo {year} {2009})}\BibitemShut
  {NoStop}%
\bibitem [{\citenamefont {Brown}\ \emph {et~al.}(2011)\citenamefont {Brown},
  \citenamefont {Ospelkaus}, \citenamefont {Colombe}, \citenamefont {Wilson},
  \citenamefont {Leibfried},\ and\ \citenamefont {Wineland}}]{Brown2011}%
  \BibitemOpen
  \bibfield  {author} {\bibinfo {author} {\bibfnamefont {K.~R.}\ \bibnamefont
  {Brown}}, \bibinfo {author} {\bibfnamefont {C.}~\bibnamefont {Ospelkaus}},
  \bibinfo {author} {\bibfnamefont {Y.}~\bibnamefont {Colombe}}, \bibinfo
  {author} {\bibfnamefont {A.~C.}\ \bibnamefont {Wilson}}, \bibinfo {author}
  {\bibfnamefont {D.}~\bibnamefont {Leibfried}}, \ and\ \bibinfo {author}
  {\bibfnamefont {D.~J.}\ \bibnamefont {Wineland}},\ }\href {\doibase
  10.1038/nature09721} {\bibfield  {journal} {\bibinfo  {journal} {Nature}\
  }\textbf {\bibinfo {volume} {471}},\ \bibinfo {pages} {196} (\bibinfo {year}
  {2011})}\BibitemShut {NoStop}%
\bibitem [{\citenamefont {Birkl}\ \emph {et~al.}(1992)\citenamefont {Birkl},
  \citenamefont {Kassner},\ and\ \citenamefont {Walther}}]{Birkl1992}%
  \BibitemOpen
  \bibfield  {author} {\bibinfo {author} {\bibfnamefont {G.}~\bibnamefont
  {Birkl}}, \bibinfo {author} {\bibfnamefont {S.}~\bibnamefont {Kassner}}, \
  and\ \bibinfo {author} {\bibfnamefont {H.}~\bibnamefont {Walther}},\ }\href
  {http://dx.doi.org/10.1038/357310a0} {\bibfield  {journal} {\bibinfo
  {journal} {Nature}\ }\textbf {\bibinfo {volume} {357}},\ \bibinfo {pages}
  {310} (\bibinfo {year} {1992})}\BibitemShut {NoStop}%
\bibitem [{\citenamefont {Szymanski}\ \emph {et~al.}(2012)\citenamefont
  {Szymanski}, \citenamefont {Dubessy}, \citenamefont {Dubost}, \citenamefont
  {Guibal}, \citenamefont {Likforman},\ and\ \citenamefont
  {Guidoni}}]{Szymanski2012}%
  \BibitemOpen
  \bibfield  {author} {\bibinfo {author} {\bibfnamefont {B.}~\bibnamefont
  {Szymanski}}, \bibinfo {author} {\bibfnamefont {R.}~\bibnamefont {Dubessy}},
  \bibinfo {author} {\bibfnamefont {B.}~\bibnamefont {Dubost}}, \bibinfo
  {author} {\bibfnamefont {S.}~\bibnamefont {Guibal}}, \bibinfo {author}
  {\bibfnamefont {J.-P.}\ \bibnamefont {Likforman}}, \ and\ \bibinfo {author}
  {\bibfnamefont {L.}~\bibnamefont {Guidoni}},\ }\href {\doibase
  10.1063/1.4705153} {\bibfield  {journal} {\bibinfo  {journal} {Applied
  Physics Letters}\ }\textbf {\bibinfo {volume} {100}},\ \bibinfo {pages}
  {171110} (\bibinfo {year} {2012})},\ \Eprint {http://arxiv.org/abs/1201.2584}
  {arXiv:1201.2584} \BibitemShut {NoStop}%
\bibitem [{\citenamefont {Porras}\ and\ \citenamefont
  {Cirac}(2004{\natexlab{b}})}]{Porras2004b}%
  \BibitemOpen
  \bibfield  {author} {\bibinfo {author} {\bibfnamefont {D.}~\bibnamefont
  {Porras}}\ and\ \bibinfo {author} {\bibfnamefont {J.}~\bibnamefont {Cirac}},\
  }\href {\doibase 10.1103/PhysRevLett.93.263602} {\bibfield  {journal}
  {\bibinfo  {journal} {Phys. Rev. Lett.}\ }\textbf {\bibinfo {volume} {93}},\
  \bibinfo {pages} {263602} (\bibinfo {year} {2004}{\natexlab{b}})}\BibitemShut
  {NoStop}%
\bibitem [{\citenamefont {Bollinger}\ \emph {et~al.}(1985)\citenamefont
  {Bollinger}, \citenamefont {Wells}, \citenamefont {Wineland},\ and\
  \citenamefont {Itano}}]{Bollinger1985}%
  \BibitemOpen
  \bibfield  {author} {\bibinfo {author} {\bibfnamefont {J.~J.}\ \bibnamefont
  {Bollinger}}, \bibinfo {author} {\bibfnamefont {J.~S.}\ \bibnamefont
  {Wells}}, \bibinfo {author} {\bibfnamefont {D.~J.}\ \bibnamefont {Wineland}},
  \ and\ \bibinfo {author} {\bibfnamefont {W.~M.}\ \bibnamefont {Itano}},\
  }\href {\doibase 10.1103/PhysRevA.31.2711} {\bibfield  {journal} {\bibinfo
  {journal} {Phys. Rev. A}\ }\textbf {\bibinfo {volume} {31}},\ \bibinfo
  {pages} {2711} (\bibinfo {year} {1985})}\BibitemShut {NoStop}%
\bibitem [{\citenamefont {Meekhof}\ \emph {et~al.}(1996)\citenamefont
  {Meekhof}, \citenamefont {Monroe}, \citenamefont {King}, \citenamefont
  {Itano},\ and\ \citenamefont {Wineland}}]{Meekhof1996}%
  \BibitemOpen
  \bibfield  {author} {\bibinfo {author} {\bibfnamefont {D.~M.}\ \bibnamefont
  {Meekhof}}, \bibinfo {author} {\bibfnamefont {C.}~\bibnamefont {Monroe}},
  \bibinfo {author} {\bibfnamefont {B.~E.}\ \bibnamefont {King}}, \bibinfo
  {author} {\bibfnamefont {W.~M.}\ \bibnamefont {Itano}}, \ and\ \bibinfo
  {author} {\bibfnamefont {D.~J.}\ \bibnamefont {Wineland}},\ }\href {\doibase
  10.1103/PhysRevLett.76.1796} {\bibfield  {journal} {\bibinfo  {journal}
  {Phys. Rev. Lett.}\ }\textbf {\bibinfo {volume} {76}},\ \bibinfo {pages}
  {1796} (\bibinfo {year} {1996})}\BibitemShut {NoStop}%
\bibitem [{\citenamefont {Johanning}\ \emph
  {et~al.}(2009{\natexlab{a}})\citenamefont {Johanning}, \citenamefont
  {Var\'{o}n},\ and\ \citenamefont {Wunderlich}}]{Johanning2009}%
  \BibitemOpen
  \bibfield  {author} {\bibinfo {author} {\bibfnamefont {M.}~\bibnamefont
  {Johanning}}, \bibinfo {author} {\bibfnamefont {A.~F.}\ \bibnamefont
  {Var\'{o}n}}, \ and\ \bibinfo {author} {\bibfnamefont {C.}~\bibnamefont
  {Wunderlich}},\ }\href
  {http://iopscience.iop.org/article/10.1088/0953-4075/42/15/154009/meta}
  {\bibfield  {journal} {\bibinfo  {journal} {J. Phys. B}\ }\textbf {\bibinfo
  {volume} {42}},\ \bibinfo {pages} {154009} (\bibinfo {year}
  {2009}{\natexlab{a}})}\BibitemShut {NoStop}%
\bibitem [{\citenamefont {Blatt}\ and\ \citenamefont {Roos}(2012)}]{Blatt2012}%
  \BibitemOpen
  \bibfield  {author} {\bibinfo {author} {\bibfnamefont {R.}~\bibnamefont
  {Blatt}}\ and\ \bibinfo {author} {\bibfnamefont {C.~F.}\ \bibnamefont
  {Roos}},\ }\href {\doibase doi:10.1038/nphys2252} {\bibfield  {journal}
  {\bibinfo  {journal} {Nat. Phys.}\ }\textbf {\bibinfo {volume} {8}},\
  \bibinfo {pages} {277} (\bibinfo {year} {2012})}\BibitemShut {NoStop}%
\bibitem [{\citenamefont {Schneider}\ \emph {et~al.}(2012)\citenamefont
  {Schneider}, \citenamefont {Porras},\ and\ \citenamefont
  {Schaetz}}]{Schneider2012}%
  \BibitemOpen
  \bibfield  {author} {\bibinfo {author} {\bibfnamefont {C.}~\bibnamefont
  {Schneider}}, \bibinfo {author} {\bibfnamefont {D.}~\bibnamefont {Porras}}, \
  and\ \bibinfo {author} {\bibfnamefont {T.}~\bibnamefont {Schaetz}},\ }\href
  {http://iopscience.iop.org/article/10.1088/0034-4885/75/2/024401/meta}
  {\bibfield  {journal} {\bibinfo  {journal} {Rep. Prog. Phys.}\ }\textbf
  {\bibinfo {volume} {75}},\ \bibinfo {pages} {024401} (\bibinfo {year}
  {2012})}\BibitemShut {NoStop}%
\bibitem [{\citenamefont {McConnell}\ \emph {et~al.}(2015)\citenamefont
  {McConnell}, \citenamefont {Bruzewicz}, \citenamefont {Chiaverini},\ and\
  \citenamefont {Sage}}]{McConnell2015}%
  \BibitemOpen
  \bibfield  {author} {\bibinfo {author} {\bibfnamefont {R.}~\bibnamefont
  {McConnell}}, \bibinfo {author} {\bibfnamefont {C.}~\bibnamefont
  {Bruzewicz}}, \bibinfo {author} {\bibfnamefont {J.}~\bibnamefont
  {Chiaverini}}, \ and\ \bibinfo {author} {\bibfnamefont {J.}~\bibnamefont
  {Sage}},\ }\href {\doibase 10.1103/PhysRevA.92.020302} {\bibfield  {journal}
  {\bibinfo  {journal} {Phys. Rev. A}\ }\textbf {\bibinfo {volume} {92}},\
  \bibinfo {pages} {1} (\bibinfo {year} {2015})}\BibitemShut {NoStop}%
\bibitem [{\citenamefont {Jurcevic}\ \emph {et~al.}(2014)\citenamefont
  {Jurcevic}, \citenamefont {Lanyon}, \citenamefont {Hauke}, \citenamefont
  {Hempel}, \citenamefont {Zoller}, \citenamefont {Blatt},\ and\ \citenamefont
  {Roos}}]{Jurcevic2014}%
  \BibitemOpen
  \bibfield  {author} {\bibinfo {author} {\bibfnamefont {P.}~\bibnamefont
  {Jurcevic}}, \bibinfo {author} {\bibfnamefont {B.~P.}\ \bibnamefont
  {Lanyon}}, \bibinfo {author} {\bibfnamefont {P.}~\bibnamefont {Hauke}},
  \bibinfo {author} {\bibfnamefont {C.}~\bibnamefont {Hempel}}, \bibinfo
  {author} {\bibfnamefont {P.}~\bibnamefont {Zoller}}, \bibinfo {author}
  {\bibfnamefont {R.}~\bibnamefont {Blatt}}, \ and\ \bibinfo {author}
  {\bibfnamefont {C.~F.}\ \bibnamefont {Roos}},\ }\href
  {http://www.nature.com/nature/journal/v511/n7508/full/nature13461.html}
  {\bibfield  {journal} {\bibinfo  {journal} {Nature}\ }\textbf {\bibinfo
  {volume} {511}},\ \bibinfo {pages} {202} (\bibinfo {year}
  {2014})}\BibitemShut {NoStop}%
\bibitem [{\citenamefont {Jurcevic}\ \emph {et~al.}(2015)\citenamefont
  {Jurcevic}, \citenamefont {Hauke}, \citenamefont {Maier}, \citenamefont
  {Hempel}, \citenamefont {Lanyon}, \citenamefont {Blatt},\ and\ \citenamefont
  {Roos}}]{Jurcevic2015}%
  \BibitemOpen
  \bibfield  {author} {\bibinfo {author} {\bibfnamefont {P.}~\bibnamefont
  {Jurcevic}}, \bibinfo {author} {\bibfnamefont {P.}~\bibnamefont {Hauke}},
  \bibinfo {author} {\bibfnamefont {C.}~\bibnamefont {Maier}}, \bibinfo
  {author} {\bibfnamefont {C.}~\bibnamefont {Hempel}}, \bibinfo {author}
  {\bibfnamefont {B.}~\bibnamefont {Lanyon}}, \bibinfo {author} {\bibfnamefont
  {R.}~\bibnamefont {Blatt}}, \ and\ \bibinfo {author} {\bibfnamefont
  {C.}~\bibnamefont {Roos}},\ }\href
  {http://journals.aps.org/prl/abstract/10.1103/PhysRevLett.115.100501}
  {\bibfield  {journal} {\bibinfo  {journal} {Phys. Rev. Lett.}\ }\textbf
  {\bibinfo {volume} {115}},\ \bibinfo {pages} {100501} (\bibinfo {year}
  {2015})}\BibitemShut {NoStop}%
\bibitem [{\citenamefont {Johanning}\ \emph
  {et~al.}(2009{\natexlab{b}})\citenamefont {Johanning}, \citenamefont {Braun},
  \citenamefont {Timoney}, \citenamefont {Elman}, \citenamefont {Neuhauser},\
  and\ \citenamefont {Wunderlich}}]{M.Johanning2009}%
  \BibitemOpen
  \bibfield  {author} {\bibinfo {author} {\bibfnamefont {M.}~\bibnamefont
  {Johanning}}, \bibinfo {author} {\bibfnamefont {A.}~\bibnamefont {Braun}},
  \bibinfo {author} {\bibfnamefont {N.}~\bibnamefont {Timoney}}, \bibinfo
  {author} {\bibfnamefont {V.}~\bibnamefont {Elman}}, \bibinfo {author}
  {\bibfnamefont {W.}~\bibnamefont {Neuhauser}}, \ and\ \bibinfo {author}
  {\bibfnamefont {C.}~\bibnamefont {Wunderlich}},\ }\href {\doibase
  10.1103/PhysRevLett.102.073004} {\bibfield  {journal} {\bibinfo  {journal}
  {Phys. Rev. Lett.}\ }\textbf {\bibinfo {volume} {102}},\ \bibinfo {pages}
  {073004} (\bibinfo {year} {2009}{\natexlab{b}})}\BibitemShut {NoStop}%
\bibitem [{\citenamefont {Kim}\ \emph {et~al.}(2009)\citenamefont {Kim},
  \citenamefont {Chang}, \citenamefont {Islam}, \citenamefont {Korenblit},
  \citenamefont {Duan},\ and\ \citenamefont {Monroe}}]{Kim2009}%
  \BibitemOpen
  \bibfield  {author} {\bibinfo {author} {\bibfnamefont {K.}~\bibnamefont
  {Kim}}, \bibinfo {author} {\bibfnamefont {M.~S.}\ \bibnamefont {Chang}},
  \bibinfo {author} {\bibfnamefont {R.}~\bibnamefont {Islam}}, \bibinfo
  {author} {\bibfnamefont {S.}~\bibnamefont {Korenblit}}, \bibinfo {author}
  {\bibfnamefont {L.~M.}\ \bibnamefont {Duan}}, \ and\ \bibinfo {author}
  {\bibfnamefont {C.}~\bibnamefont {Monroe}},\ }\href {\doibase
  10.1103/PhysRevLett.103.120502} {\bibfield  {journal} {\bibinfo  {journal}
  {Phys. Rev. Lett.}\ }\textbf {\bibinfo {volume} {103}},\ \bibinfo {pages} {1}
  (\bibinfo {year} {2009})}\BibitemShut {NoStop}%
\bibitem [{\citenamefont {'t~Hooft}(2005)}]{Hooft2005}%
  \BibitemOpen
  \bibinfo {editor} {\bibfnamefont {G.}~\bibnamefont {'t~Hooft}},\ ed.,\
  \href@noop {} {\emph {\bibinfo {title} {50 years of Yang-Mills theory}}}\
  (\bibinfo  {publisher} {World Scientific},\ \bibinfo {year}
  {2005})\BibitemShut {NoStop}%
\bibitem [{\citenamefont {Piltz}\ \emph {et~al.}()\citenamefont {Piltz},
  \citenamefont {Sriarunothai}, \citenamefont {Ivanov}, \citenamefont
  {W{\"o}lk},\ and\ \citenamefont {Wunderlich}}]{Piltz2015}%
  \BibitemOpen
  \bibfield  {author} {\bibinfo {author} {\bibfnamefont {C.}~\bibnamefont
  {Piltz}}, \bibinfo {author} {\bibfnamefont {T.}~\bibnamefont {Sriarunothai}},
  \bibinfo {author} {\bibfnamefont {S.}~\bibnamefont {Ivanov}}, \bibinfo
  {author} {\bibfnamefont {S.}~\bibnamefont {W{\"o}lk}}, \ and\ \bibinfo
  {author} {\bibfnamefont {C.}~\bibnamefont {Wunderlich}},\ }\href
  {http://arxiv.org/abs/1509.01478} {\bibinfo  {journal} {arXiv:1509.01478v1}\
  }\BibitemShut {NoStop}%
\bibitem [{\citenamefont {Piltz}\ \emph {et~al.}(2014)\citenamefont {Piltz},
  \citenamefont {Sriarunothai}, \citenamefont {Var{\'{o}}n},\ and\
  \citenamefont {Wunderlich}}]{Piltz2014}%
  \BibitemOpen
\bibfield  {journal} {  }\bibfield  {author} {\bibinfo {author} {\bibfnamefont
  {C.}~\bibnamefont {Piltz}}, \bibinfo {author} {\bibfnamefont
  {T.}~\bibnamefont {Sriarunothai}}, \bibinfo {author} {\bibfnamefont {a.~F.}\
  \bibnamefont {Var{\'{o}}n}}, \ and\ \bibinfo {author} {\bibfnamefont
  {C.}~\bibnamefont {Wunderlich}},\ }\href {\doibase 10.1038/ncomms5679}
  {\bibfield  {journal} {\bibinfo  {journal} {Nature communications}\ }\textbf
  {\bibinfo {volume} {5}},\ \bibinfo {pages} {4679} (\bibinfo {year}
  {2014})}\BibitemShut {NoStop}%
\bibitem [{\citenamefont {Monz}\ \emph {et~al.}(2011)\citenamefont {Monz},
  \citenamefont {Schindler}, \citenamefont {Barreiro}, \citenamefont {Chwalla},
  \citenamefont {Nigg}, \citenamefont {Coish}, \citenamefont {Harlander},
  \citenamefont {H{\"{a}}nsel}, \citenamefont {Hennrich},\ and\ \citenamefont
  {Blatt}}]{Monz2011}%
  \BibitemOpen
  \bibfield  {author} {\bibinfo {author} {\bibfnamefont {T.}~\bibnamefont
  {Monz}}, \bibinfo {author} {\bibfnamefont {P.}~\bibnamefont {Schindler}},
  \bibinfo {author} {\bibfnamefont {J.~T.}\ \bibnamefont {Barreiro}}, \bibinfo
  {author} {\bibfnamefont {M.}~\bibnamefont {Chwalla}}, \bibinfo {author}
  {\bibfnamefont {D.}~\bibnamefont {Nigg}}, \bibinfo {author} {\bibfnamefont
  {W.~A.}\ \bibnamefont {Coish}}, \bibinfo {author} {\bibfnamefont
  {M.}~\bibnamefont {Harlander}}, \bibinfo {author} {\bibfnamefont
  {W.}~\bibnamefont {H{\"{a}}nsel}}, \bibinfo {author} {\bibfnamefont
  {M.}~\bibnamefont {Hennrich}}, \ and\ \bibinfo {author} {\bibfnamefont
  {R.}~\bibnamefont {Blatt}},\ }\href {\doibase 10.1103/PhysRevLett.106.130506}
  {\bibfield  {journal} {\bibinfo  {journal} {Phys. Rev. Lett.}\ }\textbf
  {\bibinfo {volume} {106}},\ \bibinfo {pages} {130506} (\bibinfo {year}
  {2011})}\BibitemShut {NoStop}%
\bibitem [{\citenamefont {Lucas}\ \emph {et~al.}()\citenamefont {Lucas},
  \citenamefont {Keitch}, \citenamefont {Home}, \citenamefont {Imreh},
  \citenamefont {McDonnell}, \citenamefont {Stacey}, \citenamefont {Szwer},\
  and\ \citenamefont {Steane}}]{Lucas2007}%
  \BibitemOpen
  \bibfield  {author} {\bibinfo {author} {\bibfnamefont {D.~M.}\ \bibnamefont
  {Lucas}}, \bibinfo {author} {\bibfnamefont {B.~C.}\ \bibnamefont {Keitch}},
  \bibinfo {author} {\bibfnamefont {J.~P.}\ \bibnamefont {Home}}, \bibinfo
  {author} {\bibfnamefont {G.}~\bibnamefont {Imreh}}, \bibinfo {author}
  {\bibfnamefont {M.~J.}\ \bibnamefont {McDonnell}}, \bibinfo {author}
  {\bibfnamefont {D.~N.}\ \bibnamefont {Stacey}}, \bibinfo {author}
  {\bibfnamefont {D.~J.}\ \bibnamefont {Szwer}}, \ and\ \bibinfo {author}
  {\bibfnamefont {A.~M.}\ \bibnamefont {Steane}},\ }\href
  {http://arxiv.org/abs/0710.4421} {\ }\Eprint {http://arxiv.org/abs/0710.4421}
  {arXiv:0710.4421} \BibitemShut {NoStop}%
\bibitem [{\citenamefont {Benhelm}\ \emph {et~al.}(2008)\citenamefont
  {Benhelm}, \citenamefont {Kirchmair}, \citenamefont {Roos},\ and\
  \citenamefont {Blatt}}]{Benhelm2008}%
  \BibitemOpen
  \bibfield  {author} {\bibinfo {author} {\bibfnamefont {J.}~\bibnamefont
  {Benhelm}}, \bibinfo {author} {\bibfnamefont {G.}~\bibnamefont {Kirchmair}},
  \bibinfo {author} {\bibfnamefont {C.~F.}\ \bibnamefont {Roos}}, \ and\
  \bibinfo {author} {\bibfnamefont {R.}~\bibnamefont {Blatt}},\ }\href
  {\doibase 10.1103/PhysRevA.77.062306} {\bibfield  {journal} {\bibinfo
  {journal} {Phys. Rev. A}\ }\textbf {\bibinfo {volume} {77}},\ \bibinfo
  {pages} {062306} (\bibinfo {year} {2008})}\BibitemShut {NoStop}%
\bibitem [{\citenamefont {Britton}\ \emph {et~al.}(2012)\citenamefont
  {Britton}, \citenamefont {Sawyer}, \citenamefont {Keith}, \citenamefont
  {Wang}, \citenamefont {Freericks}, \citenamefont {Uys}, \citenamefont
  {Biercuk},\ and\ \citenamefont {Bollinger}}]{Britton2012}%
  \BibitemOpen
  \bibfield  {author} {\bibinfo {author} {\bibfnamefont {J.~W.}\ \bibnamefont
  {Britton}}, \bibinfo {author} {\bibfnamefont {B.~C.}\ \bibnamefont {Sawyer}},
  \bibinfo {author} {\bibfnamefont {A.~C.}\ \bibnamefont {Keith}}, \bibinfo
  {author} {\bibfnamefont {C.-C.~J.}\ \bibnamefont {Wang}}, \bibinfo {author}
  {\bibfnamefont {J.~K.}\ \bibnamefont {Freericks}}, \bibinfo {author}
  {\bibfnamefont {H.}~\bibnamefont {Uys}}, \bibinfo {author} {\bibfnamefont
  {M.~J.}\ \bibnamefont {Biercuk}}, \ and\ \bibinfo {author} {\bibfnamefont
  {J.~J.}\ \bibnamefont {Bollinger}},\ }\href {\doibase
  doi:10.1038/nature10981} {\bibfield  {journal} {\bibinfo  {journal} {Nature}\
  }\textbf {\bibinfo {volume} {484}},\ \bibinfo {pages} {489} (\bibinfo {year}
  {2012})}\BibitemShut {NoStop}%
\bibitem [{\citenamefont {Bohnet}\ \emph {et~al.}(2015)\citenamefont {Bohnet},
  \citenamefont {Sawyer}, \citenamefont {Britton}, \citenamefont {Wall},
  \citenamefont {Rey}, \citenamefont {Foss-Feig},\ and\ \citenamefont
  {Bollinger}}]{Bohnet2015}%
  \BibitemOpen
  \bibfield  {author} {\bibinfo {author} {\bibfnamefont {J.~G.}\ \bibnamefont
  {Bohnet}}, \bibinfo {author} {\bibfnamefont {B.~C.}\ \bibnamefont {Sawyer}},
  \bibinfo {author} {\bibfnamefont {J.~W.}\ \bibnamefont {Britton}}, \bibinfo
  {author} {\bibfnamefont {M.~L.}\ \bibnamefont {Wall}}, \bibinfo {author}
  {\bibfnamefont {A.~M.}\ \bibnamefont {Rey}}, \bibinfo {author} {\bibfnamefont
  {M.}~\bibnamefont {Foss-Feig}}, \ and\ \bibinfo {author} {\bibfnamefont
  {J.~J.}\ \bibnamefont {Bollinger}},\ }\href@noop {} {\bibfield  {journal}
  {\bibinfo  {journal} {arXiv:1512.03756 [quant-ph]}\ } (\bibinfo {year}
  {2015})}\BibitemShut {NoStop}%
\end{thebibliography}%

\end{document}